\documentclass{article}%
\usepackage[a4paper, margin=1.5in]{geometry}
\usepackage[numbers,sort&compress]{natbib}
\usepackage{array}

\usepackage{tikz}
\newcommand{\Mod}[1]{\ (\mathrm{mod}\ #1)}
\newcommand{\Mdensity}{\ensuremath{\mathbf{P}_{\text{mit}}}}
\newcommand{\Ndensity}{\ensuremath{\mathbf{P}_{\text{noisy}}}}
\newcommand{\Idensity}{\ensuremath{\mathbf{P}_{\text{ideal}}}}
\newcommand{\Udensity}{\ensuremath{\mathbf{P}_{\text{uniform}}}}

\newcommand{\Circ}{\ensuremath{\textbf{C}}}
\newcommand{\Backendinfo}{\ensuremath{\textbf{B}}}
\newcommand{\Carray}
{\ensuremath{\mathbf{C}_{\text{array}}}}
\usepackage{xspace}
\usepackage{amsfonts}
\usepackage{amsmath}
\newcommand{\Ibmalgiers}{\texttt{ibm\_algiers}\xspace}
\newcommand{\Ibmhanoi}{\texttt{ibm\_hanoi}\xspace}

\usepackage{floatrow}
\usepackage{mdframed}  
\usepackage{longtable}
\usepackage{booktabs}        
\usepackage[dvipsnames]{xcolor}  
\newfloatcommand{capbtabbox}{table}[][\FBwidth]

\pdfoutput=1

\usepackage{caption}
\usepackage{tikz}
\usepackage{verbatim}  
\usepackage{booktabs} 
\usepackage{colortbl} 
\usetikzlibrary{quantikz2} 
\usetikzlibrary{matrix}
\usepackage{multirow} 
\usepackage{algorithm}
\usepackage{algpseudocode}
\usepackage{graphicx,subcaption} 
\usepackage{hyperref}
\usepackage[capitalise]{cleveref} 
\crefname{app}{Appendix}{Appendices}

\usepackage{authblk}

\title{Deep Learning Approaches to Quantum Error Mitigation}

\author[1,4,5]{Leonardo Placidi}
\author[2]{Ifan Williams}
\author[3,6]{Enrico Rinaldi}
\author[2]{Daniel Mills}
\author[2]{Cristina C\^irstoiu}
\author[2]{Vanya Eccles}
\author[2]{Ross Duncan}

\affil[1]{Quantinuum K.K., Otemachi Financial City Grand Cube 3F 1-9-2 Otemachi, Chiyoda-ku
Tokyo, Japan, 1000004}
\affil[2]{Quantinuum Ltd., Terrington House, 13–15 Hills Road, Cambridge CB2 1NL, United Kingdom}
\affil[3]{Quantinuum Ltd., Partnership House, Carlisle Place, London SW1P 1BX, United Kingdom}
\affil[4]{The University of Osaka, Graduate School of Engineering Science, 1-2, Machikaneyama, Toyonaka, Osaka 560-8531, Japan}
\affil[5]{QIQB Center for Quantum Information and Quantum Biology, The University of Osaka, 1-2 Machikaneyama, Toyonaka, Osaka 560-0043, Japan}
\affil[6]{Center for Quantum Computing, RIKEN, 2-1 Hirosawa, Wako, Saitama 351-0198, Japan}

\usepackage{titlesec}

\begin{document}

\maketitle

\begin{abstract}
We present a systematic investigation of deep learning methods applied to quantum error mitigation of noisy output probability distributions from measured quantum circuits. 
We  compare different architectures, from fully connected neural networks to transformers, and we test different design/training modalities, identifying sequence-to-sequence, attention-based models as the most effective on our datasets.
These models consistently produce mitigated distributions that are closer to the ideal outputs when tested on both simulated and real device data obtained from IBM superconducting quantum processing units (QPU) up to five qubits. 
Across several different circuit depths, our approach outperforms other baseline error mitigation techniques. 
We perform a series of ablation studies to examine: how different input features (circuit, device properties, noisy output statistics) affect performance; cross-dataset generalization across circuit families; and transfer learning to a different IBM QPU.
We observe that generalization performance across similar devices with the same architecture works effectively, without needing to fully retrain models.
\end{abstract}

\section{Introduction}
\label{sec:introduction}

Quantum error correction enables exponential noise suppression~\cite{aharonov08, knill98, kitaev03}, and has been partially demonstrated at small scales \cite{Bluvstein_2023,acharya2024quantumerrorcorrectionsurface,wang2024fault,paetznick2024demonstrationlogicalqubitsrepeated,Ryan-Anderson:2024zyh}. However practical-scale, fault-tolerant applications remain out of reach. Quantum error mitigation (QEM) can enhance the utility of noisy quantum hardware \cite{kim2023evidence,cai2023quantum, haghshenas2025digitalquantummagnetismfrontier}, and may be used to mitigate the effect of residual logical noise when using error correction on devices with limited sizes \cite{zhang2025demonstratingquantumerrormitigation}.  QEM does not improve quantum state fidelity, but instead reduces noise bias in estimating quantities of interest. There are a variety of QEM methods, but they typically have similar workflows \cite{cirstoiu2023volumetric}  that require minimal increase in circuit size, and rely mostly on classical post-processing of noisy samples from many different circuits. As a result, QEM requires exponentially more samples  \cite{takagi2022fundamental, quek2024exponentially}.

Machine learning (ML) has been applied to QEM in several ways \cite{du2025artificialintelligencerepresentingcharacterizing}.
ML can mimic existing error mitigation schemes with reduced overhead \cite{liao2023machine}, it can guide which modified circuits are run \cite{zlokapa2020deeplearningmodelnoise, PRXQuantum.2.040330}, or can be used to train a neural network to map noisy output samples to error-mitigated estimators \cite{Czarnik_2021, liao2023machine, cantori2024synergy, sack2024large, zhukov2024quantum, babukhin2023echo, 9226505, liao2023flexible, kim2022quantum, czarnik2022improving}. For instance, Clifford data regression \cite{Czarnik_2021, czarnik2022improving} uses classically simulatable circuits derived from the target circuit to learn a linear map from noisy to ideal expectation values. Other supervised learning approaches use feed-forward neural networks to mitigate expectation values for quantum simulation \cite{zhukov2024quantum, babukhin2023echo, cantori2024synergy} or combinatorial optimization \cite{sack2024large}. 

Supervised ML has also been used for mitigating noise in the
outcome probabilities. The main motivation to address this task is its greater generality: all observables in the computational basis can be determined from the full distribution over outcomes. While we can expect scalability bottlenecks at smaller systems than in the case of mitigating expectation values, this setting provides a worst-case benchmark to stress-test ML for QEM. 
Previous studies used ML to correct complex non-linear readout noise that is not mitigated by standard state preparation and measurement (SPAM) correction methods \cite{kim2022quantum}. 
Other works \cite{9226505} used shallow random circuits to train a neural network to map inputs, including the noisy distribution, error rates, circuit depth, and the numbers of single- and two-qubit gates, to a correction for each outcome probability.
While promising, these approaches are often demonstrated on restricted datasets and simple models, leaving open questions about how architectural choices influence generalizability and robustness, particularly on time-varying hardware data.

In this article, we perform a systematic study investigating the use of deep learning models to mitigate noise in quantum circuit outcome probabilities. 
Envisioning a pipeline where models are pre-trained and then continuously refined based on expanded datasets, we train the models using noisy simulation data, fine-tune using real hardware data, and also perform transfer learning to data run on another quantum device. 
We use a dataset of over $246,000$ unique circuits, which includes extensive hardware data collected over a period of time of several months. 
This is a markedly larger and more diverse dataset than has been used in prior work, enabling our models to account for time drift and noise fluctuations.
The dataset collection is facilitated by Quantinuum Nexus~\cite{quantinuum_nexus}. 
We use two classes of 5-qubit circuits for training and testing: circuits consisting of random gates, inspired by  random circuit sampling \cite{Boixo_2018}; and circuits consisting of random Pauli gadgets \cite{Cowtan2019loc}, motivated by  dynamical simulation. The dataset generation is described in \cref{sec:data-preparation}.

We compare a much wider selection of deep learning architectures and training methods than have been investigated to date, including models based on two output types: \emph{prediction models}, which output the mitigated probability distribution, and \emph{correction models}, which output corrections to the noisy probability distributions in the form of an element-wise convolution.
For each dataset, we separately optimize the models through extensive hyperparameter searches, as described in \cref{sec:experiment-design}.
Our study shows that the prediction models perform best, and that attention-based models are particularly effective, improving the accuracy of the probability distributions for up to $\sim 80 \%$ of the circuits in our real hardware test sets. We also compare the strategy learned by our models to simple error mitigation schemes that modify the noisy probability distribution using the available device noise data (without requiring additional circuit-dependent error characterization experiments).
Our models outperform these approaches across a range of circuit depths.  The full comparative analysis results are shown in \cref{sec:results-all}.

The models take the following input features: an encoding of the circuit; the noisy probability distribution; and a characterization of the device (taken at the most recent round of device calibration). To probe how the inputs affect the performance of each model, we perform ablation studies that selectively randomize different inputs. We demonstrate that the noisy probability distribution is by far the most important feature used by the models, with circuit and backend information acting only as secondary conditioning factors. Furthermore, removing the circuit information does not lead to degradation in performance for most of the models (i.e recurrent neural networks [RNNs] or attention-based), indicating they have not learned to simply simulate the original circuit.

We also perform transfer learning studies between different classes of circuits and different hardware devices in \cref{sec:additional-study}. Among the architectures we studied, generative attention-based models are the most robust to dataset and hardware shifts. 
Models trained with measurements from one device (\Ibmalgiers{}) transfer effectively to another (\Ibmhanoi{}), without the need to retrain from scratch, as long as the hardware characterization data remains comparable. This may be explained by the importance of noisy output statistics on performance. Transferring between random and Pauli circuit types is less successful, demonstrating that the models are dependent on the circuit type due to differences in their output distributions.

Finally, we emphasize that this study is limited to quantum circuits with few qubits, for which the true outcome probability distributions are known.  This restriction allows us to isolate the effects of input features, training strategy, and model architectures in a controlled setting. Consequently, our analysis focuses on how these different architectural choices influence robustness and generalization. Some of the insights above may persist at scale, but otherwise would have been difficult to separate from scaling aspects. In particular, the exponential dimensionality of the full probability distributions, and resolving its discriminating noise features with limited finite samples, will pose significant scaling challenges. However, our results suggest guiding principles to improve the architectural design of ML-based error mitigation.  In \cref{sec:conclusion} we provide a detailed discussion of scalability constraints and suggest how to address these in future work by adapting our analysis to quantities like marginal distributions or local observables.

\section{Data}
\label{sec:data-preparation}

In this section, we discuss the data used to train our ML models, describing how our input data and labels are generated and post-processed.
The main features that we use are the following: 
\begin{itemize}
\item \Circ{} - an encoding of the circuit to be run, described in \cref{sec:data circuits}; 
\item \Backendinfo{} -  the  QPU properties and error calibration data, described in \cref{sec:data-backend-information}; 
\item \Ndensity{} - the noisy output probability distribution of the circuit.  
\end{itemize} 
\Idensity{}, a circuit's ideal probability distribution obtained from classical simulation, is used as a label in the supervised training process to compute the loss function and evaluate performance. 
\Ndensity{}, and \Idensity{}, are described in \cref{sec:data-noisy-distribution}.

\subsection{Circuits}
\label{sec:data circuits}

We use two application-motivated classes of circuits \cite{Mills2021application, cirstoiu2023volumetric}. These are both built iteratively in blocks, which we call \emph{time steps}, $t$, (which differ between circuit classes), up to a maximum time step, $T$. We vary $T$ to obtain circuits with many different depths and gate numbers: 
\begin{description}
    \item[Pauli] Synthetic versions of the product formula circuits seen in Hamiltonian simulation and quantum chemistry~\cite{berry2005ham,peruzzo13, Barkoutsos2018igw}. 
    At each time step, a Pauli gadget \cite{Cowtan2019loc} of the form $\exp(-i\alpha P)$, where $P \in \left\{ X, Y, Z, I \right\}^n$, and the angle, $\alpha$, are randomly chosen, is applied to the circuit. 
    The maximum time steps considered were $T \in \{3, 4, 5, 6, 7, 9\}$. 
    See \cref{algpauli} for the complete description.
    \item[Random] These are random circuits, seen in applications such as random circuit sampling \cite{Aaronson2016guw}. 
    At each time step, a gate is randomly selected from the IBM native gateset 
    \begin{align}
        \label{eq:gates}
        \mathcal{G} = \{X, SX, R_{z}(\alpha), CX\},
    \end{align}
    and applied to a randomly selected qubit, or qubits in the case of the two-qubit $CX$ gate. 
    The maximum time steps considered are $T \in \{48, 64, 80, 96, 112, 144\}$.  
    See \cref{algrandom} for the complete description.%
\end{description}

We use Quantinuum's \textsc{tket} compiler \cite{Sivarajah_2021} to compile each circuit onto the native connectivity and gate set of IBM devices. 
No further optimization is performed. 
Each circuit is then converted into a 3D array, \Carray{}, of shape, $(n_l,n_\text{qubits},|\mathcal{G}|) = (n_l,5,5)$. 
Each of the $n_l$ layers in the array has an associated $n_\text{qubits}\times |\mathcal{G}|$ matrix, with each qubit representing a row in this matrix (indexed starting from 1), and each operation a column.
The total number of layers depend on $T$ and on the type of circuit, due to the fact that each layer may have multiple parallel operations on $n_\text{qubits}$ qubits.
We use an encoding of each of the possible gates, as specified in \cref{tab:one_hot_encoding}.
To create a homogeneous dataset of pre-defined shapes, \Carray{} is padded with $n_\text{qubits} \times |\mathcal{G}|$ zero matrices in order to be the same shape as the \Carray{} of the maximal depth circuit with the largest $T$ value.

\begin{table}[!ht]
    \label{tab:circuit encoding}
    \centering
    \begin{tabular}{c|c}
        \textbf{Gate type} & \textbf{Encoding} \\
        \hline 
        $X$          & $(1, 0, 0, 0, 0)$  \\
        $SX$         & $(0, 1, 0, 0, 0)$  \\
        $R_{z}(\alpha)$ & $(0, 0, (\alpha\Mod{2})/2, 0, 0)$  \\
        $CX$ (Control) & $(0, 0, 0, -\text{(1+(target-qubit index))}, 0)$ \\
        $CX$ (Target) & $(0, 0, 0, \text{1+(control-qubit index)}, 0)$ \\
        Measure    & $(0, 0, 0, 0, 1)$  \\
        Idle  &  $(0, 0, 0, 0, 0)$ \\
    \end{tabular}
     \caption{\textbf{Encoding scheme for 5-qubit circuits compiled into the IBM native gateset $\mathcal{G}$ given in \cref{eq:gates}.} $\alpha$ is a real number representing an angle in units of $\pi$.}\label{tab:one_hot_encoding}
\end{table}

\subsection{Device Information}
\label{sec:data-backend-information}

We used two 27-qubit superconducting transmon devices from IBM, \Ibmalgiers{} and \Ibmhanoi{}, of the same quantum chip generation. 
The backend information, \Backendinfo{}, consists of a bundle of parameters, retrieved via a public API \cite{ibmbackendinfo}, giving the calibrated properties of the QPU at a particular time:

\begin{description}
    \item[Qubit properties:] Qubit frequencies, and the $\text{T}_1$ and $\text{T}_2$ decay times.
    \item[Instruction properties:] Gate times, gate error rates, and measurement error rates.
\end{description}

\begin{figure}[!htbp]
    \centering
    \begin{subfigure}[t]{0.3\textwidth}
        \centering
        \includegraphics[width=\linewidth]{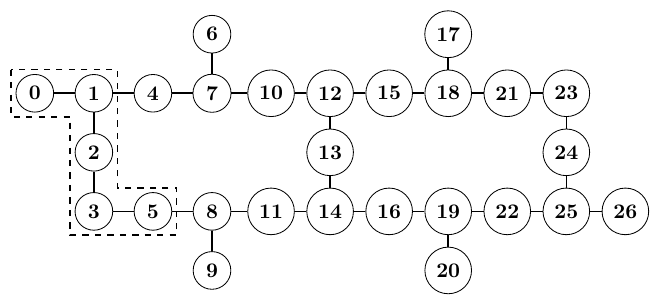}
        \caption{Topology of \Ibmalgiers{} and \Ibmhanoi{}. The dashed line encloses the five qubits used in this article.}
        \label{fig:subfig1}
    \end{subfigure}
    \hfill
    \begin{subfigure}[t]{0.65\textwidth}
        \centering
        \includegraphics[width=\linewidth]{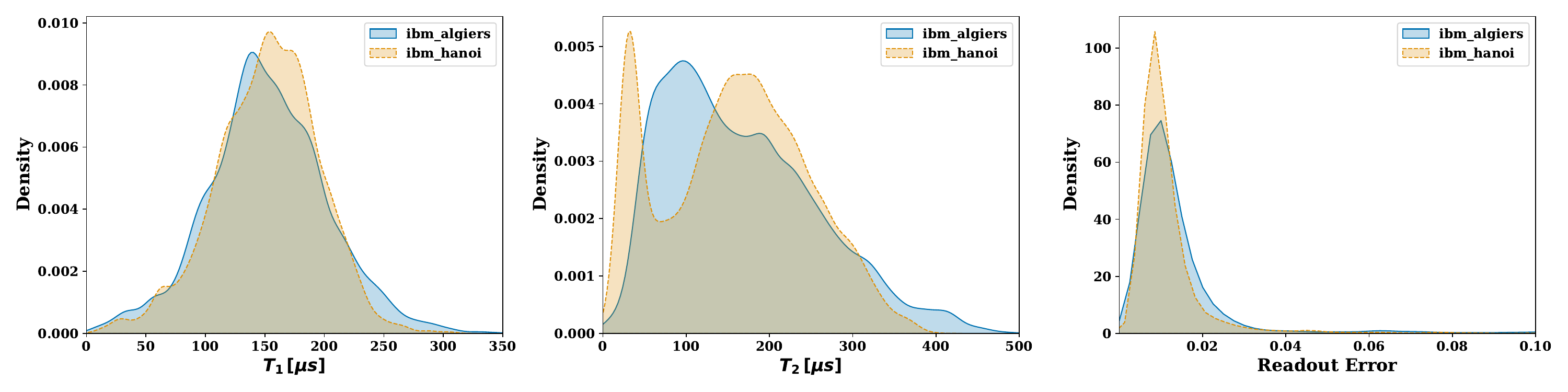}
        \caption{Probability density functions of coherence times $\text{T}_1$, $\text{T}_2$ and readout errors for \Ibmalgiers{} and \Ibmhanoi{}, averaged across qubits.}
        \label{fig:subfig2}
    \end{subfigure}
    
    \vskip 0.5cm
    \begin{subfigure}{0.8\textwidth}
        \centering
        \includegraphics[width=\linewidth]{{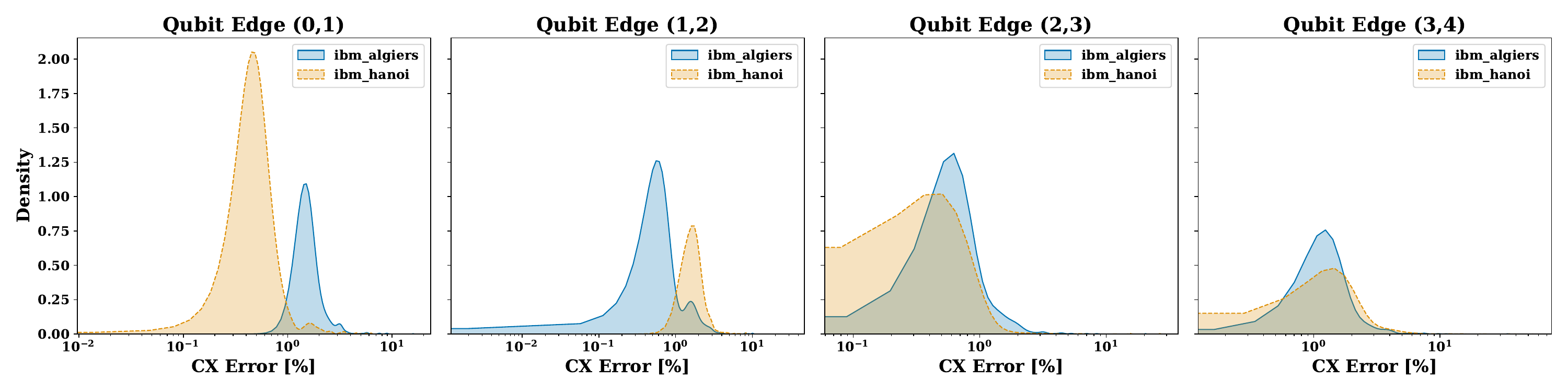}}
        \caption{Probability density functions of \(CX\) error rates between each pair of qubits using a logarithmic scale.}
        \label{fig:subfig3}
    \end{subfigure}

    \caption{\textbf{Device properties used as an input feature \Backendinfo{} for our models.} Densities are normalised across backend instances and include temporal variations in calibration data between periodic circuit executions.}
    \label{fig:mainfigure}
\end{figure}

We select a fixed subset of five qubits of the device, as shown in \cref{fig:subfig1}. 
We discard the parameters that refer to unused parts of the QPU, leaving 101 relevant parameters, which we store as a 1D array, \Backendinfo{}. 
In \cref{fig:subfig2,fig:subfig3} we show the distribution of device properties in our data set for both \Ibmalgiers{} and \Ibmhanoi{}.
These distributions are generated by collecting calibration data from the first online date of the device until the date of the data-taking period, with an approximate daily frequency.
Therefore, they are a collective representation of how device characteristics change with time and also with space (on the device).
We observe that the two-qubit error rates on edges $(0,1)$ and $(1,2)$ have distributions with only small overlap between the two devices, while other calibration data is more similarly distributed.

\subsection{Output Probability Distributions}
\label{sec:data-noisy-distribution}

We have two kinds of datasets based on the aforementioned circuit types and devices: one composed of \emph{simulated data}, denoted as ``Simulated'', and one composed of \emph{real data}, denoted as ``Real''. 
In the Simulated datasets, \Ndensity{} is obtained from a noisy classical simulation with the noise model parameters given by a randomly chosen \Backendinfo{} for \Ibmalgiers{}, collected at some random point in time. 
In the Real datasets, \Ndensity{} is obtained from running the circuit $\textbf{C}$ on a real quantum device, and \Backendinfo{} stores the relevant calibration data provided by the API at runtime.

Each circuit in the Real datasets is run three different times (repeats), with the repeats spaced out in time in a staggered fashion during the data-taking period (January-April 2024).
For Simulated datasets, the different repeats are achieved by sampling the device calibration data at different points in time (across the entire life span of the machines) when building the noise models.
This ensures that the datasets contain sufficient variability to effectively characterize the correlation between \Backendinfo{} and the true noise on the device, independent of the correlation of the circuit structure with the noiseless output. 
$15,000$ unique circuits are run per $T$ value,\footnote{Note that for \Ibmalgiers{} not all jobs completed before the machine was retired, and therefore the total numbers of unique circuits ended up being less than this for some $T$ values.} using batches of $300$ circuits with $20,000$ shots per circuit, with submission spaced out across the entire data-taking period. 
In the case of Simulated data, $8,000$ unique circuits are run per $T$ value, and we again run $20,000$ shots of each circuit on the noisy simulator, with each circuit also run for three repeats. 
See \cref{app:data} for complete details on the number of circuits of each type.

\begin{figure}[t]
\vspace{-0.2cm}
    \centering
    \begin{subfigure}[t]{0.49\textwidth}
        \includegraphics[width=1\linewidth]{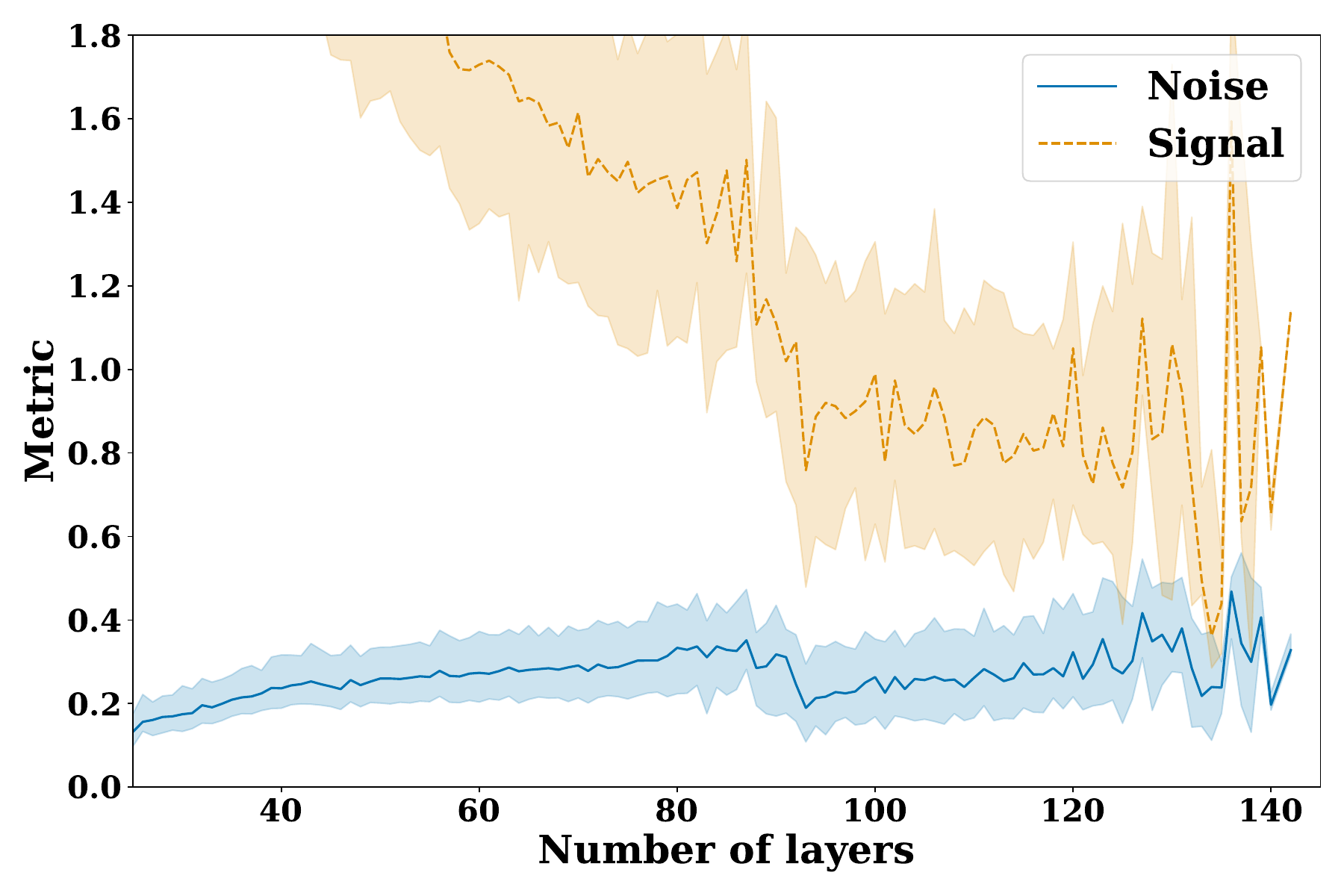}
        \subcaption{Pauli Simulated data.}
    \end{subfigure}
    \begin{subfigure}[t]{0.49\textwidth}
        \includegraphics[width=1\linewidth]{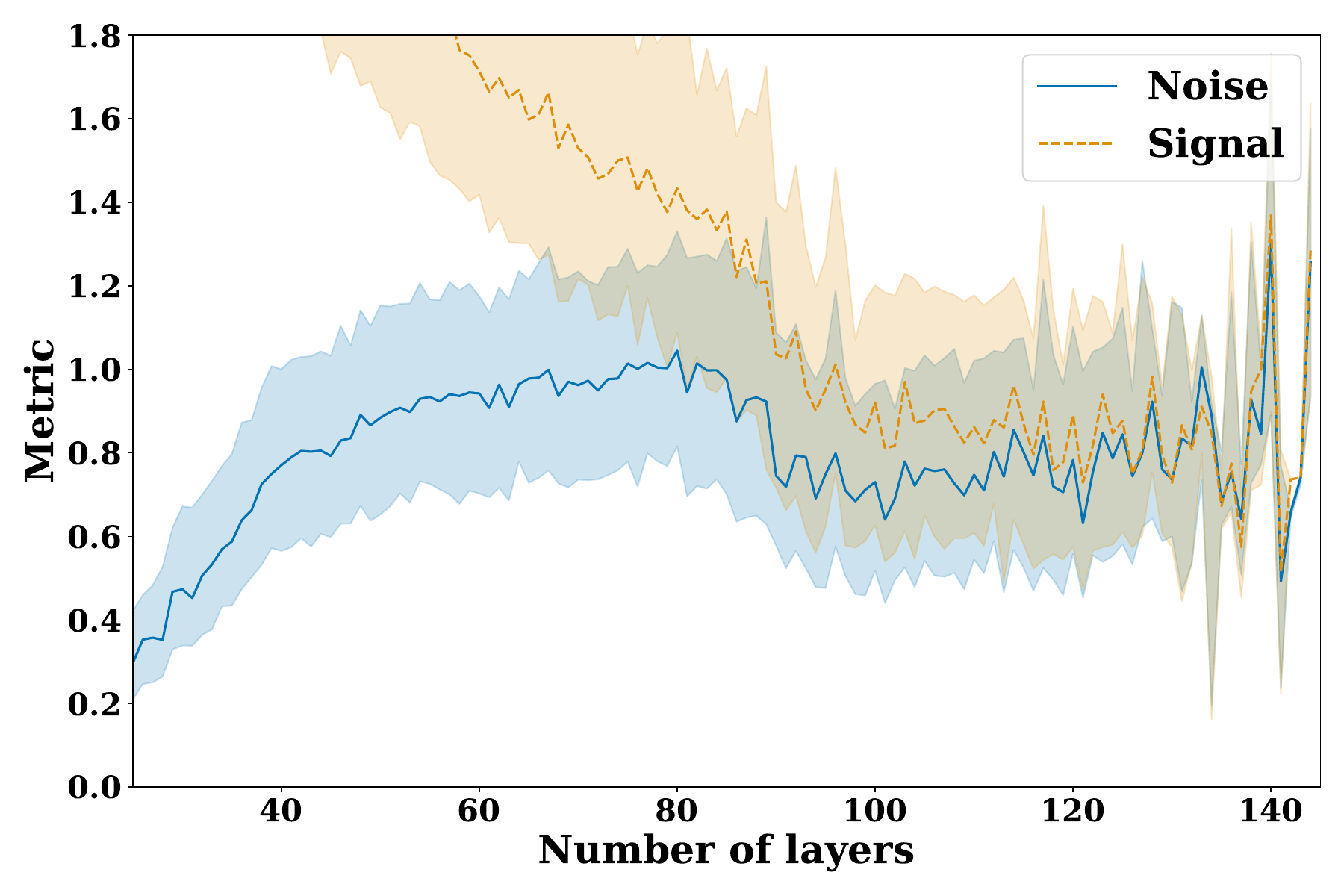}
        \subcaption{Pauli Real data.}
    \end{subfigure}
    \begin{subfigure}[t]{0.49\textwidth}
        \includegraphics[width=1\linewidth]{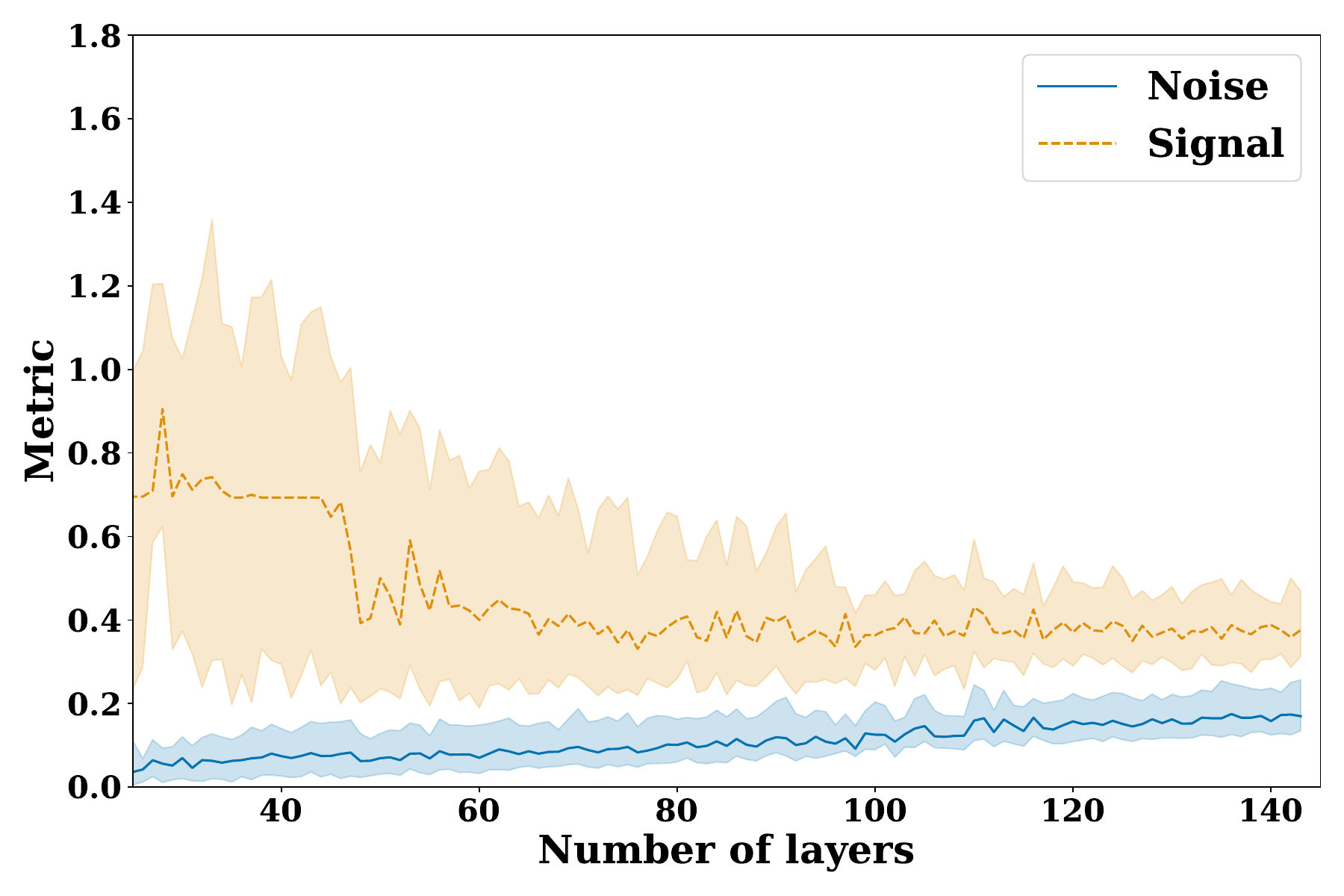}
        \subcaption{Random Simulated data.}
    \end{subfigure}
    \begin{subfigure}[t]{0.49\textwidth}
        \includegraphics[width=1\linewidth]{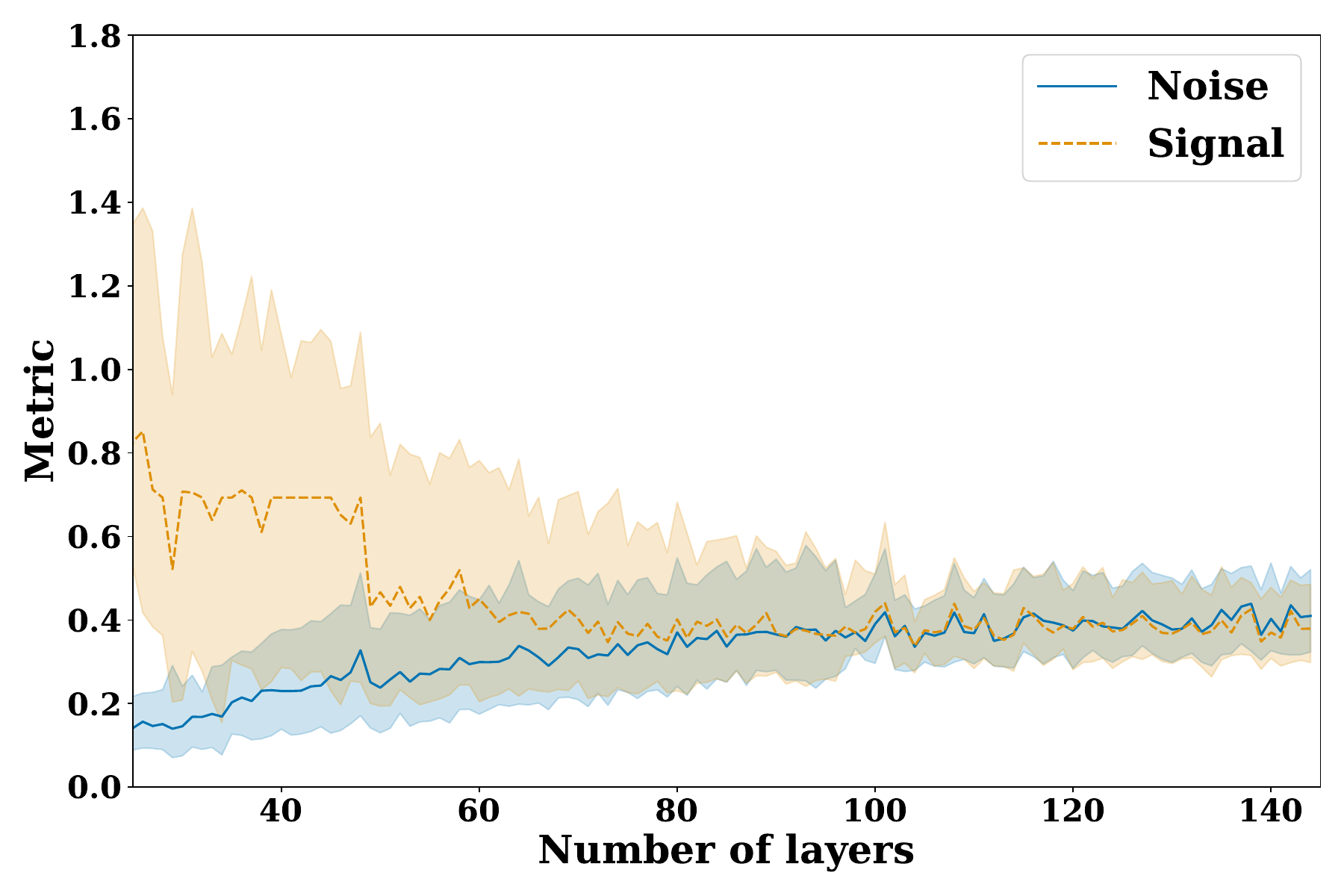}
        \subcaption{Random Real data.}
    \end{subfigure}
    \caption{\textbf{Noise and signal as a function of the number of layers in the circuit for \Ibmalgiers{} data.}}
    \label{fig:signal-to-noise}
\end{figure}

Qualitatively, we expect error mitigation to work in a regime where the circuit incurs a significant level of noise, but not too much that it completely dominates the computation. 
To quantify this in our datasets, we refer to the Kullback-Leibler \cite{kullback1959information} (KL) divergence between the ideal and noisy distribution, $\text{KL}(\Idensity{} , \Ndensity{})$,  as \emph{noise} -- and to the KL-divergence between the ideal distribution and the uniform distribution, $\text{KL}(\Idensity{} , \Udensity{})$, as  \emph{signal}. 
In \cref{fig:signal-to-noise} we show the signal and noise as a function of circuit layers, a proxy for the number of two-qubit gates, which is the dominating source of noise in the systems.

\begin{figure}[!t]
    \centering
    \begin{subfigure}[t]{0.49\textwidth}
        \includegraphics[width=0.95\linewidth]{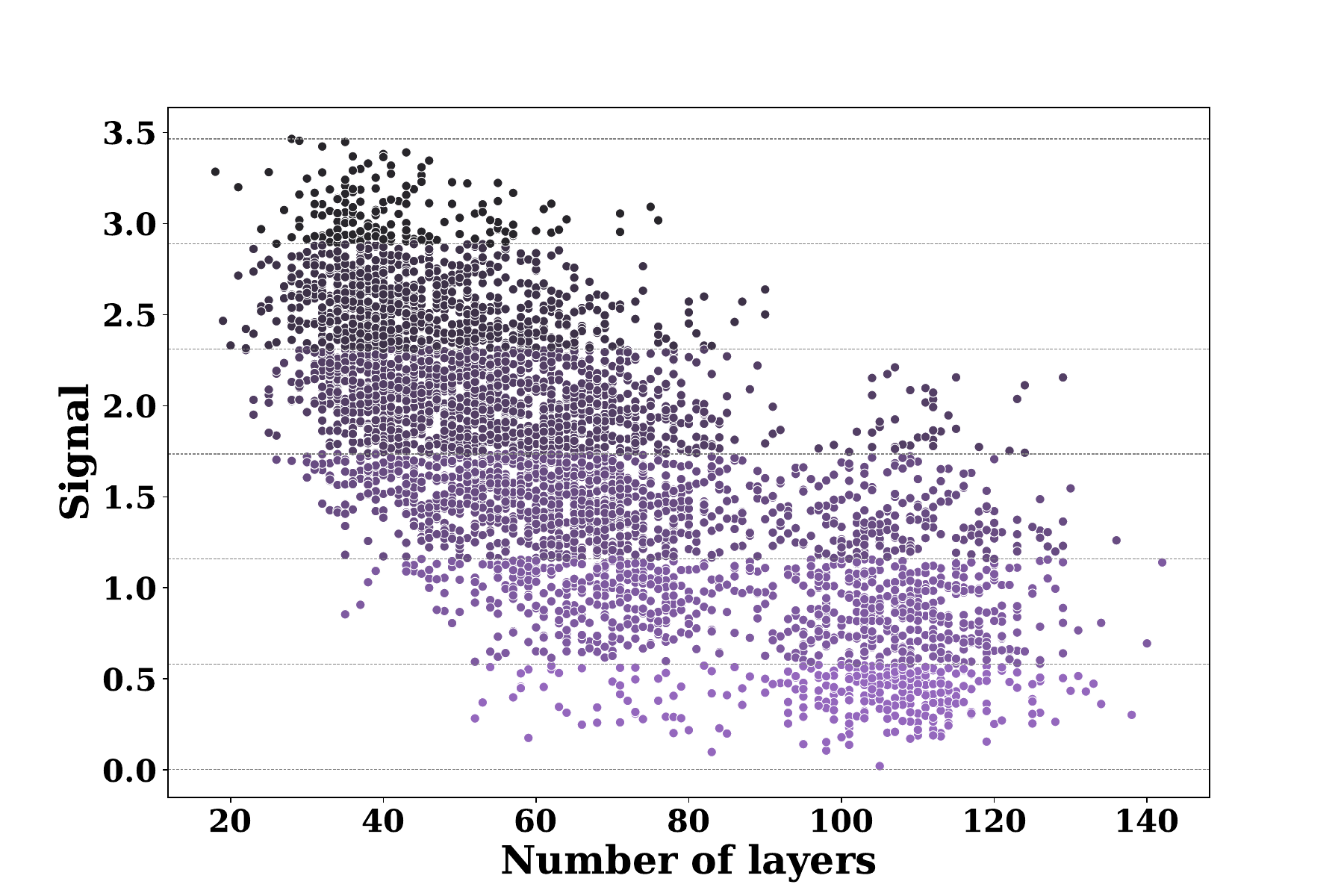}
        \subcaption{Pauli Simulated data.}
    \end{subfigure}
    \begin{subfigure}[t]{0.49\textwidth}
        \includegraphics[width=0.95\linewidth]{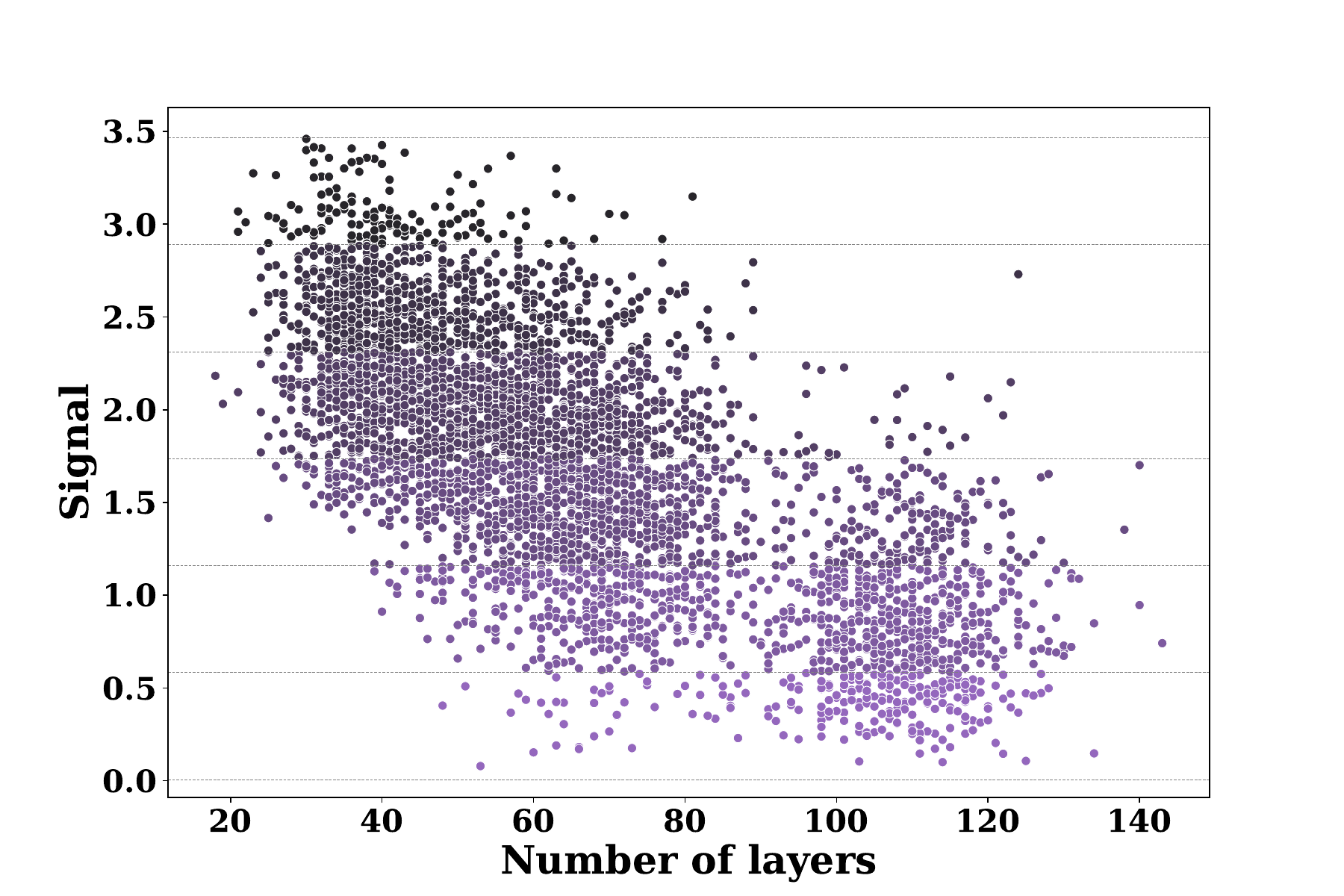}
        \subcaption{Pauli Real data.}
    \end{subfigure}
    \begin{subfigure}[t]{0.49\textwidth}
        \includegraphics[width=0.95\linewidth]{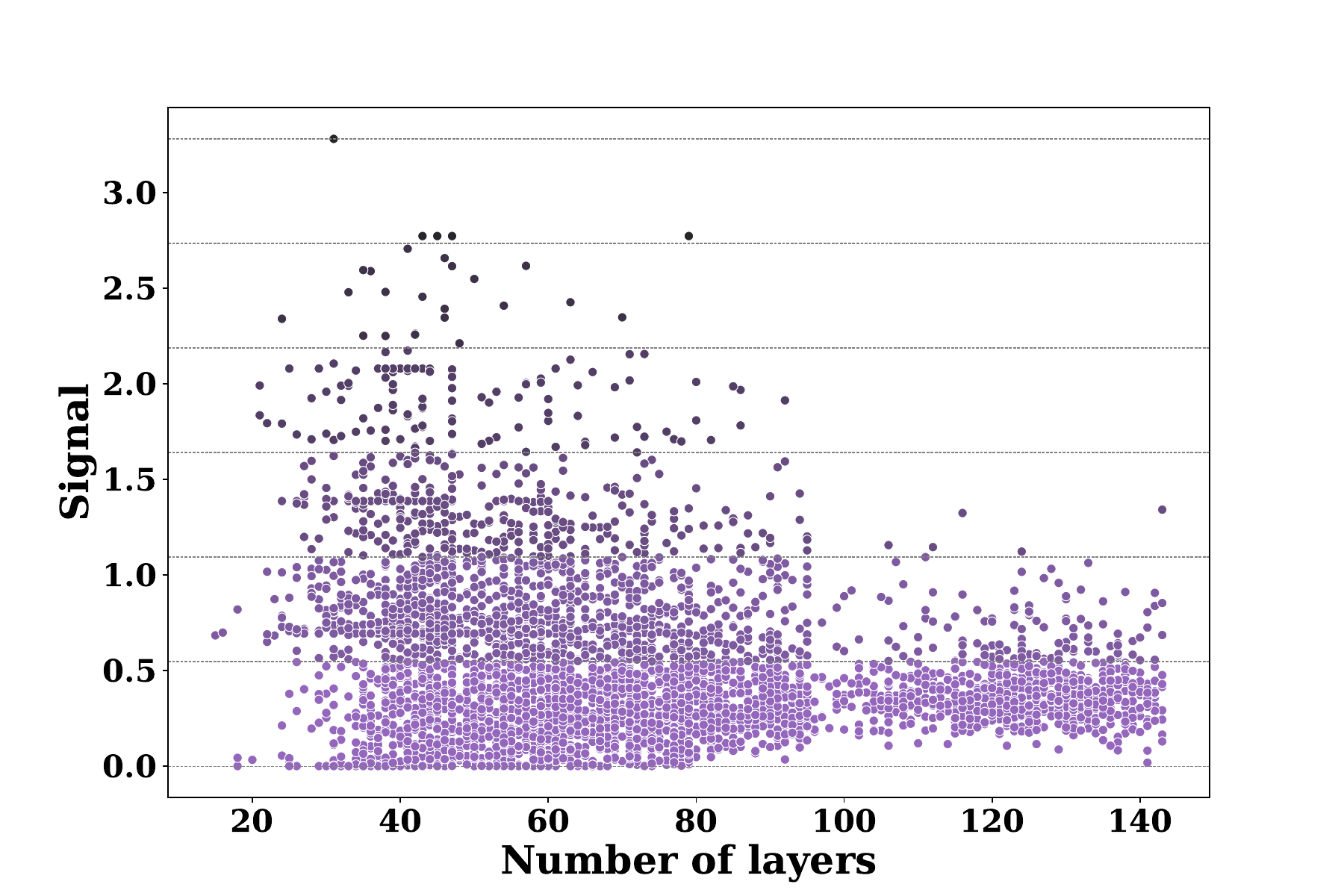}
        \subcaption{Random Simulated data.}
    \end{subfigure}
    \begin{subfigure}[t]{0.49\textwidth}
        \includegraphics[width=0.95\linewidth]{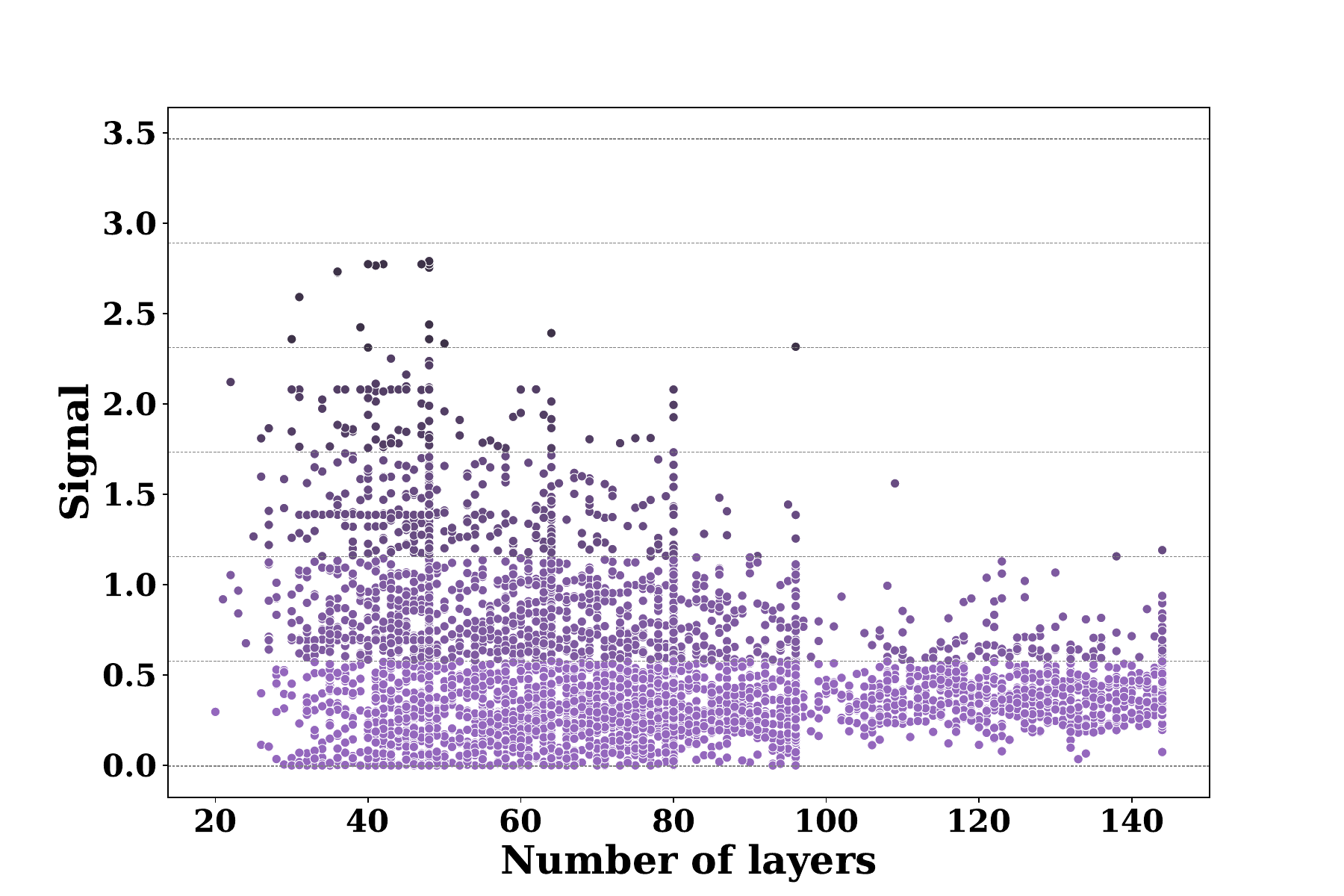}
        \subcaption{Random Real data.}
    \end{subfigure}
    \caption{\textbf{Scatter plots of signal} \textbf{against the number of layers for \Ibmalgiers{} data.} The signal value is divided into six equally-spaced bins, whose edges are represented by dashed gray lines. Here we plot 10,000 points randomly sampled from each test dataset.}
    \label{fig:signal-vs-layers}
\end{figure}

Of note in \cref{fig:signal-to-noise,fig:signal-vs-layers} is that the signal in Random circuits is smaller on average than for Pauli circuits. 
This is to be expected because Random circuits converge more quickly to a Porter-Thomas distribution \cite{mullane2020samplingrandomquantumcircuits}, whereas Pauli circuits tend to produce more peaked distributions, on average. 
We expect this to impact the performance of our mitigation methods, because at a lower signal it will be more difficult for the model to extract discriminating features in \Ndensity{}.

The second feature of note is the region where signal and noise are similar in value, which is to say the region where the ideal distribution is as comparably close to the noisy distribution as it is to the uniform distribution, or where the noisy distribution has become uniform.
In this region, we expect QEM to perform less effectively, because it will be hard to extract useful information from \Ndensity{}. 
From \cref{fig:signal-to-noise}, this threshold occurs after approximately $80$ layers (when running both types of circuits on real hardware). 
However, for Random circuits, there is still a significant overlap, even for a smaller number of layers.

Finally, from \cref{fig:signal-vs-layers} we see that the number of circuits with high signal is relatively small, and concentrated at lower numbers of layers, for both types of circuit. 
As such, while we expect the models to perform well on average for high-signal circuits, we would also expect the variance to be higher.

\section{Machine Learning Models and Training}
\label{sec:experiment-design}

\begin{figure}[!htbp]
    \centering
    \includegraphics[width=1\linewidth]{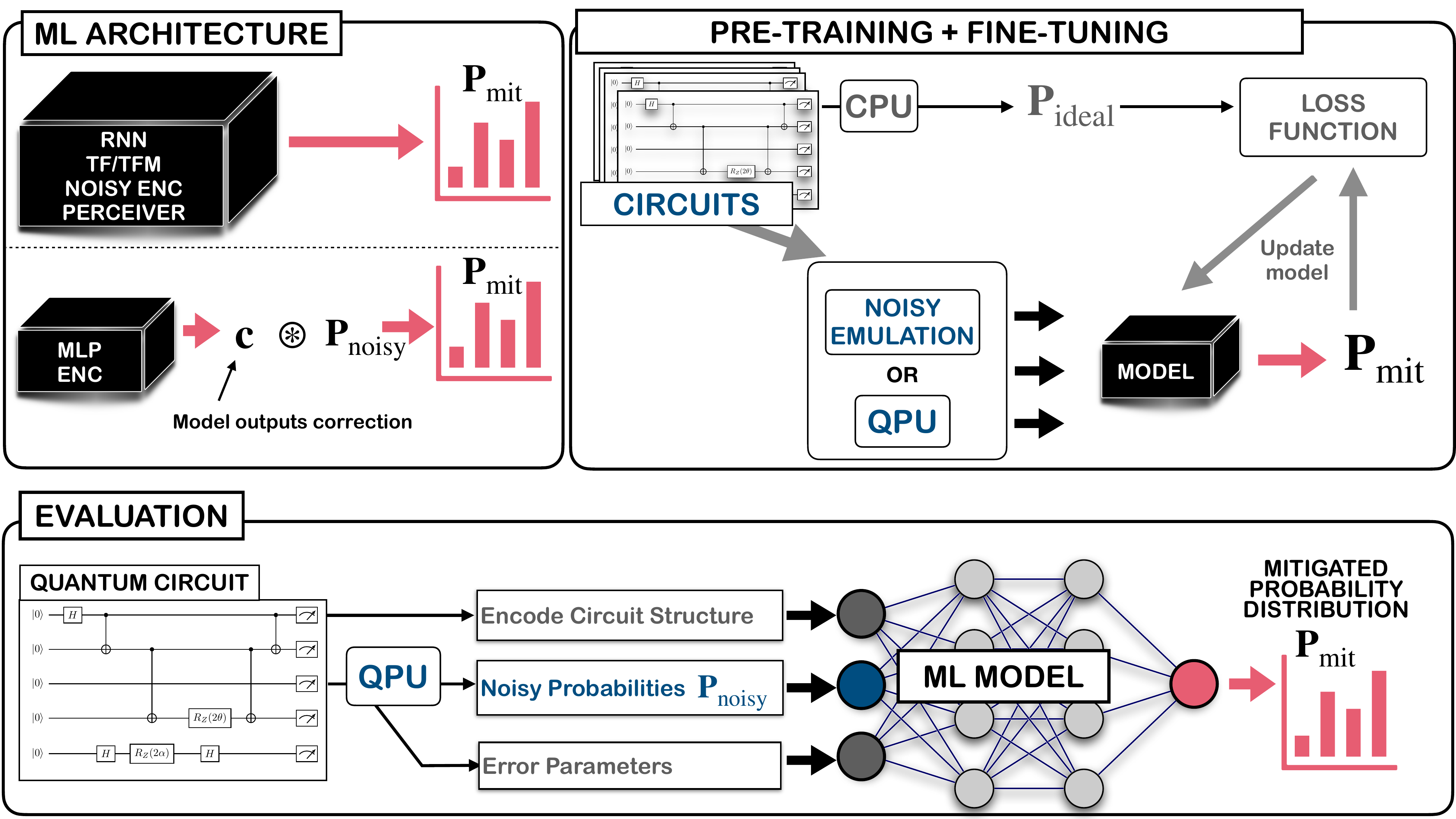}
    \caption{\textbf{ML pipeline for mitigating errors in a quantum circuit's output probability distribution.}} 
    \label{fig:pipelinefull}
\end{figure}

We detail the ML models that we investigate in \cref{sec:error mitigation with ml}, and the training pipeline that we use in \cref{sec:experiment-methodology}. 
Our overall experiment design is outlined in \cref{fig:pipelinefull}. 

\subsection{Models}
\label{sec:error mitigation with ml}

Throughout this subsection we use the following shorthand:
\begin{itemize}
    \item $\mathbf{x}_{\mathrm{CB}} \in \mathbb{R}^{3701}$ denotes the flattened circuit--backend feature vector (i.e. circuit and backend information).
    \item $\mathbf{X}_{\mathrm{CB}} \in \mathbb{R}^{n_{l} \times 132}$ denotes the embedded layer-wise circuit--backend sequence.\footnote{This representation includes a small zero padding at the end of the input vector, padding 126 real features up to a fixed size of 132.
    The padded dimensions carry no information and are ignored by the models.}
    \item $\Ndensity \in \mathbb{R}^{32}$ denotes the noisy probability distribution.
    \item $\Mdensity \in \mathbb{R}^{32}$ denotes the mitigated probability distribution.
\end{itemize}

We categorize the models we train into two main classes, prediction and correction, according to their output type.
All model architectures, training workflows, and hyperparameters are detailed in \cref{sec:models-details}.

\subsubsection*{Prediction Models}

Prediction models output the mitigated probability distribution as
\begin{equation}
\Mdensity = \text{Model} \left(\Circ, \Backendinfo{}, \Ndensity{} \right).
\end{equation}  
These models aim to learn a direct mapping between the noisy and ideal outcomes.
The following architectures are explored:

\begin{description}

    \item[Multilayer Perceptron (MLP-PREDICTION):]
    A baseline feedforward architecture~\cite{patternrecognbishop} that processes circuit, backend, and noisy output distribution features jointly.  
    The input vectors for the three features are first flattened and concatenated to give
    \begin{equation}
    \mathbf{z}
        = [\mathbf{x}_{\mathrm{CB}};\,\Ndensity]
        \in \mathbb{R}^{3701+32}.
    \end{equation}
    This is then passed through a sequence of $k$ dense layers, $h_{k}$, with batch normalization (BN) and non-linearities
    \begin{align}
    \mathbf{h}_0 &= \mathbf{z}, \nonumber \\ 
    \mathbf{h}_{k+1} &= \sigma\!\left( \mathrm{BN}\!\left( W_k \mathbf{h}_k + b_k \right) \right),
    \qquad k = 0,1,\dots,K-1 .
    \end{align}
    where $\sigma$ is an element-wise activation function, $W_k$ is the weight matrix of layer $k$, and $b_k$ its corresponding bias vector. This is then passed to a final linear layer, followed by a softmax\footnote{The softmax function is defined component-wise as
    $
    \mathrm{softmax}(\mathbf{y})_i
    =
    e^{y_i}/(\sum_j e^{y_j})$.} layer over the $32$-dimensional outcome space of the target noiseless distribution
    \begin{equation}
    \Mdensity = \mathrm{softmax}(W_{\mathrm{out}} \mathbf{h}_{\mathrm{final}} + b_{\mathrm{out}}).
    \end{equation}
    
    The MLP thus implements the direct mapping
    \begin{equation}
    (\mathbf{x}_{\mathrm{CB}},\Ndensity)
    \;\longmapsto\;
    \Mdensity,
    \end{equation}
    without sequential modelling, attention, or latent-space aggregation. 
    
    Compared to the models that follow, the MLP applies non-linear function approximations to flattened inputs only, and does not leverage the inherent structure of quantum circuits.
    Despite its architectural simplicity, the MLP provides an essential baseline for quantifying the benefit of the more structured inductive biases present in the following models. 
    Further details are given in \cref{app:mlp-models-training}.

    \item[Recurrent Neural Network (RNN):]
    A two-stream RNN architecture~\cite{rnns, patternrecognbishop} that encodes \Circ{},  \Backendinfo{}, and \Ndensity{} separately, then fuses their representations to predict a mitigated $32$-dimensional distribution. 
    
    The circuit and backend information are represented as a embedded sequence over $n_l$ layers such that $\mathbf{X}_{\mathrm{CB}} \in \mathbb{R}^{n_{l} \times 132}$,
    which is processed by the first RNN stream
    \begin{equation}
    \mathbf{H}_{\mathrm{circ}} = \mathrm{RNN}_{\mathrm{circ}}(\mathbf{X}_{\mathrm{CB}})
        \in \mathbb{R}^{n_{l} \times h},
    \end{equation}
    and its final hidden state retained $\mathbf{h}_{\mathrm{circ}} = \mathbf{H}_{\mathrm{circ}}[n_{l}] \in \mathbb{R}^{h}$.
    In parallel, the noisy output distribution is processed by a second RNN stream, which directly produces a hidden representation
    \begin{equation}
    \mathbf{h}_{\mathrm{noisy}} = \mathrm{RNN}_{\mathrm{noisy}}(\Ndensity) \in \mathbb{R}^{h}.
    \end{equation}
    This representation is concatenated with the final hidden state of the circuit–backend stream, and mapped to a probability vector via a fully connected layer, followed by a softmax layer

    \begin{equation}
    \Mdensity
        = \mathrm{softmax}\!\bigl(
            W\,[\mathbf{h}_{\mathrm{circ}};\mathbf{h}_{\mathrm{noisy}}] + b
          \bigr),
    \end{equation}
    where $W$ and $b$ are again weight matrices and bias vectors.
    Thus, the RNN implements
    \begin{equation}\label{eq:rnn}
(\mathbf{X}_{\mathrm{CB}}, \Ndensity)
        \longmapsto \Mdensity,
    \end{equation}
    in a way that explicitly models depth- or time-dependent correlations in the circuit and backend, while conditioning on the noisy output distribution.  
    Further details are provided in \cref{app:rnn-model-training}.

    \item[Transformer (TF):]
    An encoder-decoder transformer model featuring an attention mechanism \cite{NIPS2017_AttentionIsAllYouNeed}.
    The input data consists of $\mathbf{X}_{\mathrm{CB}}$ alongside 
    $\Ndensity$.
    Because the encoder and decoder operate on different feature dimensions, the architecture must be adapted to allow the model to predict a single $32$-dimensional mitigated distribution.
    
    The encoder, $\textit{Enc}$, processes the circuit sequence using standard self-attention mechanisms to capture global correlations among circuit layers, gates, and backend features.  
    The encoder output is then projected into the 32-dimensional decoder feature space via a linear
    map $f_c$
    \begin{align}
\mathbf{H}_{\mathrm{enc}} 
    &= \mathrm{Enc}(\mathbf{X}_{\mathrm{CB}}) 
    \in \mathbb{R}^{n_{l} \times d_{\mathrm{enc}}}, \nonumber \\
\tilde{\mathbf{H}}_{\mathrm{enc}} 
    &= f_c(\mathbf{H}_{\mathrm{enc}}) 
    \in \mathbb{R}^{n_{l} \times 32},
\end{align}
    where $d_{enc}$ is the latent-space dimension. The decoder, $\textit{Dec}$, takes as input the noisy output distribution as a length-one sequence, and using causal masking, generates a hidden representation conditioned on the encoded circuit
    \begin{equation}
    \mathbf{H}_{\mathrm{dec}}
        = \text{Dec}\!\left(
            \Ndensity^{(1 \times 32)},
            \tilde{\mathbf{H}}_{\mathrm{enc}}
          \right)
        \in \mathbb{R}^{1 \times 32}.
    \end{equation}
    
    A final fully-connected (FC)~\cite{patternrecognbishop} linear layer, $h$, maps this hidden state to a probability vector
    \begin{equation}
    \Mdensity
        = \text{softmax}\Bigl(
            h(\mathbf{H}_{\mathrm{dec}})
          \Bigl).
    \end{equation}
    
    In summary, the full TF model implements the same transformation as in \cref{eq:rnn},
    where the encoder and decoder operate in different feature spaces and are connected via the learned projection $f_c$. Further details are given in \cref{app:transf-model-train}.
    
    \item[Transformer with SPAM Mitigation (TFM):]
    The TFM variant extends the previous TF model by injecting \emph{SPAM-mitigated} (which will be introduced in \cref{sec:results-compare-mitigation}) intermediate distributions directly into the decoder, alongside the noisy output distribution.
    This means that the decoder receives a multi-token sequence (e.g. $[\Ndensity, m_1]$, where $m_1$ indicates the Spam-mitigated sequence), allowing the model to condition using previously mitigated data, while still leveraging the encoder’s circuit-level representation.  
    The structure of the encoder is unchanged, but the decoder input is amplified to operate on a sequence encoding that has previously been mitigated.  
    This variant is also discussed in \cref{app:transf-model-train}.

   \item[Noisy Encoder (NOISY ENC):]
    An encoder-only model that operates solely on the noisy output distribution.
    In the encoder, $\Ndensity$ is treated as a length-one sequence, embedded, and then passed through a number of self-attention layers, followed by a linear softmax layer that produces a mitigated distribution
    \begin{equation}
    \Mdensity = \text{NoisyEnc}(\Ndensity).
    \end{equation}
    This architecture does not use explicit circuit and backend information, and does not have a separate decoder, thus is solely a function of the noisy distribution. Further details are given in \cref{app:noisy-encoder-model-training}.

    \item[Perceiver (PERCEIVER):]
    A Perceiver-based encoder~\cite{jaegle2021perceiver} that maps the circuit sequence and noisy distribution to a mitigated $32$-dimensional probability vector via a
    fixed-size latent bottleneck.
    
    The circuit and backend features, $\mathbf{X}_{\mathrm{CB}}$, form a sequence as previously.
    The noisy distribution is broadcast along the sequence length and concatenated, giving an effective input, $\mathbf{X}_{\mathrm{in}}
        \in \mathbb{R}^{n_{l} \times (132 + 32)}$. Then, instead of processing the full sequence directly, the Perceiver maps $\mathbf{X}_{\mathrm{in}}$ onto a fixed number of latent vectors $\mathbf{Z}_0 \in \mathbb{R}^{M \times d}$~\footnote{We use $M=256$ latent vectors, each of dimension $d=1024$ for Pauli data, and $d=768$ for Random data.} using cross-attention
    \begin{equation}
    \mathbf{Z}_0 = 
        \mathrm{CrossAttn}
        \left(
            \text{queries}=\mathbf{Z}_{\mathrm{init}},
            \text{keys/values}=\mathbf{X}_{\mathrm{in}}
        \right),
    \end{equation}
    where $M \ll n_{l}$.
    This step compresses the large, heterogeneous input into a compact, order-agnostic latent representation. Following this, several latent self-attention blocks repeatedly process the latent array
    \begin{equation}
    \mathbf{Z}_{k+1}
        = \mathrm{LatentSelfAttnBlock}(\mathbf{Z}_k),
        \qquad k=0,\ldots,K{-}1,
    \end{equation}
    allowing information to mix globally within the latent space while keeping the computation independent of the input length.
    The final latent array is then mean-pooled and passed through a small MLP head
    \begin{equation}
    \Mdensity
        = \mathrm{softmax}\!\bigl(
            \mathrm{MLP}[\mathrm{meanpool}(\mathbf{Z}_B)
          ]\bigr).
    \end{equation}
    
    In summary, the full Perceiver model implements the same mapping as in \cref{eq:rnn}. This is done via a modality-agnostic attention bottleneck that scales linearly in the input dimension, and can naturally integrate heterogeneous features such as circuit and backend information~\cite{jaegle2021perceiver, jaegle2022perceiveriogeneralarchitecture}.
    This flexibility makes it especially suited to our multi-format inputs.
    Further details are provided in \cref{app:perceiver-model-training}.
\end{description}

\subsubsection*{Correction Models}

Correction models learn a multiplicative correction vector, $\mathbf{c} \in \mathbb{R}^{32}$, applied to the noisy distribution to obtain
\begin{equation}
\Mdensity = \text{softmax}\!\left(\mathbf{c} \cdot \Ndensity \right),
\end{equation}
where the product is element-wise. 

\begin{description}
    \item[Multilayer Perceptron (MLP-CORRECTION):]
    A compact feed-forward network that learns a multiplicative correction to the noisy output distribution~\cite{Rumelhart1986LearningRB}.  
    The circuit and backend features are flattened into a vector and passed through a stack of dense
    layers with BN and nonlinearities, similarly to the MLP-PREDICTION model.
    The correction vector
    \begin{equation}
    \label{eq:corr_mit}
    \mathbf{c}
      = \mathrm{MLP}(\mathbf{x}_{\mathrm{CB}}),
    \end{equation}
    is interpreted as a correction mask, which is applied element-wise to the noisy distribution, and the result is renormalized by a softmax layer to obtain the mitigated probabilities.
    Thus, unlike MLP-PREDICTION, which directly maps $(\mathbf{x}_{\mathrm{CB}},\Ndensity)$ to $\Mdensity$, MLP-CORRECTION first learns a correction factor, and then reshapes the noisy distribution via a simple convolution. 
    Further details are given in \cref{app:mlp-models-training}.

    \item[Encoder (ENC):]
    A transformer--convolution encoder that produces a multiplicative correction mask applied to the noisy probabilities.
    
    $\mathbf{X}_{\mathrm{CB}}$ is first augmented with positional encodings and processed by a multi-layer transformer encoder, followed by two one-dimensional convolutional layers with BN.
    After global average pooling, the network produces a correction vector $\mathbf{c} \in \mathbb{R}^{32}$.  
    The noisy distribution is not used by the encoder; instead, it is applied at the end via an element-wise product, the same transformation as in \cref{eq:corr_mit}.
    
    Thus, ENC implements a correction-based mapping, as in \cref{eq:rnn}, where the encoder learns to reshape the noisy distribution via a feature extractor.
    Further details are given in \cref{app:encod-model-train}.
\end{description}

All models in our study are trained with fixed-size input tensors, which implicitly define an upper bound on the circuit depth and feature dimensionality supported during inference.  
However, the Perceiver differs from the transformer-based models in how it scales with input length; its latent cross-attention bottleneck decouples the computational cost from the input-sequence length, allowing variable-length circuit representations without modifying the size of the model~\cite{jaegle2022perceiveriogeneralarchitecture}.  
Thus, while all models operate on fixed dimensions in practice, the Perceiver architecture provides substantially greater flexibility for heterogeneous or longer circuit descriptions.

\subsection{Methodology}
\label{sec:experiment-methodology}

We train our models with a loss function given by the KL-divergence, which measures the statistical distance between probability distributions. 
Specifically, for probability distributions, $\textbf{P}$, and, $\textbf{Q}$, on the same support, $X$, it is defined as
\begin{equation}
    \text{KL}(\textbf{P} , \textbf{Q}) = \sum_{x \in X} \textbf{P}(x) \log\left(\frac{\textbf{P}(x)}{\textbf{Q}(x)}\right).
    \label{eq:kl-divergence}
\end{equation}
The KL-divergence is sensitive to points where the target distribution assigns high probabilities, making it an appropriate measure for QEM. 
Indeed, the KL-divergence is preferred over symmetric cost functions such as the mean square error, which treats all deviations equally, regardless of their significance. 
Additionally, the KL-divergence is unbounded and therefore sensitive to large deviations.\footnote{For a study on the impact of choosing a loss in this problem setting, the reader can refer to \cref{app:perceiver-loss} of~\cref{app:perceiver}.}

Pre-training on large, diverse, and easily accessible datasets has been shown to improve model robustness  \cite{devlin2019bertpretrainingdeepbidirectional}. 
In our case, we train first on Simulated data (pre-training phase), that can be easily produced through classical simulations of noisy channels. 
Then, we fine-tune on Real data, which accounts for hardware errors, such as cross-talk, leakage, and non-Markovianity that are not captured by the noise models used to generate the Simulated data.

We use a total of $246,000$ unique quantum circuits, split by circuit class and $T$ value, details of which are given in \cref{tab:dataset_info}. 
Across all datasets, the \emph{train : validation : test} split is $50\%:12.5\%:37.5\%$. 
We choose a relatively large test set because we want to increase the statistical robustness of our performance metrics defined in \cref{eq:relative-change}.
Our training and testing procedure is as follows:
\begin{description}
    \item[Hyperparameter search:] For each model, run a hyperparameter search \cite{Akiba2019OptunaAN, Wu2019HyperparameterOF} using the Simulated datasets. 
    The hyperparameters are separately optimized for the Random and Pauli datasets, and are optimised over $50$ iterations of Bayesian optimization. 
    For each iteration, the model is trained for at most 50 epochs, with early stopping set at 10 epochs without improvement. 
    Hyperparameters are updated between iterations by randomly sampling new parameters, in order to minimize the KL-divergence between the ideal and predicted probability distributions. 
    Details of the hyperparameter search procedures for each model are given in \cref{sec:models-details}.

    \item[Pre-training:] Each model is then trained for a maximum of $2,000$ epochs using Simulated data, with an early stopping criterion of $25$ epochs. 
    This process is repeated $5$ times, with different seeds for the network's weight initializations, resulting in $5$ different trained models. 
    Using $5$ different models generated in this way allows us to evaluate the expected performance of each model class, independent of the weight initialization. 

    \item[Fine-tuning:] For each class of models and the $5$ trained instances, we perform fine-tuning based on the corresponding Real dataset. 
    Training was capped at $2,000$ epochs, with an early stopping criterion of $25$ iterations, and the learning rate was reduced by a factor of $10$ compared to the original pre-training.
\end{description}

\begin{figure}[!htbp]
    \centering
    \begin{subfigure}[t]{0.49\textwidth}
        \centering
        \includegraphics[width=1\linewidth]{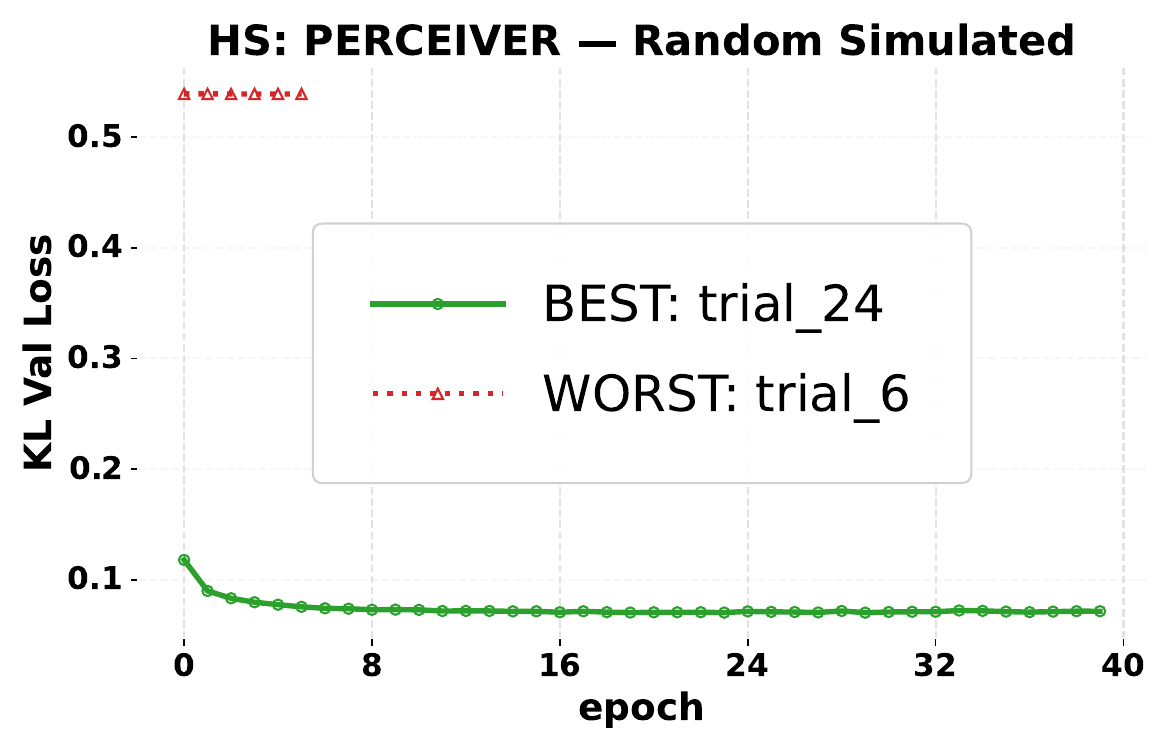}
        \subcaption{Perceiver model trained on \Ibmalgiers{} Random Simulated data.}
        \label{fig:hs1}
    \end{subfigure}%
    \hfill
    \begin{subfigure}[t]{0.49\textwidth}
        \centering
        \includegraphics[width=1\linewidth]{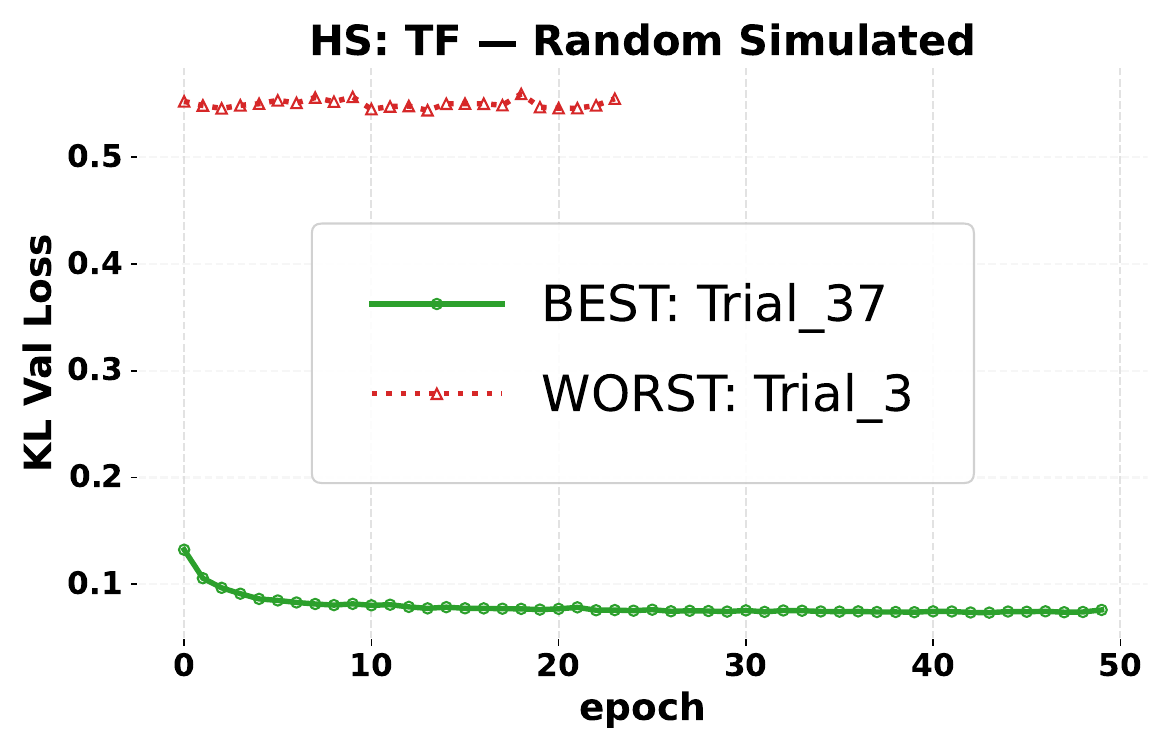}
        \subcaption{Transformer model trained on \Ibmalgiers{}  Random Simulated data.}
        \label{fig:hs2}
    \end{subfigure}
    \caption{\textbf{Example hyperparameter searches for two models for the Random Simulated dataset for \Ibmalgiers{}. Best and worst configurations in the search.} The losses on the validation set during the runs are labelled by the \emph{trial} index, given by the order they occurred during the procedure.
    }
    \label{fig:HS-show-main}
\end{figure}

We observe significant differences in performance for different choices of hyperparameters, as demonstrated in \cref{fig:HS-show-main}. 
$50$ iterations of hyperparameter search was found to suffice to give performant models, and 50 epochs always suffice to estimate a convergence interval (the point where the model's performance stabilizes and stops significantly improving). Note that we carry out distinct hyperparameter searches for each dataset, which results in only minor variations in the final optimized settings.

\section{Model Performance}
\label{sec:results-all}

In this section and in \cref{sec:results-real-algiers} we compare the performance of the different ML models studied. 
In \cref{sec:results-compare-mitigation} we compare the performance of the trained models to baseline error mitigation procedures.

To quantify model performance, we introduce the \emph{L1 Relative Change}
\begin{equation}
    \mathcal{R} \left( \Mdensity,\Idensity,\Ndensity \right) :=
    \frac{||\Idensity - \Mdensity||_1 - ||\Idensity - \Ndensity||_1}
    {||\Idensity - \Ndensity||_1}.
    \label{eq:relative-change}
\end{equation}
Here, $||{\bf{P}}||_1 = \sum_{x\in X} |\bf{P}(x)|$, is the $L1$ norm.
Error mitigation is successful if $||\Idensity - \Mdensity||_1 < ||\Idensity - \Ndensity||_1$. 
Due to the normalization, the relative change measures performance independently of the magnitude of the noise. 
A negative relative change ($-1\leq \mathcal{R}<0$) means that the error mitigation was successful at reducing the distance between the ideal and noisy distribution by a factor of $|\mathcal{R}|$. 
In our results, we also define the \emph{L1RC \% Improved} as the percentage of circuits (referring to the median seed) in the test dataset that obtained a negative L1 Relative Change after mitigation.

Using a performance metric based on the L1 norm (or equivalently, the statistical distance between probability distributions) is well motivated operationally, as it provides an upper bound on the absolute error in computing expectation values of observables in the computational basis.  
Furthermore, the statistical distance is bounded by the KL-divergence via Pinsker's inequality, and therefore models that achieve a low training cost function will also have a low statistical distance between the ideal and mitigated distributions. 
Finally, since our models are trained using KL-divergence as the loss function, the L1 Relative Change gives an unbiased metric with which to evaluate performance.

\begin{figure}[t]
\vspace{-0.2cm}
    \centering

    \begin{subfigure}{0.7\linewidth}
        \centering
        \includegraphics[width=\linewidth]{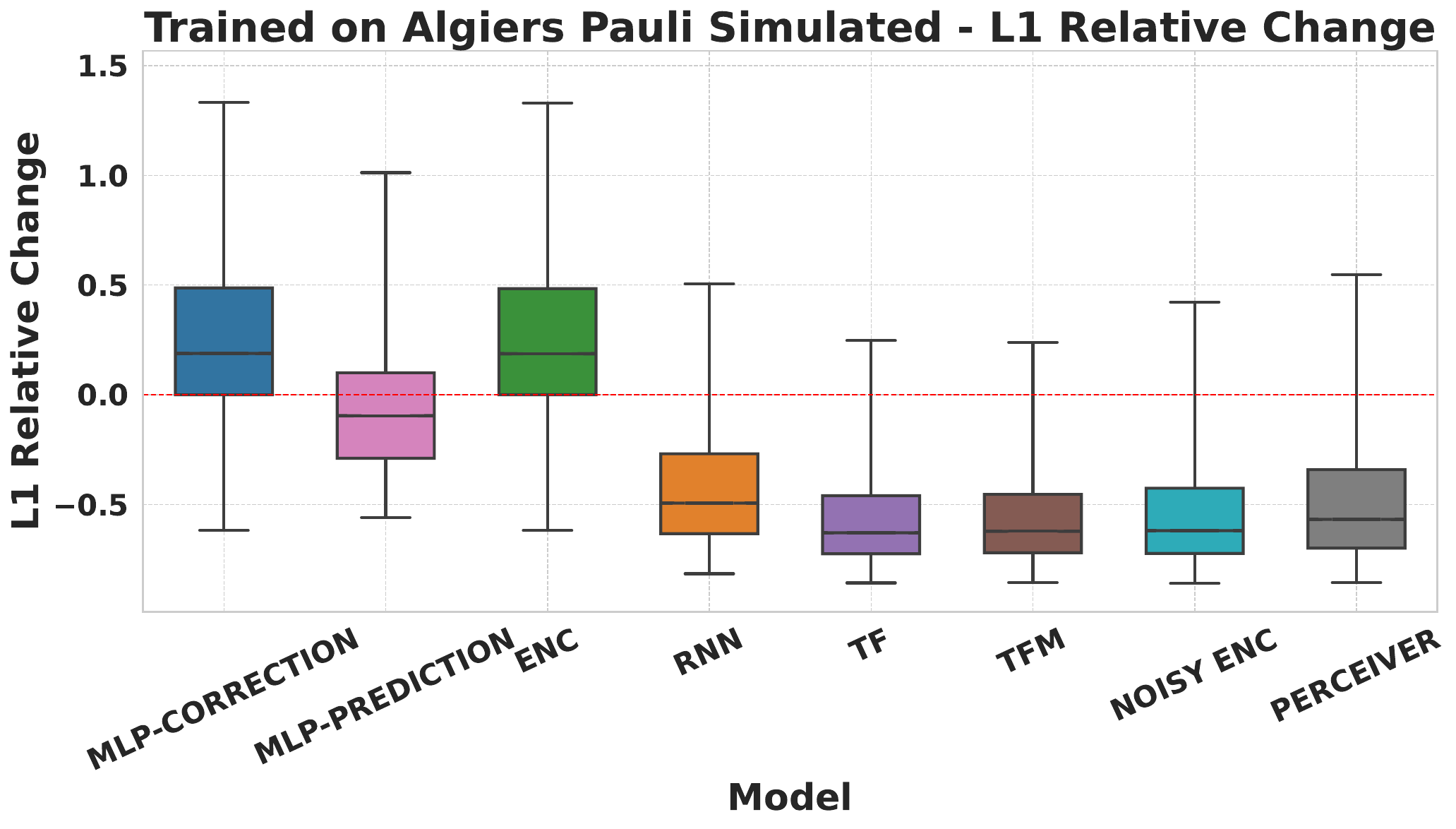}
        \caption{Pauli data.}
        \label{fig:sim-pauli}
    \end{subfigure}

    \vspace{3mm}

    \begin{subfigure}{0.7\linewidth}
        \centering
        \includegraphics[width=\linewidth]{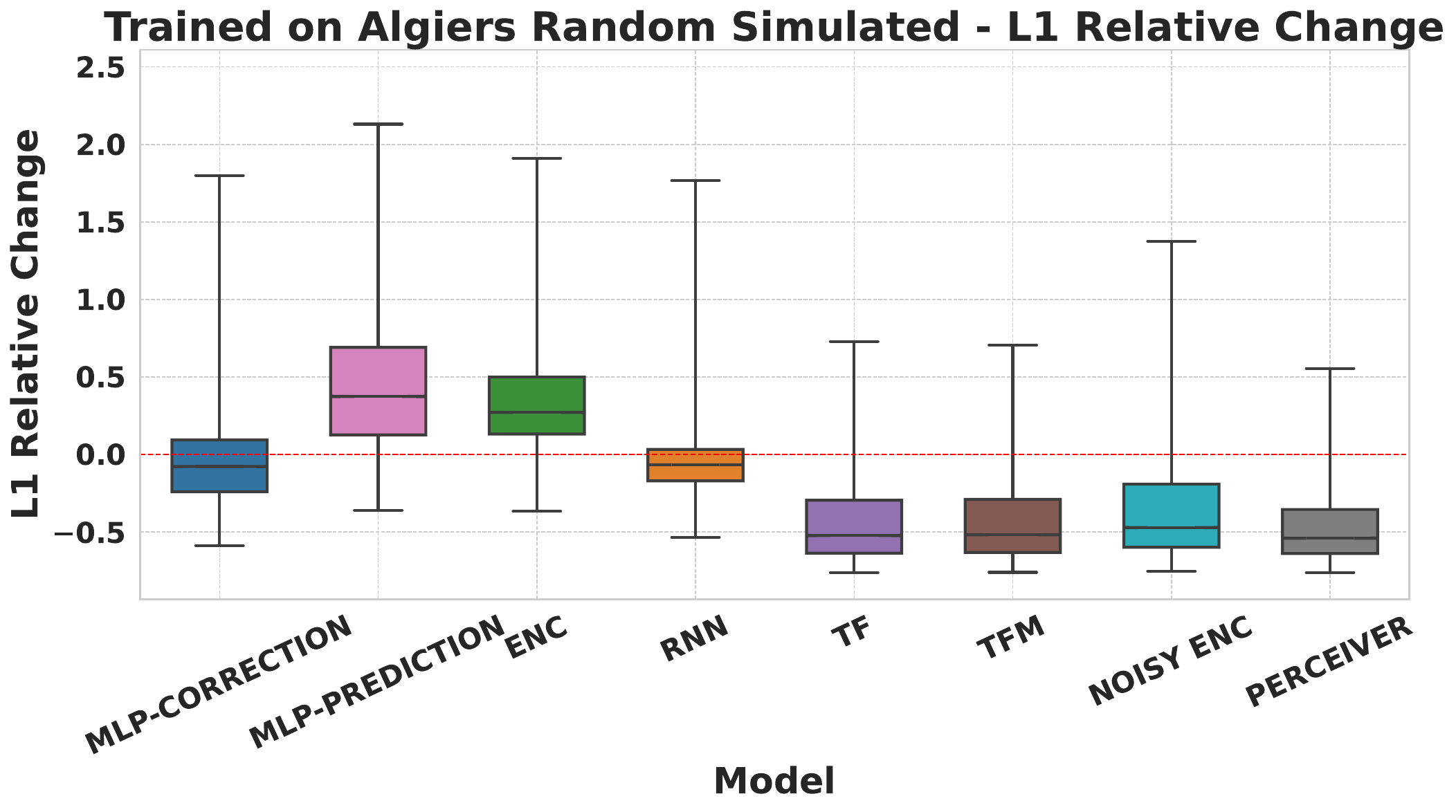}
        \caption{Random data.}
        \label{fig:sim-random}
    \end{subfigure}

    \caption{\textbf{Performance of models trained and tested on \Ibmalgiers{} Simulated data.}
    Box plots for models trained on Simulated Pauli (a) and Random (b) circuits.
    The black lines indicate medians; boxes span the 25th–75th percentiles; whiskers
    cover the 1st–99th percentiles. Values below the red dashed line ($y=0$) indicate
    successful mitigation.}
    \label{fig:simulated-results}
\end{figure}

We first pre-train all models on Simulated data from \Ibmalgiers{}.
The goal is to provide stable initializations, inject an inductive bias into the training, and reduce the amount of real data required (which is costly to generate).
Summary results are given in \cref{fig:simulated-results} and in \cref{tab:simulated-table} in \cref{app:tables-performance-appendix}. 
Attention-based prediction models (TF, TFM, NOISY ENC, PERCEIVER) tend to perform generally better than the rest, with the TF improving $94.2\%$ of Pauli circuits, and the PERCEIVER improving $91.7\%$ of Random circuits. 
This difference in performance is likely due to the inclusion of the attention mechanism, which extracts cross-correlations between elements in the layers of the circuit, backend properties, and the noisy output distribution.
The models trained on simulated data provide a strong initialization for additional fine-tuning on real data, or for different datasets afterwards.

\subsection{Real Data}
\label{sec:results-real-algiers}

\begin{figure}[!htbp]

    \centering
    \begin{subfigure}{0.7\linewidth}
        \centering
        \includegraphics[width=\linewidth]{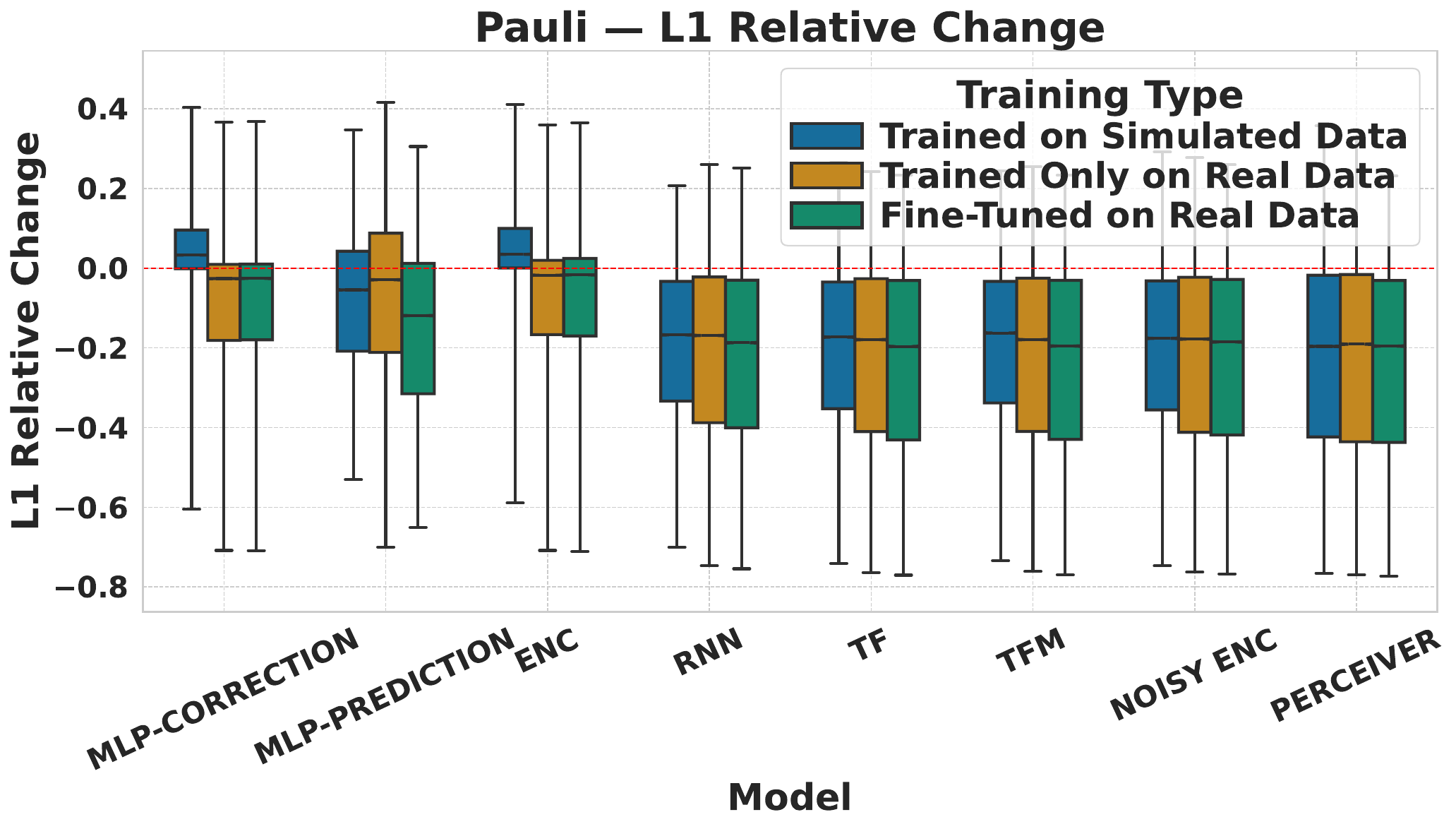}
        \caption{Pauli Real data.}
        \label{fig:real-pauli}
    \end{subfigure}

    \vspace{3mm}

    \begin{subfigure}{0.7\linewidth}
        \centering
        \includegraphics[width=\linewidth]{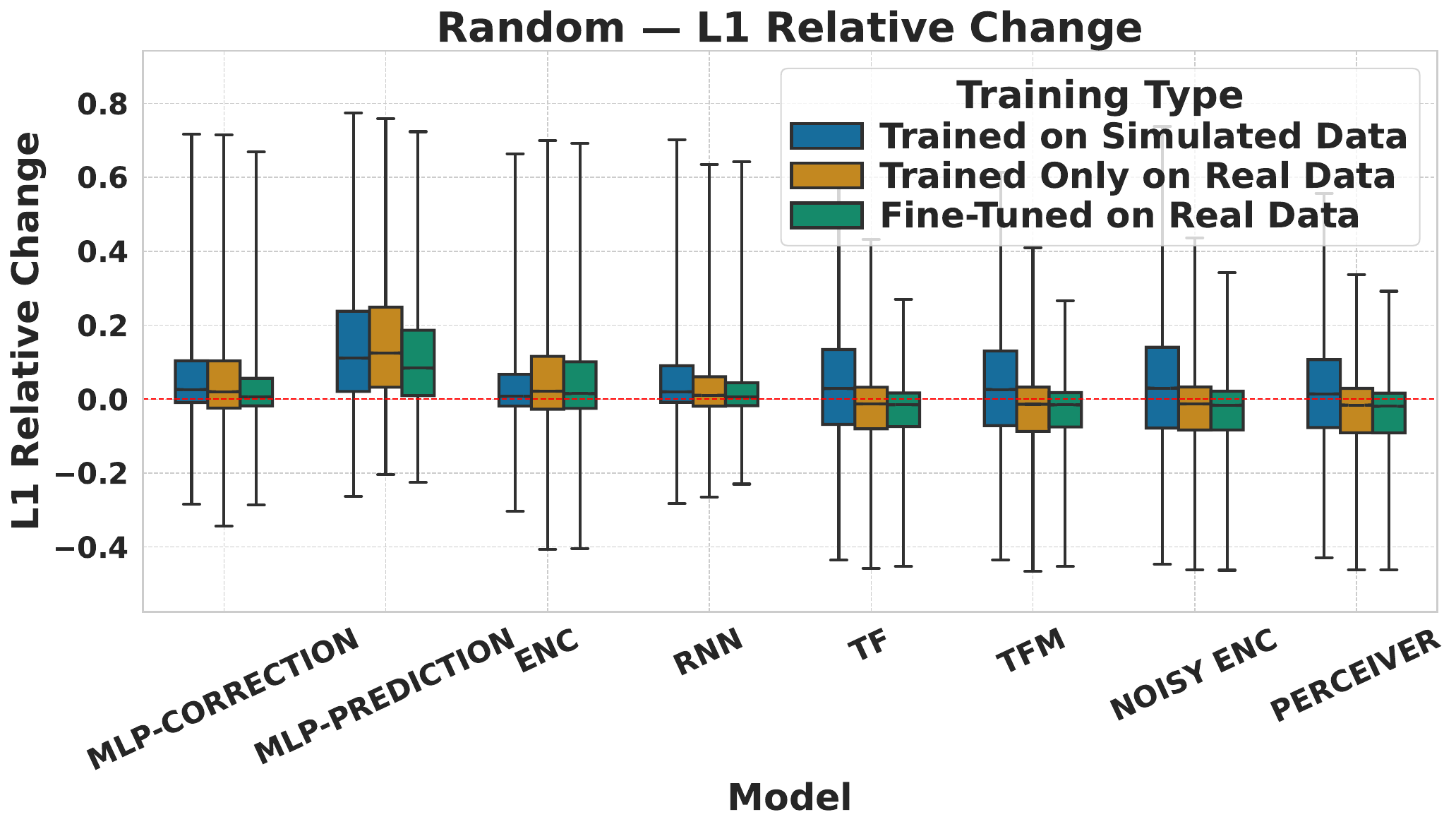}
        \caption{Random Real data.}
        \label{fig:real-random}
    \end{subfigure}

    \caption{\textbf{Performance comparison of models tested on \Ibmalgiers{} Real data.}
    Box plots for models trained on Pauli (a) and Random (b) circuits.
    Each plot compares models trained on Simulated data, Real data, and models fine-tuned on Real data.
    Boxes span the 25th–75th percentiles; whiskers indicate the 1st–99th percentiles.}
    \label{fig:real-algiers-results}
\end{figure}

Next, we consider Real data from \Ibmalgiers{}. 
We compare three approaches:
\begin{description}
    \item[Trained on Simulated Data:] Models trained on Simulated data and tested on Real data.
    \item[Trained Only on Real Data:] Models trained on Real data and tested on Real data (this does not include pre-training).
    \item[Fine-Tuned on Real Data:] Models pre-trained using Simulated data, fine-tuned on Real data, and finally tested on Real data. This represents a real-world scenario where pre-training occurs using simulated data, followed by fine-tuning on more costly real data.
\end{description}
Results are given in \cref{fig:real-algiers-results} and in \cref{tab:performance_results_algiers} in \cref{app:tables-performance-appendix}. 
For each training approach, we can see that, overall, the prediction models (RNN, TF, TFM, NOISY ENC, PERCEIVER) tend to outperform the correction models and the MLPs (MLP-CORRECTION, MLP-PREDICTION, ENC). 
This is likely because the prediction models take in strictly more input information, in particular the noisy output distributions, as compared to the correction models, which only indirectly see the noisy output distributions via the loss function. 
The MLPs also do not have access to a more explicit temporal structure because the input data is processed into a flattened vector rather than a tensor. 
We see that models with attention mechanisms show a median trend toward better performance.
This is clearly visible for Pauli circuits in \cref{fig:real-algiers-results}, and it is also reported quantitatively in \cref{tab:performance_results_algiers}

The approach of fine-tuning on Real data is typically as good as, or better than, the other two training approaches, measured both by the median L1 Relative Change, and the percentage of circuits in the test set which the model improves. 
On the other hand, training only on Simulated data is the worst approach on average, but the difference is very small. 
While unlikely, this may be because the noise effects omitted in the Simulated data have minimal impact on the resulting distributions. 
More likely is that the un-modelled noise is too complex to be captured in its entirety by the simple internal representation used by the model. 
We explore this representation further in \cref{sec:robustness-rand-inputs,sec:results-compare-mitigation}.

\begin{figure}[!htbp]
    \centering
    \begin{subfigure}[t]{0.48\textwidth}
        \centering
        \includegraphics[width=\linewidth]{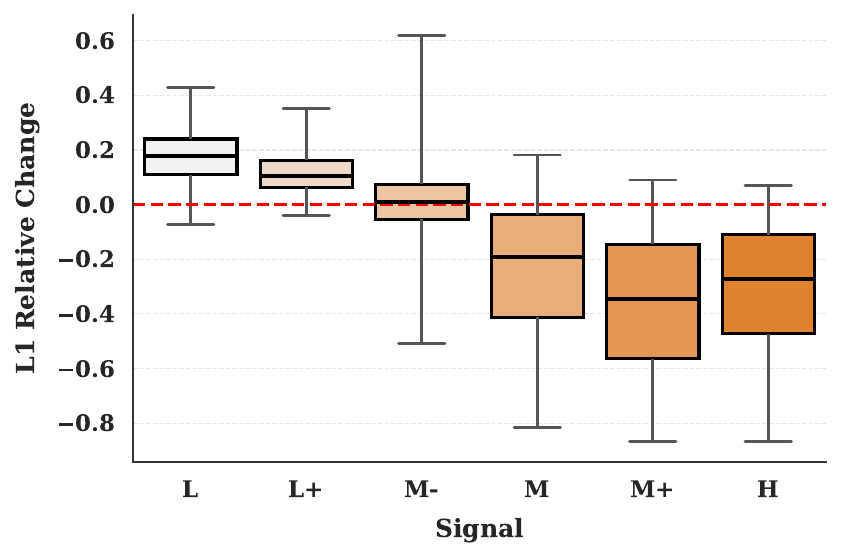}
        \subcaption{Fine-tuned on \Ibmalgiers{} Pauli Real data.}
        \label{fig:signalpaulireal}
    \end{subfigure}
    \hfill
    \begin{subfigure}[t]{0.48\textwidth}
        \centering
        \includegraphics[width=\linewidth]{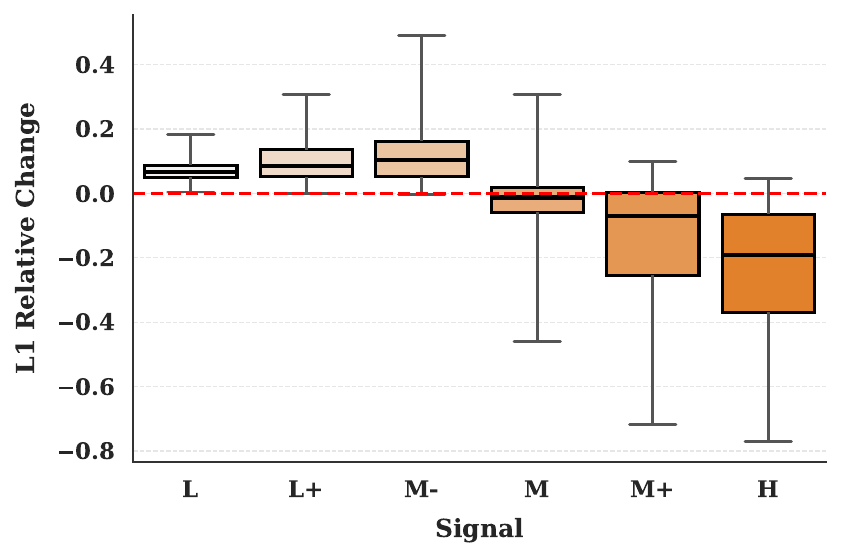}
        \subcaption{Fine-tuned on \Ibmalgiers{} Random Real data.}
        \label{fig:signalrandomreal}
    \end{subfigure}
    \caption{\textbf{L1 Relative Change for the TF model for \Ibmalgiers{} data, binned by signal.}
    Data points are grouped into signal bins, from low (L) to high (H). 
    Boxes span the 25th–75th percentiles; whiskers show the 1st–99th percentiles.}
    \label{fig:best-models-signal}
\end{figure}

\Cref{fig:best-models-signal} shows that model performance is correlated to the amount of signal in the circuits (cfr. \cref{fig:signal-to-noise,fig:signal-vs-layers}), suggesting that high-signal circuits are easier to mitigate. 
This is likely because there remains more structure in the noisy probability distributions that the model can exploit. 
Note that this also explains the strong performance of the models when considering Pauli circuits, which typically have higher signal. Hence, a smaller dataset consisting predominantly of high-signal circuits may enable a more efficient learning process.

\begin{figure}[t]
\vspace{-0.2cm}
    \centering

    \begin{subfigure}{0.48\textwidth}
        \centering
        \includegraphics[width=\linewidth]{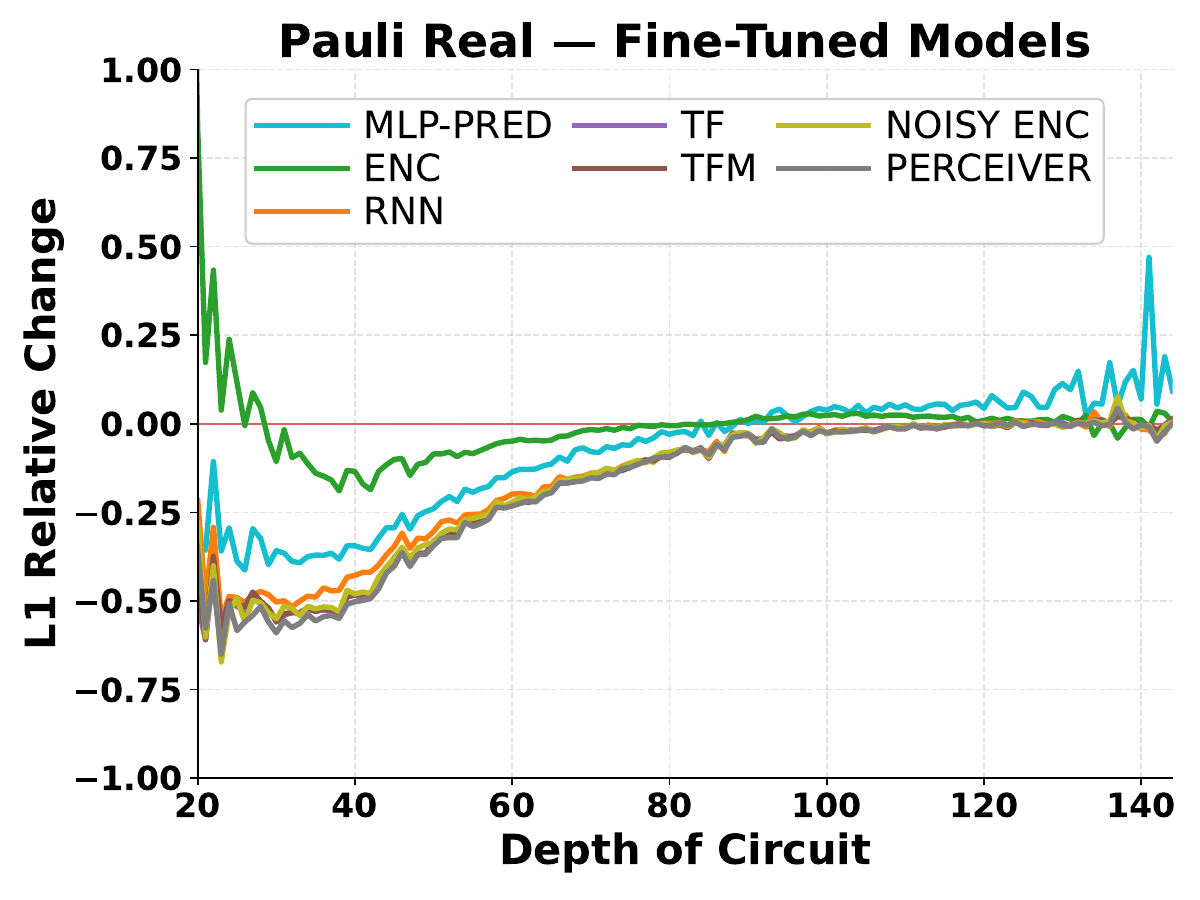}
        \caption{\Ibmalgiers{} Pauli Real data.}
        \label{fig:cp-pauli}
    \end{subfigure}
    \hfill
    \begin{subfigure}{0.48\textwidth}
        \centering
        \includegraphics[width=\linewidth]{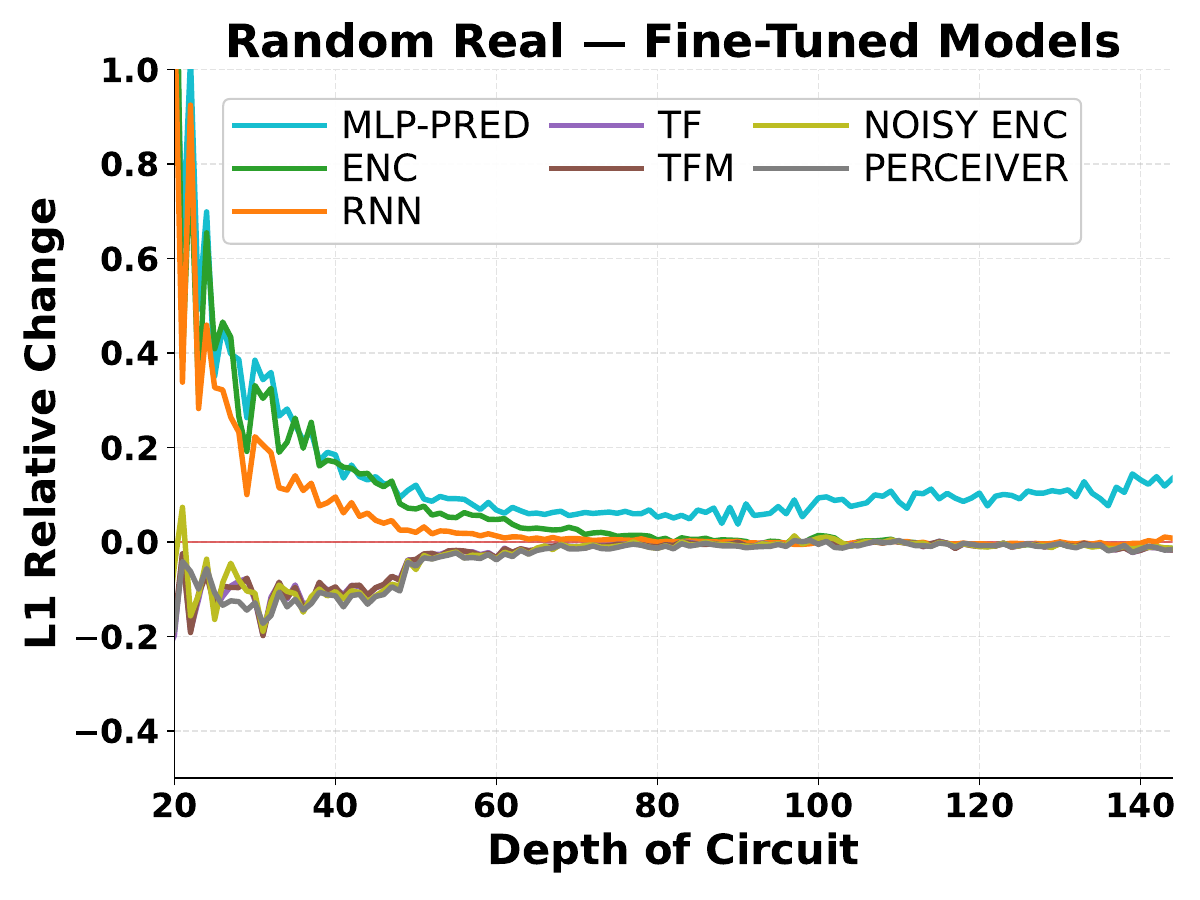}
        \caption{\Ibmalgiers{} Random Real data.}
        \label{fig:cp-random}
    \end{subfigure}

    \caption{\textbf{Median L1 Relative Change with respect to \Idensity{} against circuit depth.}
    (a) and (b) correspond to \Ibmalgiers{} Pauli Real and Random Real data, respectively.
    The plots compare the median performance of the models.}
    \label{fig:cp-models}
\end{figure}

In addition, \cref{fig:cp-models} shows the models' performances (for the configuration `Fine-Tune on Real Data') as a function of circuit depth.  For Pauli data (exception MLP-PRED) at low depths (i.e. high signal), the models tend to perform best, with the performance then flatlining at high depths (i.e. low signal). For Random data, the same trend is exhibited for those models that are generally performant (TF, TFM, NOISY ENC, PERCEIVER).

\subsection{Comparison with Standard Error Mitigation Methods}
\label{sec:results-compare-mitigation}

In this section we compare our models to different baseline error mitigation schemes. These schemes mitigate probability distributions, and to do so require only post-processing of results from unmodified circuits, similar to the ML-based approach presented in this article:
\begin{description}
    \item[SPAM:] models SPAM as an invertible binary channel, using only SPAM calibration data from $\Backendinfo{}$ and the measured distribution $\Ndensity$.
    \item[Repolarizer:] assumes the device noise can be approximated as a global depolarizing channel acting on two-qubit gates.  
    Using the two-qubit gate error rate, $e_{2q}$, and the gate count, $t_{2q}$, it inverts the analytical relation, $\Ndensity = (1-e_{2q})^{t_{2q}}\Idensity + [1-(1-e_{2q})^{t_{2q}}]\Udensity$, to recover an estimate of $\Idensity$.  
    Details are given in \cref{app:repolariser}.
    \item[Mix:] applies SPAM correction followed by Repolarizer. 
    \item[Thresholding:] post-processes $\Ndensity$ by removing probabilities below a certain threshold $\tau$ and renormalizing the remaining probabilities.  
    The threshold is optimized for each dataset after conducting a grid search over 30 equispaced values in the range $\tau \in [0, 0.5]$. Performing this search ensures the comparison to the models is a `worst-case comparison' i.e. Thresholding performs as well as it possibly could on this particular dataset. The optimal value for the Pauli Real dataset is $\tau = 0.069$, and for Random Real is $\tau = 0.0$ (the noisy output distribution remains unchanged).
    More details are given in \cref{app:thresholding}.
\end{description} 

SPAM and Repolarizer both rely on characterizations of the device done independently of the context in which they are used. In contrast, deep learning models can learn corrections directly from data gathered during circuit runs, more easily capturing correlations.


\begin{table}[!t]
\centering
{\scriptsize
\setlength{\tabcolsep}{3pt}
\renewcommand{\arraystretch}{0.9}

\caption{L1 Relative Change results for analytical methods (Spam, Repolarizer, Mix, and Thresholding) on \Ibmalgiers{} data. The table reports the median ($\pm$SE) and interquartile range (P25--P75) of the L1 relative change across datasets, with the final column showing the percentage of circuits that are improved as compared to the noisy baseline. In \textbf{bold} the best performing model per dataset, in green improvements over the noisy baseline, in red worse results than the baseline.}
\label{tab:analytical-split}

\begin{tabular}{@{}>{\raggedright\arraybackslash}p{2.9cm}lcccc@{}}
\toprule
Dataset & Model & L1RC Median & P25 & P75 & L1RC \% Improved \\
\midrule

\multicolumn{6}{@{}l}{\textbf{Simulated}}\\
\midrule

\multirow{4}{2.9cm}{\raggedright Pauli}
 & SPAM & \textcolor{ForestGreen}{-0.0930} & \textcolor{ForestGreen}{-0.1507} & \textcolor{ForestGreen}{-0.0523} & 96.6\% \\
 & REPOLARIZER & \textcolor{ForestGreen}{-0.5613} & \textcolor{ForestGreen}{-0.6592} & \textcolor{ForestGreen}{-0.4412} & 93.7\% \\
 & \textbf{MIX} & \textcolor{ForestGreen}{-0.6470} & \textcolor{ForestGreen}{-0.7333} & \textcolor{ForestGreen}{-0.5243} & 93.5\% \\
 & THRESHOLDING & \textcolor{ForestGreen}{-0.4871} & \textcolor{ForestGreen}{-0.6754} & \textcolor{ForestGreen}{-0.2272} & 85.4\% \\
\midrule

\multirow{4}{2.9cm}{\raggedright Random}
 & SPAM & \textcolor{ForestGreen}{-0.0792} & \textcolor{ForestGreen}{-0.1396} & \textcolor{ForestGreen}{-0.0416} & 93.5\% \\
 & \textbf{REPOLARIZER} & \textcolor{ForestGreen}{-0.6058} & \textcolor{ForestGreen}{-0.7216} & \textcolor{ForestGreen}{-0.4000} & 87.1\% \\
 & MIX & \textcolor{ForestGreen}{-0.5594} & \textcolor{ForestGreen}{-0.6979} & \textcolor{ForestGreen}{-0.3245} & 85.6\% \\
 & THRESHOLDING & \textcolor{ForestGreen}{-0.0921} & \textcolor{ForestGreen}{-0.3253} & \textcolor{BrickRed}{0.0000} & 63.8\% \\
\midrule

\multicolumn{6}{@{}l}{\textbf{Real}}\\
\midrule

\multirow{4}{2.9cm}{\raggedright Pauli}
 & SPAM & \textcolor{ForestGreen}{-0.0079} & \textcolor{ForestGreen}{-0.0188} & \textcolor{ForestGreen}{-0.0024} & 90.1\% \\
 & REPOLARIZER & \textcolor{ForestGreen}{-0.0876} & \textcolor{ForestGreen}{-0.1559} & \textcolor{ForestGreen}{-0.0315} & 85.8\% \\
 & MIX & \textcolor{ForestGreen}{-0.0977} & \textcolor{ForestGreen}{-0.1780} & \textcolor{ForestGreen}{-0.0338} & 85.5\% \\
 & \textbf{THRESHOLDING} & \textcolor{ForestGreen}{-0.1675} & \textcolor{ForestGreen}{-0.3589} & \textcolor{BrickRed}{0.1011} & 68.2\% \\
\midrule

\multirow{4}{2.9cm}{\raggedright Random}
 & \textbf{SPAM} & \textcolor{ForestGreen}{-0.0008} & \textcolor{ForestGreen}{-0.0095} & \textcolor{BrickRed}{0.0033} & 55.5\% \\
 & REPOLARIZER & \textcolor{BrickRed}{0.0318} & \textcolor{ForestGreen}{-0.0847} & \textcolor{BrickRed}{0.1662} & 41.6\% \\
 & MIX & \textcolor{BrickRed}{0.0383} & \textcolor{ForestGreen}{-0.0875} & \textcolor{BrickRed}{0.1828} & 40.8\% \\
 & THRESHOLDING & \textcolor{BrickRed}{0.0000} & \textcolor{BrickRed}{0.0000} & \textcolor{BrickRed}{0.0000} & 0\% \\
\bottomrule
\end{tabular}
}
\end{table}

\Cref{tab:analytical-split} compares the performance of the baseline methods to the unmitigated outputs on Simulated and Real data from \Ibmalgiers{}. The \emph{Repolarizer} and \emph{Mix} methods yield the strongest median improvements for Simulated data, demonstrating that simple global depolarization models remain effective for structured, low-depth circuits.  For Real data, the Thresholding method has the highest median improvements for the Pauli dataset, despite the relatively low percentage of improved circuits. 

\begin{figure}[!htbp]
    \centering

    \begin{subfigure}{0.48\textwidth}
        \centering
        \includegraphics[width=1\linewidth]{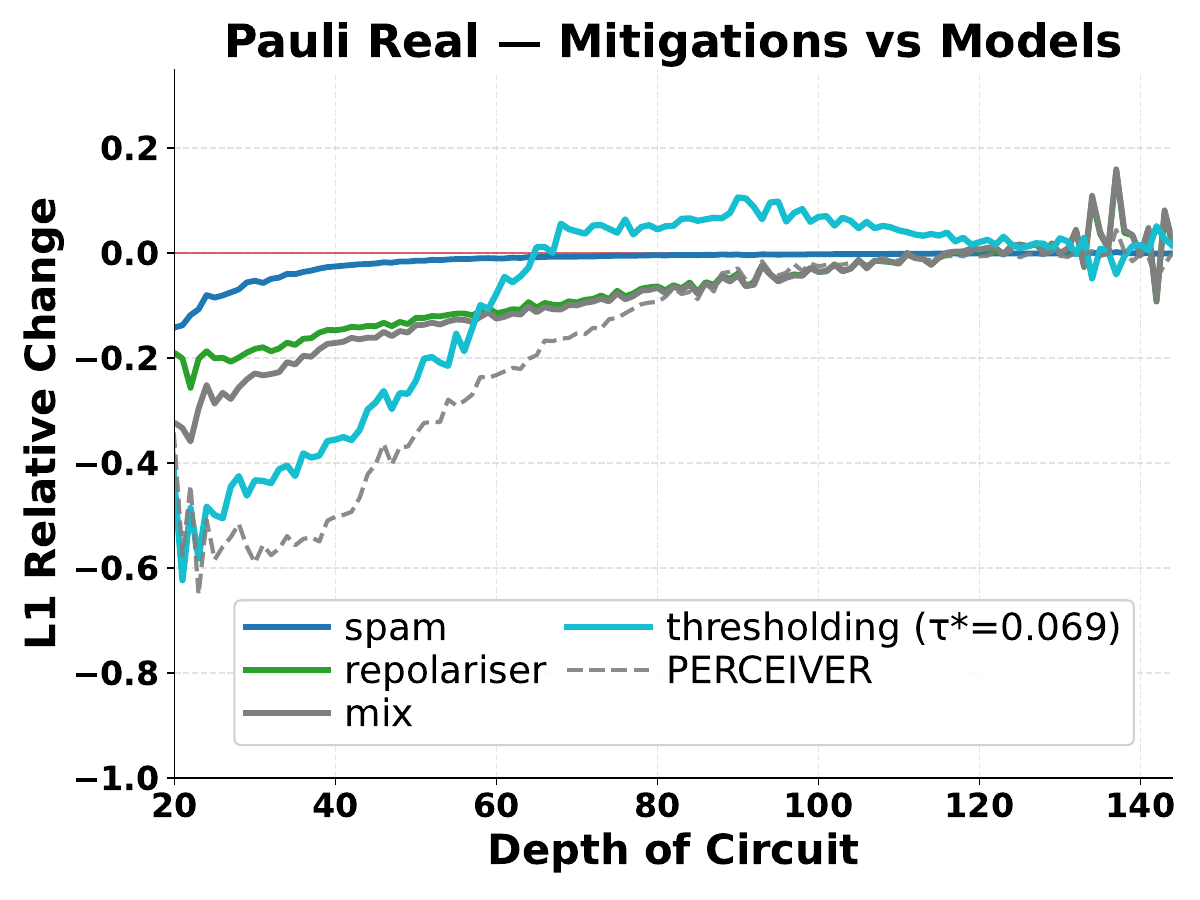}
        \caption{\Ibmalgiers{} Pauli Real circuits.}
        \label{fig:cp-mit-pauli}
    \end{subfigure}
    \hfill
    \begin{subfigure}{0.48\textwidth}
        \centering
        \includegraphics[width=1\linewidth]{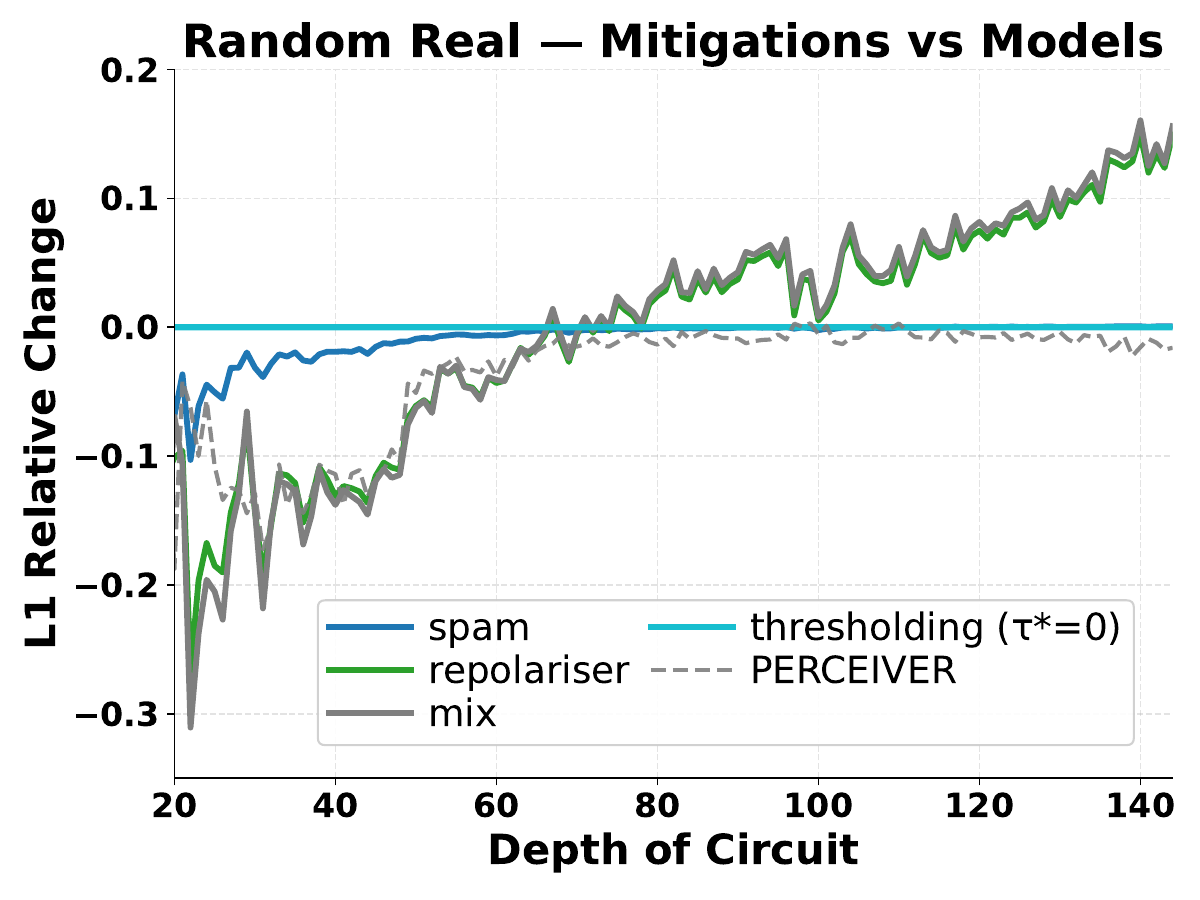}
        \caption{\Ibmalgiers{} Random Real circuits.}
        \label{fig:cp-mit-random}
    \end{subfigure}

    \caption{\textbf{Median L1 Relative Change with respect to \Idensity{} vs. circuit depth.}
    The plots compare the best model performances (dashed lines) against baseline mitigation methods
    (SPAM, Repolarizer, SPAM+Repolarizer), together with the best global thresholding applied only to the noisy distribution
    using a single dataset-wide threshold $\tau$ (\emph{Pauli Real:} $\tau=0.069$; \emph{Random Real:} $\tau=0.0$).
    Dashed horizontal guides indicate the best model’s median L1 Relative Change, for reference.}
    \label{fig:cp-mit-layers}
\end{figure}

\Cref{fig:cp-mit-layers} compares the performance of baseline mitigation to the best performing ML-based approach. 
For Pauli data, the PERCEIVER consistently achieves as good or greater median performance than the baseline mitigation techniques. 
This would indicate that the PERCEIVER model is learning a more complex representation of the effects of noise than the relatively simple representation used by the baseline mitigation techniques.
While they are not plotted here, the simpler MLP-CORRECTION and MLP-PREDICTION models perform worse than the baseline mitigation models.

For Pauli circuits, all models perform comparably to the baseline mitigation methods at shallow depths ($d < 25$).  
At intermediate depths ($25 < d < 90$), most attention-based models (TF, TFM, PERCEIVER and NOISY-ENC) consistently outperform SPAM, Repolarizer, Mix, and Thresholding. Beyond $d \approx 90$, both model-based and baseline methods perform poorly as the signal decreases.

For Random circuits, no method performs particularly well or reliably. While the PERCEIVER improves the largest fraction of circuits and has the largest median improvement, it performs markedly worse than on Pauli circuits.
This behaviour reflects the loss of learnable structure at high depths, and aligns with the signal-to-noise analysis in \cref{fig:signal-to-noise}; as the circuit depth increases, the noisy output distribution becomes statistically indistinguishable from uniform noise, leaving little residual correlation with $\Idensity$ to exploit.

As shown in \cref{tab:comparison-methods}, the best performing models are comparable to the best baseline methods on Simulated data (both Pauli and Random).
However, on Real Pauli data the advantage clearly shifts toward the ML-based models, which outperform all baselines in both median L1 relative change and fraction of improved circuits. Notably, the TF and PERCEIVER can generalize to different noise regimes and devices, without needing to rely on a pre-defined noise model.

These results collectively indicate that the baseline methods retain value as lightweight, interpretable mitigation techniques, particularly for structured, low-depth circuits -- but are ultimately limited by their fixed functional form.  
In contrast, the deep learning approaches, notably TF and PERCEIVER, can generalize across noise regimes, device generations, and circuit families without relying on a predefined noise model.  
This suggests that the learned mapping from $\Ndensity$ and circuit features to $\Idensity$ captures a richer structure that goes beyond coarse depolarization or measurement-error mitigation.

Finally, it is worth noting that the methods compared here all operate as post-processing steps and require no circuit modification.  
We therefore exclude noise-tailoring strategies such as dynamical decoupling~\cite{Viola_1999}, randomized compiling~\cite{Wallman_2016}, and Pauli twirling~\cite{PhysRevLett.119.180509,Li_2017}, which rely on circuit-level transformations.  
Such techniques could, however, be incorporated into future training datasets to further enhance the generalization of ML-based error mitigation techniques.


\begin{table}[H]
\centering
{\scriptsize
\setlength{\tabcolsep}{4pt}
\renewcommand{\arraystretch}{0.9}

\begin{tabular}{@{}l l l c c@{}}
\toprule
Dataset & Type & Best method & L1RC Median & L1RC \% Improved \\
\midrule

\multirow[t]{2}{*}{Simulated Pauli}
  & Model    & TF                 & \textcolor{ForestGreen}{-0.6318} & 94.2\% \\
  & Analytic & \textbf{MIX}       & \textcolor{ForestGreen}{-0.6470} & 93.5\% \\
\midrule

\multirow[t]{2}{*}{Simulated Random}
  & Model    & PERCEIVER          & \textcolor{ForestGreen}{-0.5450} & 91.7\% \\
  & Analytic & \textbf{REPOLARISER} & \textcolor{ForestGreen}{-0.6058} & 87.1\% \\
\midrule

\multirow[t]{2}{*}{Real Pauli}
  & Model    & \textbf{PERCEIVER} & \textcolor{ForestGreen}{-0.2062} & 83.2\% \\
  & Analytic & THRESHOLDING       & \textcolor{ForestGreen}{-0.1675} & 68.2\% \\
\midrule

\multirow[t]{2}{*}{Real Random}
  & Model    & \textbf{PERCEIVER} & \textcolor{ForestGreen}{-0.0210} & 64.4\% \\
  & Analytic & SPAM               & \textcolor{ForestGreen}{-0.0008} & 55.5\% \\
\bottomrule
\end{tabular}
}
\caption{Summary of the best performing learned models and analytical methods for each dataset class (Simulated/Real, Pauli/Random). The table reports the median ($\pm$SE) and interquartile range (P25--P75) of the L1 relative change across datasets, with the final column showing the percentage of circuits that are improved as compared to the noisy baseline. In \textbf{bold} the best performing model per dataset, in green improvements over the noisy baseline.}
\label{tab:comparison-methods}
\end{table}

\section{Ablation Studies and Generalization}
\label{sec:additional-study}

This section provides an ablation study to assess the sensitivity of the model performance to the inputs, and a study of the generalization properties of trained models. 
We begin with the ablation study in \cref{sec:robustness-rand-inputs}, which reveals how each architecture depends on its individual inputs. 
We then assess two forms of
generalization that test the model performance on distributions that have not been seen during training:

\begin{description}
    \item[Cross-dataset generalization across circuit types:]
    models trained on one circuit class are tested on the other in \cref{sec:cross-testing}.

    \item[Generalization to a different QPU:] models trained on \Ibmalgiers{} are fine-tuned and evaluated on
    data generated using \Ibmhanoi{} in \cref{sec:results-transfer-learning}.
\end{description}

\subsection{Randomized Inputs}
\label{sec:robustness-rand-inputs}
 
We consider five scenarios:
\emph{Standard} (no ablation); \emph{Random Backend} (randomizing the backend information); \emph{Random Noisy} (randomizing the noisy probability distribution); \emph{Random Circuit} (randomizing the circuit information); and \emph{Random Circuit\&Backend} (joint randomization of circuit and backend information).
\footnote{NOISY-ENC only has the noisy distribution as input, thus only \emph{Standard} and \emph{Random Noisy} are relevant. TFM is excluded, since it shares architecture with the TF.} The study is based on the fine-tuned models for \Ibmalgiers{} Real data (`Fine-Tuned on Real Data' models in \cref{tab:performance_results_algiers} in \cref{app:tables-performance-appendix}).

\begin{figure}[t]
\vspace{-0.2cm}
    \centering

    \begin{subfigure}{0.82\linewidth}
        \centering
        \includegraphics[width=\linewidth]{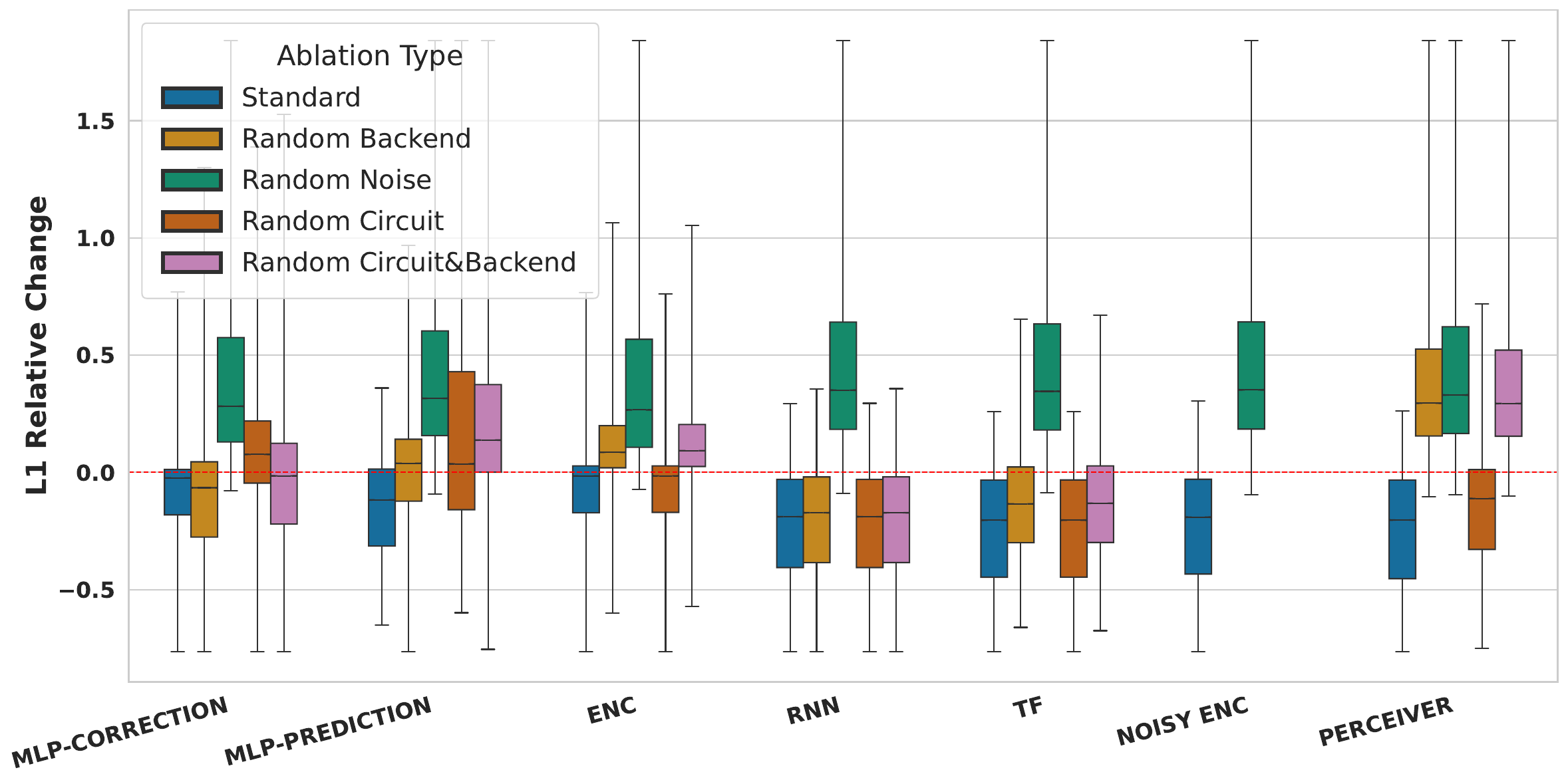}
        \caption{Pauli Real circuits randomized.}
        \label{fig:ablation-pauli}
    \end{subfigure}

    \vspace{3mm}

    \begin{subfigure}{0.82\linewidth}
        \centering
        \includegraphics[width=\linewidth]{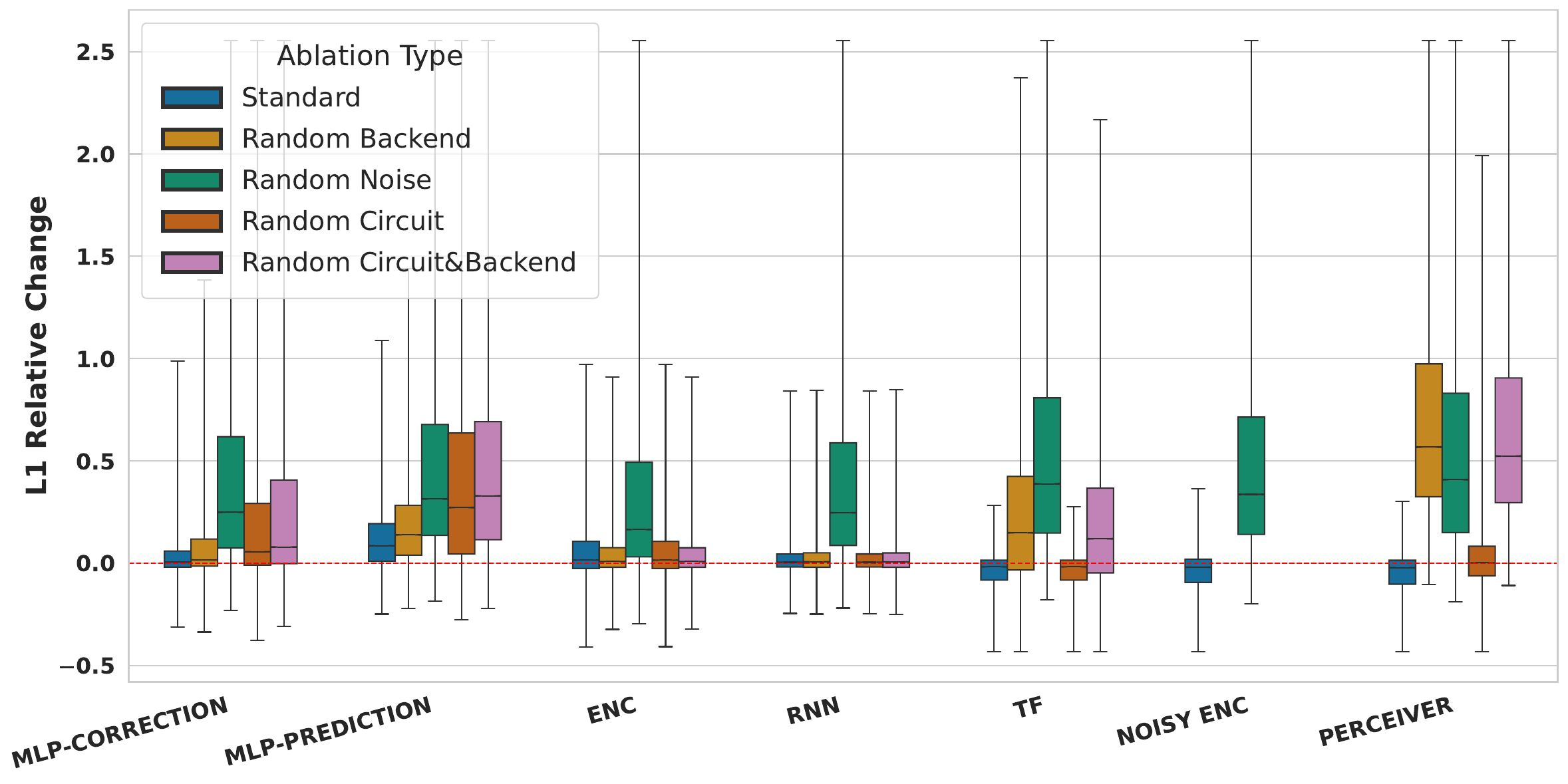}
        \caption{Random Real circuits randomized.}
        \label{fig:ablation-random}
    \end{subfigure}

    \caption{\textbf{L1 Relative Change under input randomization.}
    (a) and (b) correspond to \Ibmalgiers{} Pauli and Random data, respectively.
    Boxes show the interquartile range; whiskers indicate the 1st–99th percentiles;
    medians are shown in black; the dashed line at $y=0$ marks parity with the noisy baseline.}
    \label{fig:ablation}
\end{figure}

\cref{fig:ablation} summarizes the results for Pauli and Random data, respectively, and the numerical results are given in \cref{tab:pauli_randomization,tab:random_randomization} in \cref{app:tables-performance-appendix}, respectively.
We see that replacing the noisy distribution severely affects the performance in general, confirming that this feature is the most important for training. 
In contrast, \emph{Random Circuit} has little impact: for TF (and often RNN), the performance in the \emph{Random Circuit} scenario remains close to \emph{Standard}, indicating that circuit descriptors are at best only weakly exploited by these architectures. This also demonstrates that the model is not just directly simulating the circuit. Sensitivity to the other information is model-dependent: under \emph{Random Backend} and \emph{Circuit\&Backend}, the performance of TF and RNN on Pauli does not substantially degrade compared to the \emph{Standard} settings, whereas ENC and PERCEIVER appear sensitive to these inputs.

To summarize, the noisy distribution is thus the most important feature for all models. Attention-based models (TF) are the most robust to input randomization, with the RNN also fairly robust for Pauli data.

\subsection{Cross-dataset generalization}
\label{sec:cross-testing}

\begin{figure}[t] 
\vspace{-0.2cm}
    \centering
    \begin{minipage}[b]{0.48\textwidth}
        \centering
        \includegraphics[width=\linewidth]{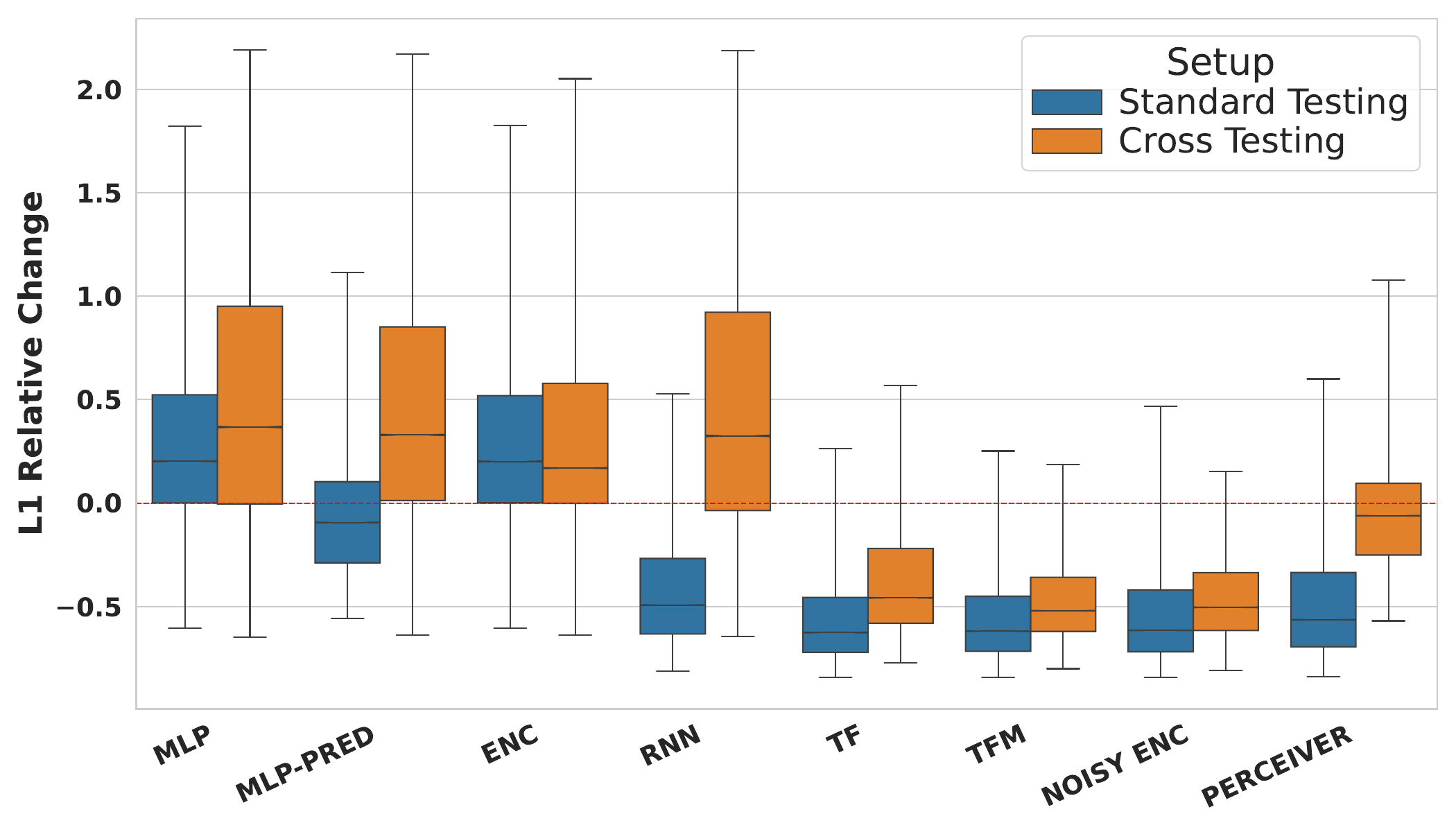}
        \subcaption{\Ibmalgiers{} Pauli Simulated data.}
        \label{fig:pauli-cross-sim}
    \end{minipage}
    \hfill
    \begin{minipage}[b]{0.48\textwidth}
        \centering
        \includegraphics[width=\linewidth]{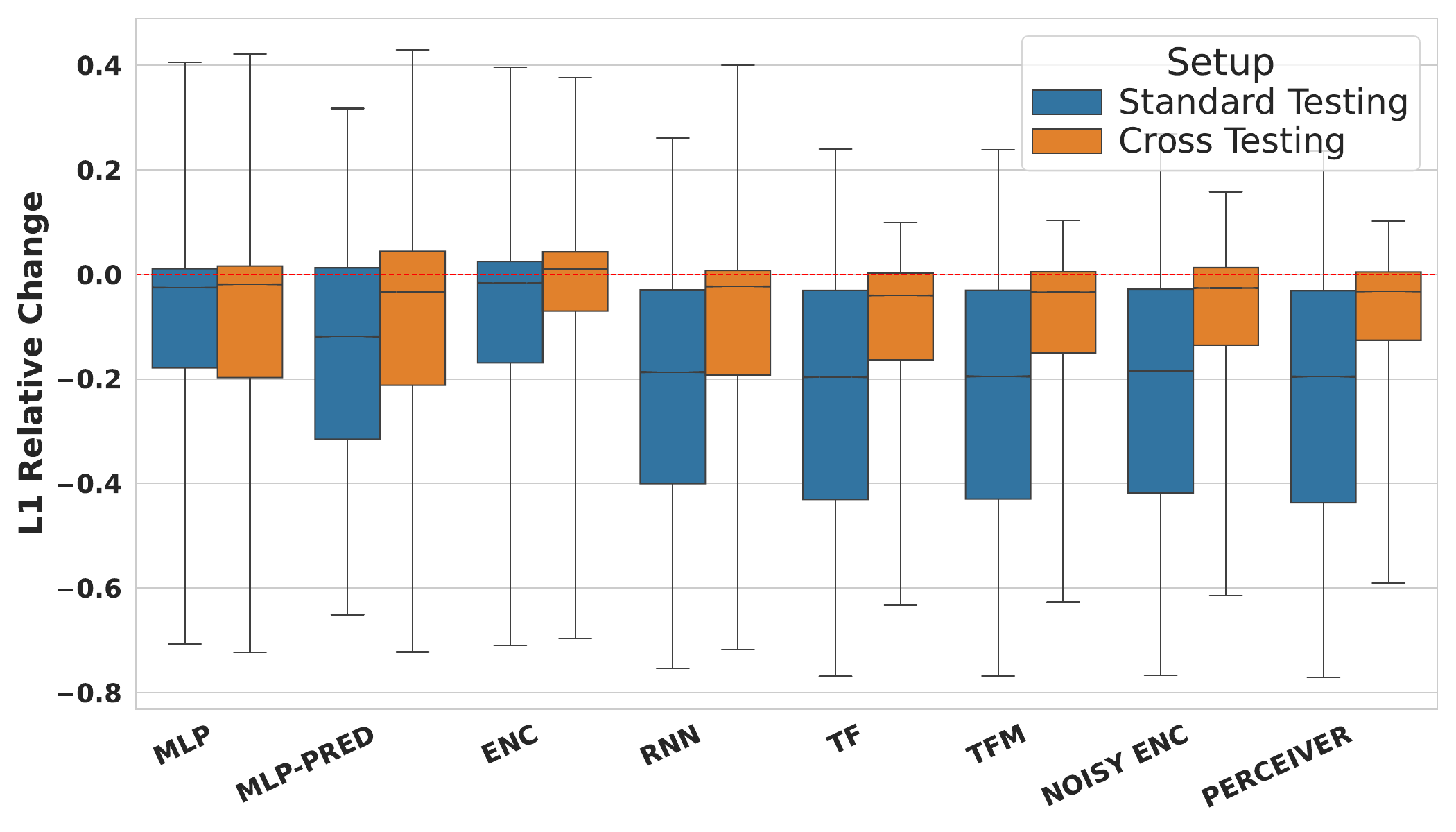}
        \subcaption{\Ibmalgiers{} Pauli Real data.}
        \label{fig:pauli-cross-real}
    \end{minipage}
    \vspace{0.3cm}
    \begin{minipage}[b]{0.48\textwidth}
        \centering
        \includegraphics[width=\linewidth]{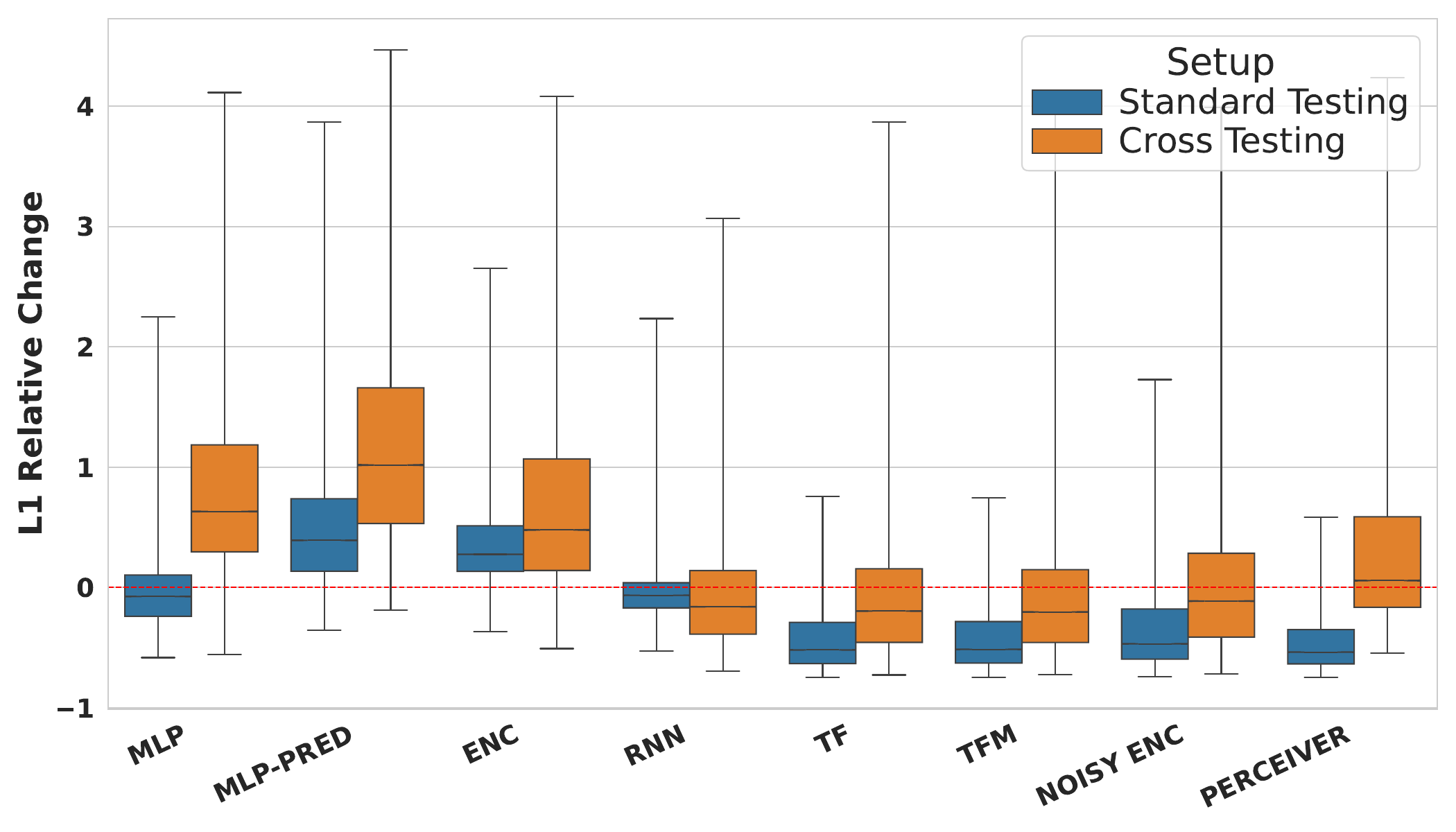}
        \subcaption{\Ibmalgiers{} Random Simulated data.}
        \label{fig:random-cross-sim}
    \end{minipage}
    \hfill
    \begin{minipage}[b]{0.48\textwidth}
        \centering
        \includegraphics[width=\linewidth]{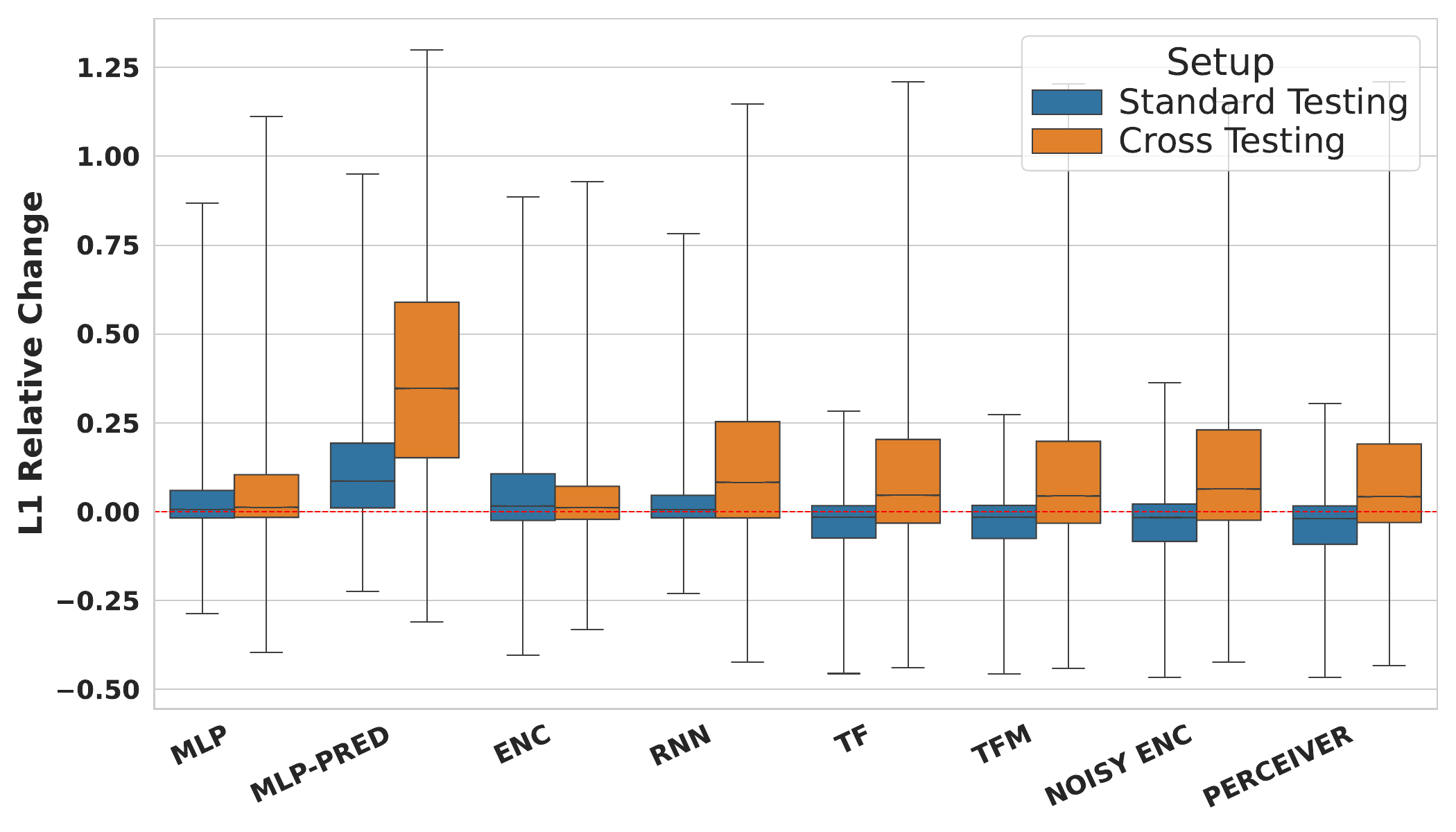}
        \subcaption{\Ibmalgiers{} Random Real data.}
        \label{fig:random-cross-real}
    \end{minipage}

    \caption{\textbf{Cross-dataset generalization study.}
    Each plot compares the $\mathrm{L1}$ relative change for models trained and tested on their native dataset class (Standard Testing) vs. those trained on the opposite dataset class and tested on the native dataset class (Cross Testing).
    (a,c) Testing on Pauli data; (b,d) Testing on Random data.
    The black lines represent median values, boxes span the 25th–75th percentiles, and whiskers cover the 1st–99th percentiles.  
    Values below the red dashed line ($y=0$) indicate successful mitigation.}
    \label{fig:cross-testing}
\end{figure}

We first evaluate the cross-dataset generalization of the models trained and fine-tuned on \Ibmalgiers{}.  
Each model trained on one circuit type (Pauli or Random) is tested on the other, while keeping the noise and device information fixed.  
\Cref{fig:cross-testing} summarizes the results for both Simulated and Real datasets and numerical results can be read in \cref{tab:cross_testing_l1_results_sim,tab:cross_testing_l1_results_real} in \cref{app:tables-performance-appendix}.

Across all architectures, performance generally degrades when cross testing, as expected when the model encounters out-of-distribution data.  
However, the attention-based models (TF, TFM, NOISY ENC, PERCEIVER) consistently show a trend towards better performance.  
The Perceiver, which employs cross- and self-attention over latent representations, performs comparably to the transformer variants, and often exhibits smaller variance across seeds.  
In contrast, the MLP and RNN architectures have a stronger dependence on the training distribution, with noticeable degradation when exposed to new circuit types.

Interestingly, models trained on Random circuits display a smaller overall improvement, but suffer less from domain shifts when tested on Pauli circuits.  
This asymmetry suggests that Random circuits, being structurally more diverse, provide a more noise-dominated, but less feature-specific learning signal, whereas Pauli circuits enable stronger, but more domain-dependent  learning.  
Overall, the results indicate that while full cross-generalization between circuit types is limited, architectures with built-in feature attention can partially bridge the gap.  
This finding supports the adoption of a mixture-of-experts strategy or hierarchical training scheme that leverages multiple circuit-type specialists.

\subsection{Generalization to a different QPU}
\label{sec:results-transfer-learning}

In this section, we test whether models trained with data from \Ibmalgiers{} can generalize to \Ibmhanoi{}.  
The hardware characteristics of \Ibmhanoi{} and \Ibmalgiers{} are comparable, allowing us to examine robustness under realistic device-level variability for each architecture.  
Three different configurations are considered:
\begin{description}
    \item[Only Predict on Hanoi:] Models pre-trained on \Ibmalgiers{} Simulated data and fine-tuned on \Ibmalgiers{} Real data, then evaluated directly on \Ibmhanoi{} Real data.
    \item[Fine-Tuned on Real Hanoi:] Models pre-trained on \Ibmalgiers{} Simulated data and fine-tuned exclusively on \Ibmhanoi{} Real data.
    \item[Transfer Learning Algiers to Hanoi:] Models pre-trained on \Ibmalgiers{} Simulated data, fine-tuned on \Ibmalgiers{} Real data, and then further fine-tuned on \Ibmhanoi{} Real data.
\end{description}

\begin{figure}[t] 
\vspace{-0.2cm}
    \centering
    \begin{subfigure}[t]{0.82\textwidth}
        \centering
        \includegraphics[width=\linewidth]{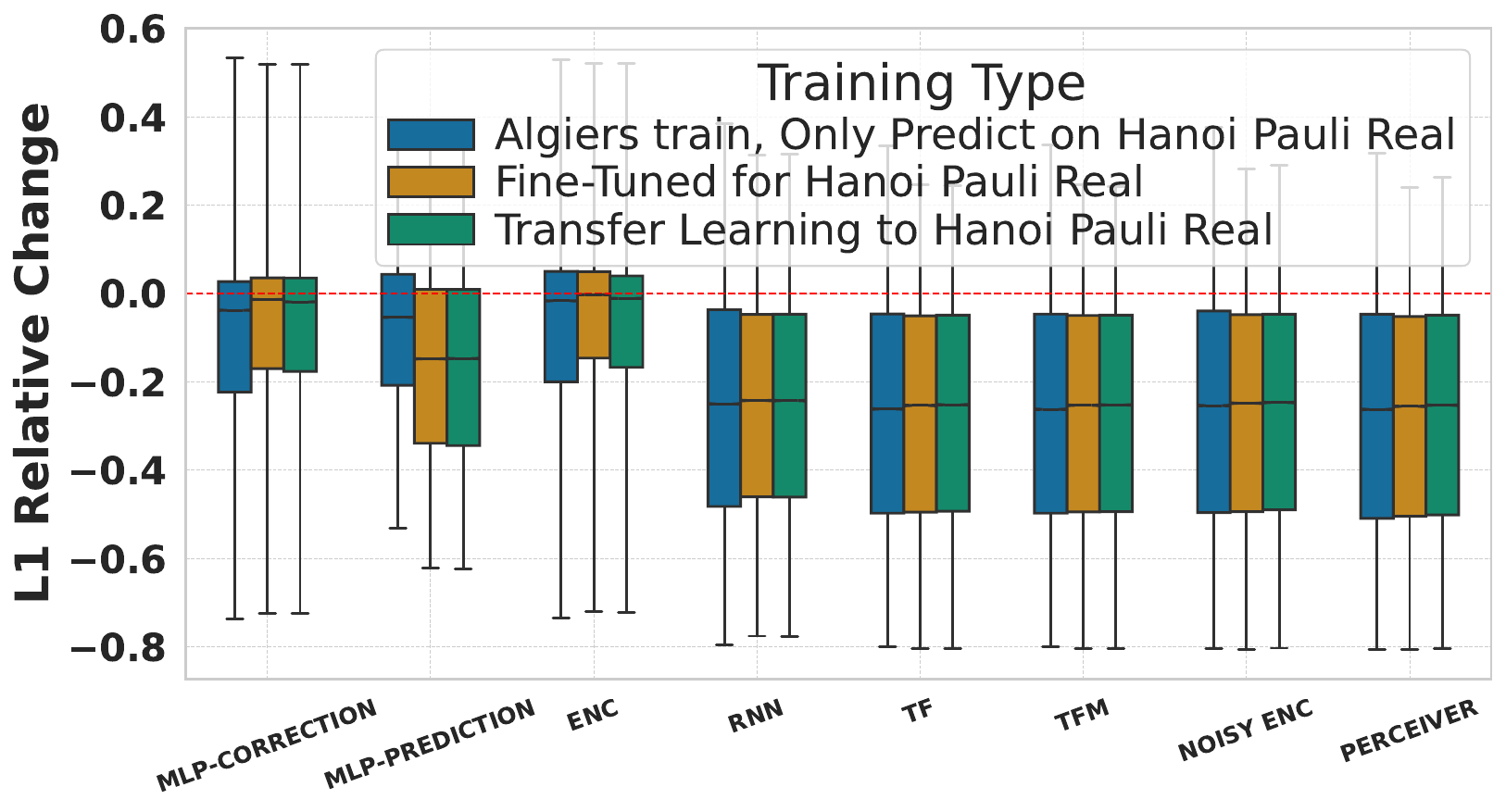}
        \subcaption{\Ibmhanoi{} Pauli Real circuits.}
        \label{fig:hanoi-pauli}
    \end{subfigure}
    \hfill
    \begin{subfigure}[t]{0.82\textwidth}
        \centering
        \includegraphics[width=\linewidth]{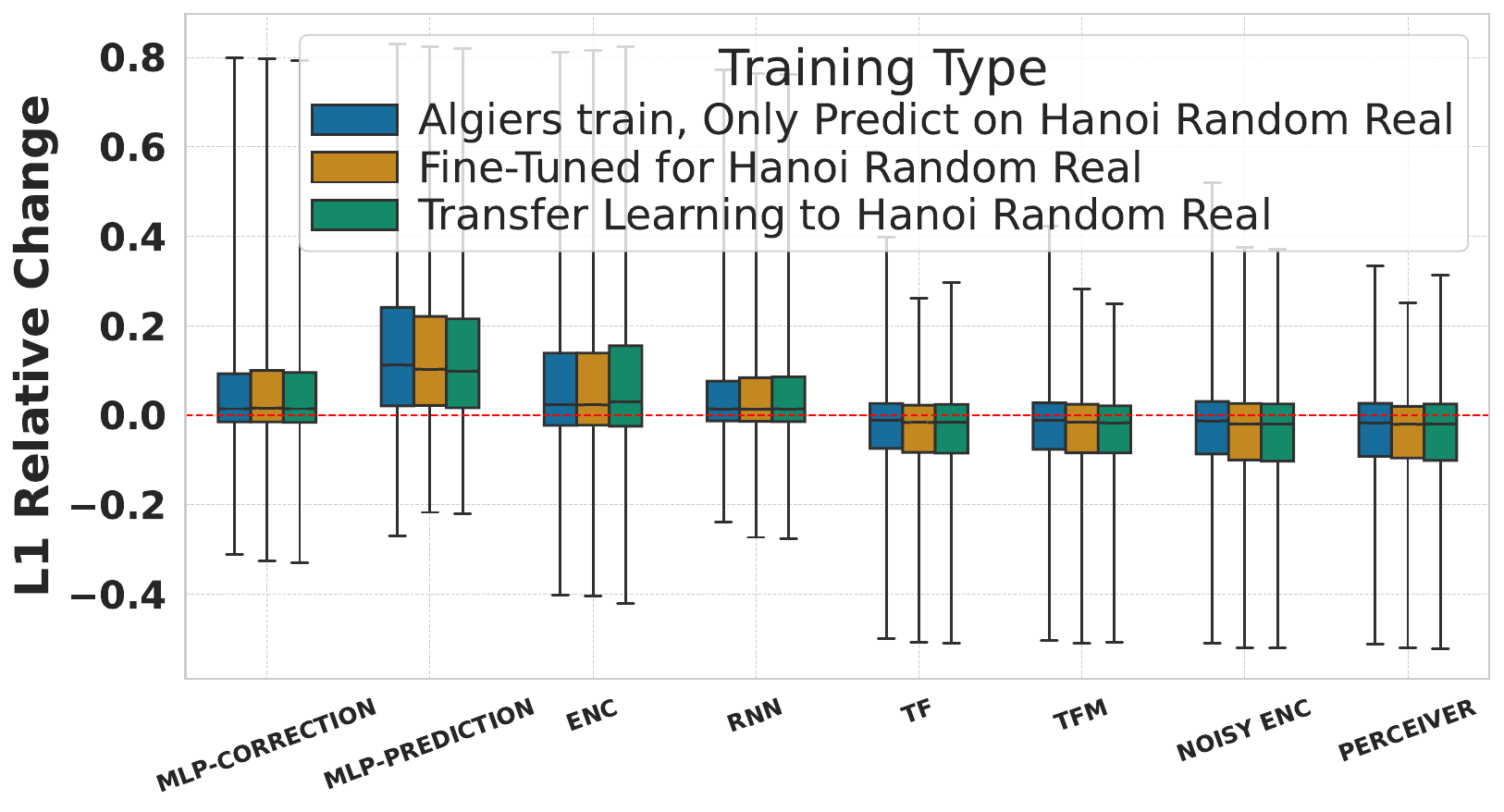}
        \subcaption{\Ibmhanoi{} Random Real circuits.}
        \label{fig:hanoi-random}
    \end{subfigure}

    \caption{\textbf{Transfer learning across devices: evaluation on \Ibmhanoi{} Real data.} 
    Box plots show the distribution of $\mathrm{L1}$ relative change values: the median (black line), interquartile range (box), and 1st–99th percentile whiskers.
    Values below the red dashed line ($y=0$) indicate successful mitigation.}
    \label{fig:hanoi-multi-boxes}
\end{figure}

\Cref{fig:hanoi-multi-boxes} and \cref{tab:different_qpus} in \cref{app:tables-performance-appendix} summarize the performances for both Pauli and Random datasets.  
We observe that all configurations achieve comparable performance to those reported for \Ibmalgiers{}, with no significant degradation.  
The transfer-learning setup offers slightly more stable results, particularly for attention-based models, confirming that the learned representations can be generalized.

The fact that earlier results indicate that the models rely primarily on the noisy distributions provided as input is likely why the performance when generalizing to a different QPU works well in this case.
Since both backends share similar noise profiles, the models’ predictions generalize effectively without requiring retraining from scratch.  
This result suggests that for near-term hardware, transfer learning can serve as a practical route to deploy models across multiple devices, within the same technology generation.

\section{Conclusions and Outlook}
\label{sec:conclusion}

In this article, we performed a systematic and comprehensive evaluation of the use of deep learning models to mitigate raw noisy outputs from quantum computers.
We collected a diverse dataset consisting of measurement shots from hundreds of thousands of circuits, using both classical simulations and real noisy QPUs available via the cloud, together with information about circuit operations and backend calibration data.
The underlying assumption is that there are patterns and correlations between measurement shots, gate sequences, and machine information, which can be leveraged to mitigate noise in a fully data-driven ML framework.
To test this assumption, we then proposed several deep learning architectures, ranging from simple FC networks to complex sequence-to-sequence generative models, to mitigate the noisy distributions.
For each, and for each of the circuit datasets, we investigated the effects of hyperparameter tuning, the effects of pre-training on simulation before deploying on real data, and the effects of the various inputs on the performances of the models.
Our benchmarking procedure is the largest available in the literature, assessing the robustness, transferability, and limitations of data-driven QEM, and it acts as a stepping stone for future large-scale efforts in applying ML to QEM~\cite{Alexeev2025_AI_for_quantum, du2025artificialintelligencerepresentingcharacterizing}.

Our results show that, across circuit types and circuit depths, deep learning models mitigating noisy distributions can match and even surpass the performance of standard baselines, such as SPAM correction, and the Repolarizer and Thresholding methods. These standard QEM baselines operate on the noisy output distributions directly, but there are other mitigation approaches that rely on carefully characterizing the noise affecting pre-determined observables \cite{aharonov2025reliablehighaccuracyerrormitigation, kim2023evidence, van2023probabilistic}.
Since we focused on mitigating distributions, we did not consider methods that mitigate expectation values of specific observables, which is one of the limitations of our study.
On the one hand, mitigating distributions automatically reduces noise bias on observables. On the other hand, the support of a noisy distribution can grow exponentially with the number of qubits, leading to poor estimation of probabilities due to finite shot sampling.
To account for this, we expect that future studies will be able to adapt our findings to the case where only marginal distributions are considered, or where generative models mitigate individual bitstrings by leveraging correlations.

Another limitation of the presented framework is the need for accurate ground-truth data, because we rely on a supervised setup to compute the training loss function; if that ground-truth data is limited to small circuits simulated via statevector methods, this hinders the scalability of the approaches.
Tensor network methods, Clifford-based methods, or mirror benchmarking methods could all be used to help extend the circuit datasets to larger sizes.
Another avenue for improvement is to investigate online-learning methods, where new data from quantum hardware is collected and used to adjust the weights of pre-trained models, or at mixture-of-experts models, where each expert is a highly optimized model on a given class of circuits and QPUs.
This would also reduce the impact of noise drift.

Finally, we emphasize the importance of open, large-scale datasets spanning multiple hardware platforms, circuit classes, and qubit counts. Such resources would enable reproducible benchmarking, and accelerate the discovery of resource-efficient mitigation methods applicable to noisy intermediate-scale quantum computers.

\paragraph{Acknowledgements}

The authors would like to thank Osaka University for providing the A100 GPUs used to generate the results in this article.
We would also like to thank Frederic Rapp and Michael Foss-Feig for their careful review of the manuscript.
Finally, we would also like to thank Keisuke Fujii for valuable discussions during the development of this article.


\bibliographystyle{unsrtnat} 
\bibliography{bib}

\clearpage

\appendix

\section{Datasets}

\subsection{Data Generation}
\label[app]{app:data}

Here we present further details about the training data, as discussed in \cref{sec:data-preparation}. 
\Cref{tab:dataset_info} lists the number of circuits of each type. 
Each circuit is run for $3$ different repeats, and for $20,000$ shots. \Cref{algpauli,algrandom} describe how the Pauli and Random circuits are generated.

\begin{table}[!ht]
    \centering
        \caption{The datasets used in this article.}
    \label{tab:dataset_info}
    \begin{tabular}{c|c|c|c|c}
        \textbf{Circuit class} & \textbf{Noise} & \textbf{QPU} &  $T$  & \textbf{Number of circuits} \\
        \hline
        Random & Real & \Ibmalgiers & 48 & 14,100 \\
         &  &  & 64 & 14,100 \\
         &  & & 80 & 14,100 \\
         &  & & 96 & 14,100 \\
         &  & & 144 & 14,100 \\ 
         &  & \Ibmhanoi{} & 48 & 15,000 \\
         &  &  & 64 & 15,000 \\
         &  & & 80 & 15,000 \\
         &  & & 96 & 15,000 \\
         &  & & 144 & 15,000 \\
          \cline{2-5}
          & Simulated & \Ibmalgiers{} & 48 & 8,000\\
         &  & & 64 & 8,000 \\
         &  & & 80 & 8,000 \\
         &  & & 96 & 8,000 \\
          &  & & 112 & 8,000 \\
         &  & & 144 & 8,000 \\
        \hline
         Pauli & Real & \Ibmalgiers{} & 3 & 14,400 \\
         &  &  & 4 & 14,400 \\
         &  & & 5 & 14,400 \\
         &  & & 6 & 14,400 \\
         &  & & 9 & 14,400 \\
          & & \Ibmhanoi & 3 & 15,000 \\
         &  &  & 4 & 15,000 \\
         &  & & 5 & 15,000 \\
         &  & & 6 & 15,000 \\
         &  & & 9 & 15,000 \\
         \cline{2-5}
          & Simulated & \Ibmalgiers{} & 3 & 8,000 \\
         &  & & 4 & 8,000 \\
         &  & & 5 & 8,000 \\
         &  & & 6 & 8,000 \\
          &  & & 7 & 8,000 \\
         &  & & 9 & 8,000 \\
    \end{tabular}
\end{table}

\begin{algorithm}[!ht]
    \caption{Pauli circuit generation protocol}
    \label{algpauli}
    \begin{algorithmic}[1]
        \Require Number of qubits, $N$; Maximum number of time steps (gadgets), $T$;
        \State Initialize $\Sigma \leftarrow \{I,X,Y,Z\}^N$
        \State Initialize $\mathcal{C} \leftarrow \text{empty circuit}$
        \For{$i$ = 1 to $N$}
        \State $q_i \leftarrow \ket0$
        \EndFor
        \For{$t$ = 1 to $T$}
        \State $P \leftarrow $ uniform random choice from $\Sigma$
        \State $\alpha \leftarrow $ uniform random choice from  $[0, 2\pi)$
        \State $\mathcal{C} \leftarrow \mathcal{C} ; \exp(-i\alpha P)[q_1,\ldots,q_N]$
        \EndFor
        \For{$i$ = 1 to $N$}
        \State $\mathcal{C} \leftarrow \mathcal{C} ; \text{Measure}[q_i]$
        \EndFor
        \State \Return $\mathcal{C}$
    \end{algorithmic}
\end{algorithm}

\begin{algorithm}[!ht]
    \caption{Random circuit generation protocol}
    \label{algrandom}
    \begin{algorithmic}[1]
        \Require Number of qubits, $N$; Maximum number of time steps (gates), $T$;
        \State Initialize $\mathcal{G} \leftarrow \{ R_z, X, SX, CX \}$
        \State Initialize $\mathcal{C} \leftarrow \text{empty circuit}$
        \For{$i$ = 1 to $N$}
        \State $q_i \leftarrow \ket0$
        \EndFor
        \For{$t$ = 1 to $T$}
        \State $i \leftarrow $ uniform random choice from $\{1,\ldots,N\}$
        \State $G \leftarrow $ uniform random choice from $\mathcal{G}$
        \If{$G = CX$}
        \State $j \leftarrow $ uniform random choice from  $\{1,...,N\}
        \setminus \{i\}$
        \State $\mathcal{C} \leftarrow \mathcal{C} ; CX[q_i,q_j]$ 
        \ElsIf{$G=R_{z}$} 
        \State $\alpha \leftarrow $ uniform random choice from  $[0, 2\pi)$
        \State $\mathcal{C} \leftarrow \mathcal{C} ; R_Z(\alpha)[q_i]$ 
        \Else
        \State $\mathcal{C} \leftarrow \mathcal{C} ; G[q_i]$ 
        \EndIf
        \EndFor
        \For{$i$ = 1 to $N$}
        \State $\mathcal{C} \leftarrow \mathcal{C} ; \text{Measure}[q_i]$
        \EndFor
        \State \Return $\mathcal{C}$
    \end{algorithmic}
\end{algorithm}

\subsection{Datasets Pre-processing}
\label[app]{app:post-processed-datasets}

Here we describe how the input data is preprocessed for each model before being passed to the model.

\paragraph{MLPs:}

The input to the MLP for each data point is constructed by flattening the padded Circuit,  $\Carray{} \in \mathbb{R}^{5200}$, to form a 1D array. 
This is then concatenated with the 1D backend information array \Backendinfo, forming a 1D input array $\mathbf{D}(\Carray, \Backendinfo) \in \mathbb{R}^{5301}$.
The noisy probability density 1D array, $\Ndensity{} \in \mathbb{R}^{32}$, is also an input.
The target of the model is the ideal probability density 1D array, $\Idensity{} \in \mathbb{R}^{32}$. 
The inputs and targets for each data point $i$ can be summarized as follows
\begin{align}
    &\text{Inputs}_\text{MLP} = \{ [\mathbf{D}^{i}(\Carray^{i}, \Backendinfo^{i})], [\Ndensity^{i}]) \mid i = 1, \dots, N \} , \nonumber \\
    &\text{Target}_\text{MLP} = \{ ( [\Idensity]) \mid i = 1, \dots, N \}. 
\end{align}

\paragraph{Encoder, RNN, Transformers and Perceiver:}

For each data point, the input to the Encoder, RNN, transformer-based and Perceiver models have the same structure. 
For each layer in \Carray{}, embedding vectors $\mathbf{E}$ are created. 
These are formed by first flattening the $5\times5$ matrix representing the gate operations for that layer, defined in \cref{tab:circuit encoding}, into a 1D array.

If a particular layer contains no operations (i.e. the entire matrix is zeros due to padding), a vector of zeros is added to the embedding, and a mask value of $-2$ is assigned to mark this as a padded time step. When the time step does contain operations, the non-zero elements of the matrix are identified, and the \Backendinfo{} properties associated with the relevant qubits and operations are extracted, with the non-relevant elements assigned mask values of zero. 
These backend properties are concatenated with the flattened gate-operation matrix to form $\mathbf{E} \in \mathbb{R}^{126}$. 

This process is repeated for all time steps, creating a list of embedding vectors $\{ \mathbf{E}_1, \mathbf{E}_2, \dots, \mathbf{E}_{n_l} \}$ for each circuit, along with corresponding masks. 
The noisy probability density 1D array, \Ndensity{}, forms a second input. 

The target of the model is the ideal probability density 1D array, \Idensity. 
The inputs and targets for each data point $i$ can be summarized as follows
\begin{align}
    &\text{Inputs}_\text{ENC, RNN, TFM} = \{ [\{ \mathbf{E}_1, \mathbf{E}_2, \dots, \mathbf{E}_t \}^{i}(\Carray^{i},\Backendinfo^{i})], [\Ndensity]) \mid i = 1, \dots, N \} , \nonumber \\
    &\text{Target}_\text{ENC, RNN, TFM} = \{ ( [\Idensity]) \mid i = 1, \dots, N \}. 
\end{align}

\paragraph{Noisy Encoder:}

This model takes as input only the noisy probability distribution, \Ndensity{}, and directly predicts a mitigated distribution. 
No circuit or backend information is used. 

\section{Model Architectures and Training}
\label{sec:models-details}

This section summarizes the architectures, hyperparameter optimization (HPO), training protocols, and fine-tuning strategies for all models.

For all architectures, we perform a systematic hyperparameter search using Optuna~\cite{Akiba2019OptunaAN}. 
Each study consists of 50 trials and uses Optuna’s \textit{TPE} \cite{bergstra2011} (Tree–structured Parzen Estimator) sampler. 
All studies are executed using PyTorch Lightning~\cite{NEURIPS2019_bdbca288}, with distributed data-parallel training on GPU nodes, early-stopping based on validation loss, and full experiment tracking via Weights \& Biases~\cite{wandb}. 
For reproducibility, each study is stored on disk via an SQLite Optuna backend, enabling resumability and ensuring that the best-performing hyperparameters across all completed trials can be deterministically recovered.

All experiments were repeated over 5 random seeds to account for randomness in initialization and training.

\subsection{MLP Models}
\label{app:mlp-models-training}

\paragraph{Architecture.}
Two FC models are used: \textbf{MLP-Correction}, which applies element-wise correction to noisy probabilities, and \textbf{MLP-Prediction}, which learns a full mapping from concatenated circuit/backend features and noisy outputs to the mitigated distribution. 

\begin{table}[!ht]
\caption{Summary of the MLP model.}
\scriptsize
\setlength{\tabcolsep}{2.5pt}
\renewcommand{\arraystretch}{1.05}
\begin{tabular}{@{}lc@{}}
\toprule
 & \textbf{MLP-Correction} and \textbf{MLP-Prediction} \\
\midrule
Input & $(\mathbf{x}_{\mathrm{CB}}, \Ndensity)$ \\
Architecture & FC layers + BN + activation  \\
Output & 32-dimension softmax \\
Loss & KL-divergence \\
Preprocessing & Quantile + MinMax scaling  \\
\bottomrule
\end{tabular}
\end{table}

\paragraph{HPO, Training and Fine-Tuning.}
HPO over learning rate (LR), number of layers, batch size, weight decay (WD), activation function, and choice of optimizer. Training uses Adam~\cite{kingma2015adam}, Pytorch's ReduceLROnPlateau, early stopping, and up to $2,000$ epochs. Fine-tuning reuses Simulated-data checkpoints with the learning rate reduced by a factor of 10.
\begin{table}[!ht]
\caption{Best configurations for the MLP model.}
\scriptsize
\setlength{\tabcolsep}{2.5pt}
\renewcommand{\arraystretch}{1.05}
\begin{tabular}{lcccccc}
\toprule
Dataset & LR & Layers & Batch size & WD & Activation & Optimizer \\
\midrule
\multicolumn{7}{c}{\textbf{MLP-Correction}} \\
Pauli  & $4.66\!\times\!10^{-3}$ & [3701,256] & 128 & $4.31\!\times\!10^{-4}$ & ReLU & Adam \\
Random & $2.98\!\times\!10^{-5}$ & [3701,512,512] & 64 & $2.64\!\times\!10^{-4}$ & ReLU & Adam \\
\midrule
\multicolumn{7}{c}{\textbf{MLP-Prediction}} \\
Pauli  & $9.59\!\times\!10^{-5}$ & [3733,512,512] & 128 & $3.92\!\times\!10^{-3}$ & Tanh & Adam \\
Random & $1.00\!\times\!10^{-5}$ & [3733,256] & 32 & $3.45\!\times\!10^{-5}$ & Tanh & Adam \\
\bottomrule
\end{tabular}
\end{table}

\subsection{RNN Model}
\label{app:rnn-model-training}

\paragraph{Architecture.}
Two RNN branches process embedded circuit/backend sequences and noisy output sequences. Final hidden states are concatenated and projected to a 32-dimensional softmax distribution.

\begin{table}[!ht]
\caption{Summary of the RNN model.}
\scriptsize
\setlength{\tabcolsep}{2.5pt}
\renewcommand{\arraystretch}{1.05}
\begin{tabular}{@{}ll@{}}
\toprule
input & $\mathbf{X}_{\mathrm{CB}}$ and  $\Ndensity$ \\
Hidden sizes & 64, 128, 256 \\
Output & Linear + softmax (32-dimensional) \\
Loss & KL-divergence \\
\bottomrule
\end{tabular}
\end{table}

\paragraph{HPO, Training and Fine-Tuning.}
HPO over LR, hidden size, batch size, and WD. Training used Adam, ReduceLROnPlateau, early stopping, and up to $2,000$ epochs. Fine-tuning reuses simulated data checkpoints with the LR divided by 10.

\begin{table}[!ht]
\caption{Best configurations for the RNN model.}
\scriptsize
\setlength{\tabcolsep}{2.5pt}
\renewcommand{\arraystretch}{1.05}
\begin{tabular}{lcccc}
\toprule
Dataset & LR & Hidden size & Batch size & WD \\
\midrule
Pauli  & $3.43\!\times\!10^{-3}$ & 64  & 128 & $1.12\!\times\!10^{-5}$ \\
Random & $1.02\!\times\!10^{-4}$ & 128 & 128 & $1.83\!\times\!10^{-5}$ \\
\bottomrule
\end{tabular}

\end{table}

\subsection{Encoder Model}
\label{app:encod-model-train}

\paragraph{Architecture.}
Transformer encoder with positional encoding, followed by 1D convolutions, BN, and residual correction to the noisy input.

\paragraph{HPO, Training and Fine-Tuning.}
HPO over LR, hidden size, number of layers, heads, dropout, batch size, and WD. Training used Adam, KL-divergence loss, Pytorch's ReduceLROnPlateau, and early stopping. Fine-tuning reduced the LR by a factor of 10.

\begin{table}[!ht]
\caption{Best configurations for the Encoder model.}
\scriptsize
\setlength{\tabcolsep}{2.5pt}
\renewcommand{\arraystretch}{1.05}
\begin{tabular}{lccccccc}
\toprule
Dataset & LR & Hidden size & Layers & Heads & Dropout & Batch size & WD \\
\midrule
Pauli  & $7.37\!\times\!10^{-3}$ & 100 & 2 & 2 & 0.1 & 128 & $1.02\!\times\!10^{-5}$ \\
Random & $5.95\!\times\!10^{-2}$ & 100 & 6 & 4 & 0.2 & 32  & $1.00\!\times\!10^{-5}$ \\
\bottomrule
\end{tabular}
\end{table}

\subsection{Transformer Models}
\label{app:transf-model-train}

\paragraph{Architecture.}
Standard encoder–decoder Transformer. TF only uses \Ndensity{} as input; TFM augments the decoder input of \Ndensity{} with SPAM-mitigated data. Output is the final decoder token (32-dimension softmax).

\paragraph{HPO, Training, Fine-Tuning. }
HPO over LR, hidden size, number of layers, heads, dropout, batch size, depth sequence (for TF), and WD.
AdamW~\cite{loshchilov2019decoupled} optimization, ReduceLROnPlateau scheduling, and early stopping during HPO (patience 4) were used. Fine-tuning on Real data used the LR divided by 10 and up to $400$ epochs.

\begin{table}[!ht]
\caption{Best configurations for the Transformer model.}
\scriptsize
\setlength{\tabcolsep}{2.5pt}
\renewcommand{\arraystretch}{1.05}
\begin{tabular}{lcccccccc}
\toprule
Dataset & LR & Hidden size & Layers & Heads & Dropout & Batch size & Depth & WD \\
\midrule
\multicolumn{9}{c}{\textbf{TF}} \\
Pauli  & $7.31\!\times\!10^{-4}$ & 1024 & 1 & 2 & 0.01 & 64 & 1 & $3.08\!\times\!10^{-5}$ \\
Random & $1.89\!\times\!10^{-4}$ & 512  & 2 & 4 & 0.01 & 32 & 1 & $6.25\!\times\!10^{-3}$ \\
\midrule
\multicolumn{9}{c}{\textbf{TFM}} \\
Pauli  & $7.39\!\times\!10^{-4}$ & 2048 & 1 & 4 & 0.01 & 64 & NA & $1.89\!\times\!10^{-3}$ \\
Random & $5.50\!\times\!10^{-4}$ & 1024 & 2 & 4 & 0.01 & 128 & NA & $2.43\!\times\!10^{-5}$ \\
\bottomrule
\end{tabular}
\end{table}

\subsection{Noisy Encoder}
\label{app:noisy-encoder-model-training}

\paragraph{Architecture.}
Transformer encoder operating on a single noisy probability vector (32-dimension) with a learnable positional embedding. Output projected to a softmax distribution.

\begin{table}[!ht]
\caption{Summary of the Noisy Encoder model.}
\scriptsize
\setlength{\tabcolsep}{2.5pt}
\renewcommand{\arraystretch}{1.05}
\begin{tabular}{@{}ll@{}}
\toprule
Input & \Ndensity \\
Encoder & Transformer (1-3 layers) \\
Output & FC + softmax \\
Loss & KL-divergence \\
\bottomrule
\end{tabular}
\end{table}

\paragraph{HPO, Training, Fine-Tuning.}
HPO over transformer depth, hidden size, heads, dropout, LR, batch size, and WD. Training uses AdamW with early stopping; fine-tuning reduces the LR by a factor of 10.

\begin{table}[ht!]
\caption{Best configurations for the Noisy Encoder model.}
\scriptsize
\setlength{\tabcolsep}{2.5pt}
\renewcommand{\arraystretch}{1.05}
\begin{tabular}{lccccccc}
\toprule
Dataset & LR & Hidden size & Layers & Heads & Dropout & Batch size & WD \\
\midrule
Pauli  & $2.63\!\times\!10^{-4}$ & 1024 & 3 & 2 & 0.01 & 128 & $1.06\!\times\!10^{-5}$ \\
Random & $1.25\!\times\!10^{-4}$ & 2048 & 3 & 4 & 0.01 & 128 & $1.14\!\times\!10^{-4}$ \\
\bottomrule
\end{tabular}

\end{table}

\subsection{Perceiver Model}
\label[app]{app:perceiver-model-training}

\paragraph{Architecture.}
Perceiver with noisy-concatenated inputs $(\mathbf{x}_{\mathrm{CB}} + \Ndensity$), latent bottleneck, alternating cross- and self-attention blocks, and MLP head producing a 32-dimensional softmax output.

\paragraph{HPO, Training and Fine-Tuning.}
HPO over latent size, hidden size, attention heads, blocks (Blk), attention heads (heads), dropout, attention sequence per block (S/Blk), temperature, LR, and WD. Training uses AdamW, KL-divergence loss, ReduceLROnPlateau, early stopping, and five random seeds. Fine-tuning on Real data reuses Simulated checkpoints, with the LR divided by 10 and an identical training protocol.

\begin{table}[ht!]
\caption{Best configurations for Perceiver model.}
\scriptsize
\setlength{\tabcolsep}{2.5pt}
\renewcommand{\arraystretch}{1.05}

\begin{tabular}{@{}lccccccccc@{}}
\toprule
Dataset & Hidden size & Latent size & Blk & Heads & S/Blk & Dropout & Temperature & LR & WD \\
\midrule
Pauli (Real)  & 1024 & 256 & 5 & (4,8)  & 1 & 0.155 & 1.17 & $1.25\times10^{-5}$ & $9.13\times10^{-3}$ \\
Random (Real) &  768 & 256 & 4 & (8,16) & 1 & 0.049 & 0.86 & $1.93\times10^{-5}$ & $3.09\times10^{-5}$ \\
\bottomrule
\end{tabular}

\end{table}

\newpage
\section{Reference Tables}
\label[app]{app:tables-performance-appendix}

In this section, we include tables detailing the results discussed in the main text (which are often visualized as boxplots).

The data corresponding to \cref{fig:simulated-results} are given in \cref{tab:simulated-table}.


{\scriptsize
\setlength{\tabcolsep}{3pt}
\renewcommand{\arraystretch}{0.9}
\setlength{\LTleft}{0pt}
\setlength{\LTright}{0pt}

\begin{longtable}{@{\extracolsep{\fill}}>{\raggedright\arraybackslash}m{2.9cm}lcccc@{}}
\label{tab:simulated-table}\\

\toprule
Dataset & Model & L1RC Median ($\pm$SE) & P25 & P75 & L1RC \% Improved \\
\midrule
\endfirsthead

\toprule
Dataset & Model & L1RC Median ($\pm$SE) & P25 & P75 & L1RC \% Improved \\
\midrule
\endhead

\bottomrule
\endfoot

\bottomrule
\caption{Performance of models trained on Simulated \Ibmalgiers{} data.  The table reports the median ($\pm$SE) and interquartile range (P25--P75) of the L1 relative change across datasets, with the final column showing the percentage of circuits that are improved, as compared to the noisy baseline. In \textbf{bold} the best performing model per dataset, in green improvements over the noisy baseline, in red worse results than the baseline.}\\
\endlastfoot

\Needspace{12\baselineskip}
\multirow[t]{8}{2.9cm}{\parbox[t]{2.9cm}{\raggedright Pauli}}
 & MLP\text{-}CORRECTION & \textcolor{BrickRed}{0.2039 $\pm$ 0.0019} & \textcolor{BrickRed}{0.0004} & \textcolor{BrickRed}{0.5328} & 24.5\% \\
 & MLP\text{-}PREDICTION & \textcolor{ForestGreen}{-0.0863 $\pm$ 0.0125} & \textcolor{ForestGreen}{-0.2836} & \textcolor{BrickRed}{0.1150} & 61.3\% \\
 & ENC & \textcolor{BrickRed}{0.2030 $\pm$ 0.0011} & \textcolor{BrickRed}{0.0013} & \textcolor{BrickRed}{0.5267} & 23.9\% \\
 & RNN & \textcolor{ForestGreen}{-0.4875 $\pm$ 0.0045} & \textcolor{ForestGreen}{-0.6272} & \textcolor{ForestGreen}{-0.2643} & 90.2\% \\
 & \textbf{TF} & \textcolor{ForestGreen}{-0.6318 $\pm$ 0.0024} & \textcolor{ForestGreen}{-0.7296} & \textcolor{ForestGreen}{-0.4637} & 94.2\% \\
 & TFM & \textcolor{ForestGreen}{-0.6277 $\pm$ 0.0017} & \textcolor{ForestGreen}{-0.7262} & \textcolor{ForestGreen}{-0.4608} & 94.1\% \\
 & NOISY\text{-}ENC & \textcolor{ForestGreen}{-0.6263 $\pm$ 0.0017} & \textcolor{ForestGreen}{-0.7320} & \textcolor{ForestGreen}{-0.4362} & 93.1\% \\
 & PERCEIVER & \textcolor{ForestGreen}{-0.5736 $\pm$ 0.0006} & \textcolor{ForestGreen}{-0.7032} & \textcolor{ForestGreen}{-0.3503} & 91.4\% \\
\addlinespace[0.3mm] 


\midrule
\multirow[t]{8}{2.9cm}{\parbox[t]{2.9cm}{\raggedright Random}}
 & MLP\text{-}CORRECTION & \textcolor{ForestGreen}{-0.0765 $\pm$ 0.0021} & \textcolor{ForestGreen}{-0.2419} & \textcolor{BrickRed}{0.1035} & 63.1\% \\
 & MLP\text{-}PREDICTION & \textcolor{BrickRed}{0.3979 $\pm$ 0.0017} & \textcolor{BrickRed}{0.1365} & \textcolor{BrickRed}{0.7583} & 11.9\% \\
 & ENC & \textcolor{BrickRed}{0.2773 $\pm$ 0.0010} & \textcolor{BrickRed}{0.1353} & \textcolor{BrickRed}{0.5162} & 9.6\% \\
 & RNN & \textcolor{ForestGreen}{-0.0652 $\pm$ 0.0004} & \textcolor{ForestGreen}{-0.1699} & \textcolor{BrickRed}{0.0395} & 67.5\% \\
 & TF & \textcolor{ForestGreen}{-0.5273 $\pm$ 0.0030} & \textcolor{ForestGreen}{-0.6464} & \textcolor{ForestGreen}{-0.2950} & 88.7\% \\
 & TFM & \textcolor{ForestGreen}{-0.5244 $\pm$ 0.0033} & \textcolor{ForestGreen}{-0.6418} & \textcolor{ForestGreen}{-0.2985} & 89.3\% \\
 & NOISY\text{-}ENC & \textcolor{ForestGreen}{-0.4791 $\pm$ 0.0052} & \textcolor{ForestGreen}{-0.6052} & \textcolor{ForestGreen}{-0.1911} & 84.6\% \\
 & \textbf{PERCEIVER} & \textcolor{ForestGreen}{-0.5450 $\pm$ 0.0030} & \textcolor{ForestGreen}{-0.6414} & \textcolor{ForestGreen}{-0.3628} & 91.7\% \\

\end{longtable}
}

\newpage

The data corresponding to \cref{fig:real-algiers-results} are given in \cref{tab:performance_results_algiers}.

{\scriptsize
\setlength{\tabcolsep}{3pt}
\renewcommand{\arraystretch}{0.9}
\setlength{\LTleft}{0pt}
\setlength{\LTright}{0pt}

\begin{longtable}{@{\extracolsep{\fill}}>{\raggedright\arraybackslash}m{2.9cm}lcccc@{}}
\label{tab:performance_results_algiers}\\

\toprule
Dataset & Model & L1RC Median ($\pm$SE) & P25 & P75 & L1RC \% Improved \\
\midrule
\endfirsthead

\toprule
Dataset & Model & L1RC Median ($\pm$SE) & P25 & P75 & L1RC \% Improved \\
\midrule
\endhead

\bottomrule
\endfoot

\bottomrule
\caption{Performance of models trained on Pauli and Random Real datasets. The table reports the median ($\pm$SE) and interquartile range (P25--P75) across datasets, with the final column showing the percentage of circuits that are improved, as compared to the noisy baseline. In \textbf{bold} the best performing model per dataset, in green improvements over the noisy baseline, in red worse results than the baseline.}\\
\endlastfoot

\multicolumn{6}{l}{\textbf{Pauli}}\\
\midrule

\Needspace{12\baselineskip}
\multirow[t]{8}{2.9cm}{\raggedright Fine-Tuned\\ on Real Data}
 & MLP\text{-}CORRECTION & \textcolor{ForestGreen}{-0.0239 $\pm$ 0.0003} & \textcolor{ForestGreen}{-0.1794} & \textcolor{BrickRed}{0.0133} & 64.4\% \\
 & MLP\text{-}PREDICTION & \textcolor{ForestGreen}{-0.1186 $\pm$ 0.0012} & \textcolor{ForestGreen}{-0.3170} & \textcolor{BrickRed}{0.0144} & 71.9\% \\
 & ENC                  & \textcolor{ForestGreen}{-0.0157 $\pm$ 0.0008} & \textcolor{ForestGreen}{-0.1740} & \textcolor{BrickRed}{0.0272} & 58.8\% \\
 & RNN                  & \textcolor{ForestGreen}{-0.1890 $\pm$ 0.0010} & \textcolor{ForestGreen}{-0.4049} & \textcolor{ForestGreen}{-0.0304} & 81.3\% \\
 & \textbf{TF}          & \textcolor{ForestGreen}{-0.2039 $\pm$ 0.0005} & \textcolor{ForestGreen}{-0.4469} & \textcolor{ForestGreen}{-0.0319} & 82.2\% \\
 & TFM                  & \textcolor{ForestGreen}{-0.2030 $\pm$ 0.0004} & \textcolor{ForestGreen}{-0.4458} & \textcolor{ForestGreen}{-0.0311} & 82.4\% \\
 & NOISY ENC            & \textcolor{ForestGreen}{-0.1945 $\pm$ 0.0029} & \textcolor{ForestGreen}{-0.4372} & \textcolor{ForestGreen}{-0.0308} & 81.8\% \\
 & PERCEIVER            & \textcolor{ForestGreen}{-0.2062 $\pm$ 0.0027} & \textcolor{ForestGreen}{-0.4542} & \textcolor{ForestGreen}{-0.0348} & 83.2\% \\
\midrule

\Needspace{12\baselineskip}
\multirow[t]{8}{2.9cm}{\raggedright Trained Only\\ on Real Data}
 & MLP\text{-}CORRECTION & \textcolor{ForestGreen}{-0.0248 $\pm$ 0.0005} & \textcolor{ForestGreen}{-0.1844} & \textcolor{BrickRed}{0.0117} & 65.0\% \\
 & MLP\text{-}PREDICTION & \textcolor{ForestGreen}{-0.0216 $\pm$ 0.0034} & \textcolor{ForestGreen}{-0.2029} & \textcolor{BrickRed}{0.0988} & 54.5\% \\
 & ENC                  & \textcolor{ForestGreen}{-0.0167 $\pm$ 0.0022} & \textcolor{ForestGreen}{-0.1696} & \textcolor{BrickRed}{0.0220} & 60.2\% \\
 & RNN                  & \textcolor{ForestGreen}{-0.1703 $\pm$ 0.0018} & \textcolor{ForestGreen}{-0.3920} & \textcolor{ForestGreen}{-0.0201} & 79.5\% \\
 & TF                   & \textcolor{ForestGreen}{-0.1810 $\pm$ 0.0020} & \textcolor{ForestGreen}{-0.4202} & \textcolor{ForestGreen}{-0.0267} & 81.9\% \\
 & TFM                  & \textcolor{ForestGreen}{-0.1820 $\pm$ 0.0027} & \textcolor{ForestGreen}{-0.4204} & \textcolor{ForestGreen}{-0.0233} & 80.5\% \\
 & NOISY ENC            & \textcolor{ForestGreen}{-0.1808 $\pm$ 0.0007} & \textcolor{ForestGreen}{-0.4206} & \textcolor{ForestGreen}{-0.0226} & 80.2\% \\
 & \textbf{PERCEIVER}   & \textcolor{ForestGreen}{-0.1964 $\pm$ 0.0009} & \textcolor{ForestGreen}{-0.4496} & \textcolor{ForestGreen}{-0.0134} & 77.5\% \\
\midrule

\Needspace{12\baselineskip}
\multirow[t]{8}{2.9cm}{\raggedright Trained on\\ Simulated Data}
 & MLP\text{-}CORRECTION & \textcolor{BrickRed}{0.0346 $\pm$ 0.0008} & \textcolor{ForestGreen}{-0.0010} & \textcolor{BrickRed}{0.1028} & 25.8\% \\
 & MLP\text{-}PREDICTION & \textcolor{ForestGreen}{-0.0512 $\pm$ 0.0036} & \textcolor{ForestGreen}{-0.2075} & \textcolor{BrickRed}{0.0482} & 62.6\% \\
 & ENC                  & \textcolor{BrickRed}{0.0356 $\pm$ 0.0017} & \textcolor{BrickRed}{0.0005} & \textcolor{BrickRed}{0.1061} & 24.6\% \\
 & RNN                  & \textcolor{ForestGreen}{-0.1673 $\pm$ 0.0005} & \textcolor{ForestGreen}{-0.3344} & \textcolor{ForestGreen}{-0.0324} & 81.9\% \\
 & TF                   & \textcolor{ForestGreen}{-0.1751 $\pm$ 0.0014} & \textcolor{ForestGreen}{-0.3566} & \textcolor{ForestGreen}{-0.0356} & 81.4\% \\
 & TFM                  & \textcolor{ForestGreen}{-0.1625 $\pm$ 0.0015} & \textcolor{ForestGreen}{-0.3420} & \textcolor{ForestGreen}{-0.0318} & 81.3\% \\
 & NOISY ENC            & \textcolor{ForestGreen}{-0.1779 $\pm$ 0.0003} & \textcolor{ForestGreen}{-0.3587} & \textcolor{ForestGreen}{-0.0327} & 80.5\% \\
 & \textbf{PERCEIVER}   & \textcolor{ForestGreen}{-0.1969 $\pm$ 0.0014} & \textcolor{ForestGreen}{-0.4333} & \textcolor{ForestGreen}{-0.0186} & 77.6\% \\
\addlinespace[0.6mm]

\multicolumn{6}{l}{\textbf{Random}}\\
\midrule

\Needspace{12\baselineskip}
\multirow[t]{8}{2.9cm}{\raggedright Fine-Tuned\\ on Real Data}
 & MLP\text{-}CORRECTION & \textcolor{BrickRed}{0.0064 $\pm$ 0.0002} & \textcolor{ForestGreen}{-0.0175} & \textcolor{BrickRed}{0.0599} & 42.5\% \\
 & MLP\text{-}PREDICTION & \textcolor{BrickRed}{0.0862 $\pm$ 0.0009} & \textcolor{BrickRed}{0.0111} & \textcolor{BrickRed}{0.1937} & 21.8\% \\
 & ENC                  & \textcolor{BrickRed}{0.0161 $\pm$ 0.0001} & \textcolor{ForestGreen}{-0.0246} & \textcolor{BrickRed}{0.1074} & 40.0\% \\
 & RNN                  & \textcolor{BrickRed}{0.0059 $\pm$ 0.0002} & \textcolor{ForestGreen}{-0.0174} & \textcolor{BrickRed}{0.0462} & 43.1\% \\
 & TF                   & \textcolor{ForestGreen}{-0.0166 $\pm$ 0.0000} & \textcolor{ForestGreen}{-0.0830} & \textcolor{BrickRed}{0.0168} & 62.7\% \\
 & TFM                  & \textcolor{ForestGreen}{-0.0168 $\pm$ 0.0000} & \textcolor{ForestGreen}{-0.0818} & \textcolor{BrickRed}{0.0167} & 63.0\% \\
 & NOISY ENC            & \textcolor{ForestGreen}{-0.0185 $\pm$ 0.0003} & \textcolor{ForestGreen}{-0.0956} & \textcolor{BrickRed}{0.0210} & 62.2\% \\
 & \textbf{PERCEIVER}   & \textcolor{ForestGreen}{-0.0210 $\pm$ 0.0001} & \textcolor{ForestGreen}{-0.1019} & \textcolor{BrickRed}{0.0160} & 64.4\% \\
\midrule

\Needspace{12\baselineskip}
\multirow[t]{8}{2.9cm}{\raggedright Trained Only\\ on Real Data}
 & MLP\text{-}CORRECTION & \textcolor{BrickRed}{0.0213 $\pm$ 0.0003} & \textcolor{ForestGreen}{-0.0240} & \textcolor{BrickRed}{0.1122} & 37.7\% \\
 & MLP\text{-}PREDICTION & \textcolor{BrickRed}{0.1290 $\pm$ 0.0005} & \textcolor{BrickRed}{0.0337} & \textcolor{BrickRed}{0.2603} & 17.2\% \\
 & ENC                  & \textcolor{BrickRed}{0.0232 $\pm$ 0.0004} & \textcolor{ForestGreen}{-0.0270} & \textcolor{BrickRed}{0.1238} & 38.1\% \\
 & RNN                  & \textcolor{BrickRed}{0.0111 $\pm$ 0.0001} & \textcolor{ForestGreen}{-0.0191} & \textcolor{BrickRed}{0.0631} & 40.3\% \\
 & TF                   & \textcolor{ForestGreen}{-0.0141 $\pm$ 0.0003} & \textcolor{ForestGreen}{-0.0877} & \textcolor{BrickRed}{0.0331} & 58.0\% \\
 & TFM                  & \textcolor{ForestGreen}{-0.0159 $\pm$ 0.0003} & \textcolor{ForestGreen}{-0.0972} & \textcolor{BrickRed}{0.0332} & 58.5\% \\
 & NOISY ENC            & \textcolor{ForestGreen}{-0.0139 $\pm$ 0.0003} & \textcolor{ForestGreen}{-0.0917} & \textcolor{BrickRed}{0.0333} & 58.2\% \\
 & \textbf{PERCEIVER}   & \textcolor{ForestGreen}{-0.0175 $\pm$ 0.0002} & \textcolor{ForestGreen}{-0.0976} & \textcolor{BrickRed}{0.0281} & 60.0\% \\
\midrule

\Needspace{12\baselineskip}
\multirow[t]{8}{2.9cm}{\raggedright Trained on\\ Simulated Data}
 & MLP\text{-}CORRECTION & \textcolor{BrickRed}{0.0276 $\pm$ 0.0005} & \textcolor{ForestGreen}{-0.0085} & \textcolor{BrickRed}{0.1125} & 31.0\% \\
 & MLP\text{-}PREDICTION & \textcolor{BrickRed}{0.1158 $\pm$ 0.0008} & \textcolor{BrickRed}{0.0225} & \textcolor{BrickRed}{0.2519} & 19.5\% \\
 & \textbf{ENC}         & \textcolor{BrickRed}{0.0083 $\pm$ 0.0001} & \textcolor{ForestGreen}{-0.0190} & \textcolor{BrickRed}{0.0710} & 41.9\% \\
 & RNN                  & \textcolor{BrickRed}{0.0209 $\pm$ 0.0001} & \textcolor{ForestGreen}{-0.0082} & \textcolor{BrickRed}{0.0974} & 31.7\% \\
 & TF                   & \textcolor{BrickRed}{0.0258 $\pm$ 0.0013} & \textcolor{ForestGreen}{-0.0742} & \textcolor{BrickRed}{0.1289} & 42.0\% \\
 & TFM                  & \textcolor{BrickRed}{0.0256 $\pm$ 0.0018} & \textcolor{ForestGreen}{-0.0784} & \textcolor{BrickRed}{0.1301} & 42.5\% \\
 & NOISY ENC            & \textcolor{BrickRed}{0.0315 $\pm$ 0.0009} & \textcolor{ForestGreen}{-0.0835} & \textcolor{BrickRed}{0.1498} & 41.1\% \\
 & PERCEIVER            & \textcolor{BrickRed}{0.0130 $\pm$ 0.0005} & \textcolor{ForestGreen}{-0.0838} & \textcolor{BrickRed}{0.1080} & 45.7\% \\

\end{longtable}
}

\newpage

The data corresponding to \cref{fig:ablation} are given in \cref{tab:pauli_randomization,tab:random_randomization}.


{\scriptsize
\setlength{\tabcolsep}{3pt}
\renewcommand{\arraystretch}{0.9}
\setlength{\LTleft}{0pt}
\setlength{\LTright}{0pt}

\begin{longtable}{@{\extracolsep{\fill}}>{\raggedright\arraybackslash}m{2.9cm}lcccc@{}}
\label{tab:pauli_randomization}\\

\toprule
Input Randomization & Model & L1RC Median ($\pm$SE) & P25 & P75 & L1RC \% Improved \\
\midrule
\endfirsthead

\toprule
Input Randomization & Model & L1RC Median ($\pm$SE) & P25 & P75 & L1RC \% Improved \\
\midrule
\endhead

\bottomrule
\endfoot

\bottomrule
\caption{Randomization study for Pauli data. The table reports the median ($\pm$SE) and interquartile range (P25--P75) of the L1 relative change across datasets, with the final column showing the percentage of circuits that are improved, as compared to the noisy baseline. In \textbf{bold} the best performing model per dataset, in green improvements over the noisy baseline, in red worse results than the baseline.}\\
\endlastfoot

\Needspace{10\baselineskip}
\multirow[t]{7}{2.9cm}{\raggedright Standard}
 & MLP\text{-}CORRECTION & \textcolor{ForestGreen}{-0.0239 $\pm$ 0.0003} & \textcolor{ForestGreen}{-0.1794} & \textcolor{BrickRed}{0.0409} & 61.1\% \\
 & ENC              & \textcolor{ForestGreen}{-0.0115 $\pm$ 0.0013} & \textcolor{ForestGreen}{-0.1740} & \textcolor{BrickRed}{0.0670} & 54.2\% \\
 & RNN              & \textcolor{ForestGreen}{-0.1890 $\pm$ 0.0010} & \textcolor{ForestGreen}{-0.4049} & \textcolor{ForestGreen}{-0.0304} & 79.9\% \\
 & TF               & \textcolor{ForestGreen}{-0.2039 $\pm$ 0.0005} & \textcolor{ForestGreen}{-0.4469} & \textcolor{ForestGreen}{-0.0319} & 82.2\% \\
 & TFM              & \textcolor{ForestGreen}{-0.2030 $\pm$ 0.0004} & \textcolor{ForestGreen}{-0.4458} & \textcolor{ForestGreen}{-0.0311} & 82.3\% \\
 & NOISY ENC         & \textcolor{ForestGreen}{-0.1945 $\pm$ 0.0029} & \textcolor{ForestGreen}{-0.4372} & \textcolor{ForestGreen}{-0.0308} & 80.8\% \\
 & \textbf{PERCEIVER}& \textcolor{ForestGreen}{-0.2062 $\pm$ 0.0027} & \textcolor{ForestGreen}{-0.4542} & \textcolor{ForestGreen}{-0.0348} & 83.2\% \\
\midrule

\Needspace{9\baselineskip}
\multirow[t]{6}{2.9cm}{\raggedright Random Backend}
 & MLP\text{-}CORRECTION & \textcolor{ForestGreen}{-0.0662 $\pm$ 0.0004} & \textcolor{ForestGreen}{-0.2759} & \textcolor{BrickRed}{0.0448} & 64.7\% \\
 & ENC              & \textcolor{BrickRed}{0.1158 $\pm$ 0.0199} & \textcolor{BrickRed}{0.0421} & \textcolor{BrickRed}{0.2298} & 10.7\% \\
 & \textbf{RNN}     & \textcolor{ForestGreen}{-0.1858 $\pm$ 0.0129} & \textcolor{ForestGreen}{-0.4062} & \textcolor{ForestGreen}{-0.0284} & 81.2\% \\
 & TF               & \textcolor{ForestGreen}{-0.1529 $\pm$ 0.0213} & \textcolor{ForestGreen}{-0.3250} & \textcolor{BrickRed}{0.0143} & 73.0\% \\
 & TFM              & \textcolor{ForestGreen}{-0.1611 $\pm$ 0.0203} & \textcolor{ForestGreen}{-0.3371} & \textcolor{BrickRed}{0.0071} & 74.1\% \\
 & PERCEIVER        & \textcolor{BrickRed}{0.2996 $\pm$ 0.0153} & \textcolor{BrickRed}{0.1625} & \textcolor{BrickRed}{0.5200} & 4.6\% \\
\midrule

\Needspace{10\baselineskip}
\multirow[t]{7}{2.9cm}{\raggedright Random Noise}
 & MLP\text{-}CORRECTION & \textcolor{BrickRed}{0.2820 $\pm$ 0.0002} & \textcolor{BrickRed}{0.1298} & \textcolor{BrickRed}{0.5748} & 4.7\% \\
 & ENC              & \textcolor{BrickRed}{0.2667 $\pm$ 0.0004} & \textcolor{BrickRed}{0.1082} & \textcolor{BrickRed}{0.5678} & 5.8\% \\
 & RNN              & \textcolor{BrickRed}{0.3518 $\pm$ 0.0014} & \textcolor{BrickRed}{0.1845} & \textcolor{BrickRed}{0.6407} & 3.7\% \\
 & TF               & \textcolor{BrickRed}{0.3465 $\pm$ 0.0007} & \textcolor{BrickRed}{0.1805} & \textcolor{BrickRed}{0.6320} & 3.7\% \\
 & \textbf{TFM}     & \textcolor{ForestGreen}{-0.1875 $\pm$ 0.0009} & \textcolor{ForestGreen}{-0.4211} & \textcolor{ForestGreen}{-0.0266} & 81.6\% \\
 & NOISY ENC         & \textcolor{BrickRed}{0.3530 $\pm$ 0.0024} & \textcolor{BrickRed}{0.1861} & \textcolor{BrickRed}{0.6420} & 3.8\% \\
 & PERCEIVER        & \textcolor{BrickRed}{0.3315 $\pm$ 0.0016} & \textcolor{BrickRed}{0.1665} & \textcolor{BrickRed}{0.6228} & 4.2\% \\
\midrule

\Needspace{9\baselineskip}
\multirow[t]{6}{2.9cm}{\raggedright Random Circuit}
 & MLP\text{-}CORRECTION & \textcolor{BrickRed}{0.0741 $\pm$ 0.0045} & \textcolor{ForestGreen}{-0.0506} & \textcolor{BrickRed}{0.2210} & 31.9\% \\
 & ENC              & \textcolor{ForestGreen}{-0.0148 $\pm$ 0.0009} & \textcolor{ForestGreen}{-0.1709} & \textcolor{BrickRed}{0.0273} & 58.4\% \\
 & RNN              & \textcolor{ForestGreen}{-0.1890 $\pm$ 0.0009} & \textcolor{ForestGreen}{-0.4049} & \textcolor{ForestGreen}{-0.0304} & 81.3\% \\
 & \textbf{TF}      & \textcolor{ForestGreen}{-0.2038 $\pm$ 0.0007} & \textcolor{ForestGreen}{-0.4476} & \textcolor{ForestGreen}{-0.0325} & 82.5\% \\
 & TFM              & \textcolor{ForestGreen}{-0.2028 $\pm$ 0.0005} & \textcolor{ForestGreen}{-0.4468} & \textcolor{ForestGreen}{-0.0310} & 82.4\% \\
 & PERCEIVER        & \textcolor{ForestGreen}{-0.1124 $\pm$ 0.0036} & \textcolor{ForestGreen}{-0.3272} & \textcolor{BrickRed}{0.0141} & 71.7\% \\
\midrule

\Needspace{9\baselineskip}
\multirow[t]{6}{2.9cm}{\raggedright Random Circuit\&Backend}
 & MLP\text{-}CORRECTION & \textcolor{ForestGreen}{-0.0151 $\pm$ 0.0018} & \textcolor{ForestGreen}{-0.2204} & \textcolor{BrickRed}{0.1172} & 56.4\% \\
 & ENC              & \textcolor{BrickRed}{0.1161 $\pm$ 0.0173} & \textcolor{BrickRed}{0.0421} & \textcolor{BrickRed}{0.2302} & 10.7\% \\
 & \textbf{RNN}     & \textcolor{ForestGreen}{-0.1859 $\pm$ 0.0130} & \textcolor{ForestGreen}{-0.4061} & \textcolor{ForestGreen}{-0.0289} & 81.2\% \\
 & TF               & \textcolor{ForestGreen}{-0.1521 $\pm$ 0.0239} & \textcolor{ForestGreen}{-0.3239} & \textcolor{BrickRed}{0.0160} & 72.8\% \\
 & TFM              & \textcolor{ForestGreen}{-0.1574 $\pm$ 0.0242} & \textcolor{ForestGreen}{-0.3277} & \textcolor{BrickRed}{0.0140} & 73.2\% \\
 & PERCEIVER        & \textcolor{BrickRed}{0.3010 $\pm$ 0.0127} & \textcolor{BrickRed}{0.1576} & \textcolor{BrickRed}{0.5286} & 4.1\% \\

\end{longtable}
}
\newpage

{\scriptsize
\setlength{\tabcolsep}{3pt}
\renewcommand{\arraystretch}{0.9}
\setlength{\LTleft}{0pt}
\setlength{\LTright}{0pt}

\begin{longtable}{@{\extracolsep{\fill}}>{\raggedright\arraybackslash}m{2.9cm}lcccc@{}}
\label{tab:random_randomization}\\

\toprule
Input Randomization & Model & L1RC Median ($\pm$SE) & P25 & P75 & L1RC \% Improved \\
\midrule
\endfirsthead

\toprule
Input Randomization & Model & L1RC Median ($\pm$SE) & P25 & P75 & L1RC \% Improved \\
\midrule
\endhead

\bottomrule
\endfoot

\bottomrule
\caption{Randomization study for Random data. The table reports the median ($\pm$SE) and interquartile range (P25--P75) of the L1 relative change across datasets, with the final column showing the percentage of circuits that are improved, as compared to the noisy baseline. In \textbf{bold} the best performing model per dataset, in green improvements over the noisy baseline, in red worse results than the baseline.}\\
\endlastfoot

\Needspace{9\baselineskip}
\multirow[t]{6}{2.9cm}{\raggedright Standard}
 & MLP\text{-}CORRECTION & \textcolor{BrickRed}{0.0064 $\pm$ 0.0002} & \textcolor{ForestGreen}{-0.0175} & \textcolor{BrickRed}{0.0599} & 42.5\% \\
 & ENC               & \textcolor{BrickRed}{0.0161 $\pm$ 0.0001} & \textcolor{ForestGreen}{-0.0246} & \textcolor{BrickRed}{0.1074} & 40.0\% \\
 & RNN               & \textcolor{BrickRed}{0.0059 $\pm$ 0.0002} & \textcolor{ForestGreen}{-0.0174} & \textcolor{BrickRed}{0.0462} & 43.1\% \\
 & TF                & \textcolor{ForestGreen}{-0.0166 $\pm$ 0.0000} & \textcolor{ForestGreen}{-0.0871} & \textcolor{BrickRed}{0.0168} & 62.7\% \\
 & TFM               & \textcolor{ForestGreen}{-0.0168 $\pm$ 0.0000} & \textcolor{ForestGreen}{-0.0818} & \textcolor{BrickRed}{0.0167} & 63.0\% \\
 & \textbf{PERCEIVER} & \textcolor{ForestGreen}{-0.0210 $\pm$ 0.0001} & \textcolor{ForestGreen}{-0.1019} & \textcolor{BrickRed}{0.0160} & 64.4\% \\
\midrule

\Needspace{9\baselineskip}
\multirow[t]{6}{2.9cm}{\raggedright Random Backend}
 & MLP\text{-}CORRECTION & \textcolor{BrickRed}{0.0157 $\pm$ 0.0017} & \textcolor{ForestGreen}{-0.0132} & \textcolor{BrickRed}{0.1133} & 36.6\% \\
 & ENC               & \textcolor{BrickRed}{0.0095 $\pm$ 0.0003} & \textcolor{ForestGreen}{-0.0195} & \textcolor{BrickRed}{0.0782} & 41.3\% \\
 & \textbf{RNN}      & \textcolor{BrickRed}{0.0072 $\pm$ 0.0013} & \textcolor{ForestGreen}{-0.0184} & \textcolor{BrickRed}{0.0484} & 42.4\% \\
 & TF                & \textcolor{BrickRed}{0.1417 $\pm$ 0.0397} & \textcolor{ForestGreen}{-0.0383} & \textcolor{BrickRed}{0.3944} & 29.4\% \\
 & TFM               & \textcolor{BrickRed}{0.1217 $\pm$ 0.0320} & \textcolor{ForestGreen}{-0.0528} & \textcolor{BrickRed}{0.3666} & 31.0\% \\
 & PERCEIVER         & \textcolor{BrickRed}{0.5575 $\pm$ 0.0505} & \textcolor{BrickRed}{0.3281} & \textcolor{BrickRed}{0.9473} & 2.3\% \\
\midrule

\Needspace{9\baselineskip}
\multirow[t]{6}{2.9cm}{\raggedright Random Noise}
 & MLP\text{-}CORRECTION & \textcolor{BrickRed}{0.2505 $\pm$ 0.0021} & \textcolor{BrickRed}{0.0764} & \textcolor{BrickRed}{0.6221} & 13.0\% \\
 & ENC               & \textcolor{BrickRed}{0.1655 $\pm$ 0.0004} & \textcolor{BrickRed}{0.0321} & \textcolor{BrickRed}{0.4947} & 18.6\% \\
 & RNN               & \textcolor{BrickRed}{0.2480 $\pm$ 0.0009} & \textcolor{BrickRed}{0.0883} & \textcolor{BrickRed}{0.5893} & 11.6\% \\
 & TF                & \textcolor{BrickRed}{0.3936 $\pm$ 0.0039} & \textcolor{BrickRed}{0.1496} & \textcolor{BrickRed}{0.8174} & 8.4\% \\
 & \textbf{TFM}      & \textcolor{BrickRed}{0.0007 $\pm$ 0.0005} & \textcolor{ForestGreen}{-0.0549} & \textcolor{BrickRed}{0.0609} & 49.6\% \\
 & PERCEIVER         & \textcolor{BrickRed}{0.4076 $\pm$ 0.0041} & \textcolor{BrickRed}{0.1553} & \textcolor{BrickRed}{0.8278} & 9.2\% \\
\midrule

\Needspace{9\baselineskip}
\multirow[t]{6}{2.9cm}{\raggedright Random Circuit}
 & MLP\text{-}CORRECTION & \textcolor{BrickRed}{0.0355 $\pm$ 0.0274} & \textcolor{ForestGreen}{-0.0128} & \textcolor{BrickRed}{0.1849} & 32.6\% \\
 & ENC               & \textcolor{BrickRed}{0.0160 $\pm$ 0.0001} & \textcolor{ForestGreen}{-0.0245} & \textcolor{BrickRed}{0.1072} & 40.0\% \\
 & RNN               & \textcolor{BrickRed}{0.0059 $\pm$ 0.0002} & \textcolor{ForestGreen}{-0.0174} & \textcolor{BrickRed}{0.0463} & 43.1\% \\
 & TF                & \textcolor{ForestGreen}{-0.0165 $\pm$ 0.0000} & \textcolor{ForestGreen}{-0.0829} & \textcolor{BrickRed}{0.0168} & 62.7\% \\
 & \textbf{TFM}      & \textcolor{ForestGreen}{-0.0167 $\pm$ 0.0000} & \textcolor{ForestGreen}{-0.0828} & \textcolor{BrickRed}{0.0173} & 62.7\% \\
 & PERCEIVER         & \textcolor{BrickRed}{0.0007 $\pm$ 0.0017} & \textcolor{ForestGreen}{-0.0617} & \textcolor{BrickRed}{0.0753} & 49.6\% \\
\midrule

\Needspace{9\baselineskip}
\multirow[t]{6}{2.9cm}{\raggedright Random Circuit\&Backend}
 & MLP\text{-}CORRECTION & \textcolor{BrickRed}{0.0830 $\pm$ 0.0150} & \textcolor{BrickRed}{0.0000} & \textcolor{BrickRed}{0.3556} & 25.0\% \\
 & ENC               & \textcolor{BrickRed}{0.0095 $\pm$ 0.0003} & \textcolor{ForestGreen}{-0.0194} & \textcolor{BrickRed}{0.0781} & 41.3\% \\
 & \textbf{RNN}      & \textcolor{BrickRed}{0.0071 $\pm$ 0.0013} & \textcolor{ForestGreen}{-0.0185} & \textcolor{BrickRed}{0.0482} & 42.6\% \\
 & TF                & \textcolor{BrickRed}{0.1575 $\pm$ 0.0269} & \textcolor{ForestGreen}{-0.0301} & \textcolor{BrickRed}{0.4181} & 28.3\% \\
 & TFM               & \textcolor{BrickRed}{0.1278 $\pm$ 0.0322} & \textcolor{ForestGreen}{-0.0421} & \textcolor{BrickRed}{0.3725} & 30.1\% \\
 & PERCEIVER         & \textcolor{BrickRed}{0.5205 $\pm$ 0.0420} & \textcolor{BrickRed}{0.2970} & \textcolor{BrickRed}{0.8934} & 2.9\% \\

\end{longtable}
}
\newpage

The data corresponding to \cref{fig:ablation} are given in \cref{tab:cross_testing_l1_results_sim,tab:cross_testing_l1_results_real}. 

{\scriptsize
\setlength{\tabcolsep}{3pt}
\renewcommand{\arraystretch}{0.9}
\setlength{\LTleft}{0pt}
\setlength{\LTright}{0pt}

\begin{longtable}{@{\extracolsep{\fill}}>{\raggedright\arraybackslash}m{2.9cm}lcccc@{}}
\label{tab:cross_testing_l1_results_sim}\\

\toprule
Dataset & Model & L1RC Median ($\pm$SE) & P25 & P75 & L1RC \% Improved \\
\midrule
\endfirsthead

\toprule
Dataset & Model & L1RC Median ($\pm$SE) & P25 & P75 & L1RC \% Improved \\
\midrule
\endhead

\bottomrule
\endfoot

\bottomrule
\caption{Standard vs Cross testing for L1 Relative Change (Simulated). `Standard' indicates training and testing on the same class, `Cross' indicates training on one class and testing on the other. The table reports the median ($\pm$SE) and interquartile range (P25--P75) of the L1 relative change across datasets, with the final column showing the percentage of circuits that are improved, as compared to the noisy baseline. In \textbf{bold} the best performing model per dataset, in green improvements over the noisy baseline, in red worse results than the baseline. }\\
\endlastfoot

\Needspace{12\baselineskip}
\multirow[t]{8}{2.9cm}{\raggedright Algiers Pauli Simulated (Standard)}
 & MLP            & \textcolor{BrickRed}{0.2047 $\pm$ 0.0010} & \textcolor{BrickRed}{0.0006} & \textcolor{BrickRed}{0.5333} & 24.4\% \\
 & MLP\text{-}PRED& \textcolor{ForestGreen}{-0.0935 $\pm$ 0.0008} & \textcolor{ForestGreen}{-0.2884} & \textcolor{BrickRed}{0.1045} & 62.3\% \\
 & ENC            & \textcolor{BrickRed}{0.2041 $\pm$ 0.0010} & \textcolor{BrickRed}{0.0012} & \textcolor{BrickRed}{0.5296} & 23.9\% \\
 & RNN            & \textcolor{ForestGreen}{-0.4935 $\pm$ 0.0006} & \textcolor{ForestGreen}{-0.6330} & \textcolor{ForestGreen}{-0.2681} & 90.4\% \\
 & \textbf{TF}    & \textcolor{ForestGreen}{-0.6326 $\pm$ 0.0004} & \textcolor{ForestGreen}{-0.7295} & \textcolor{ForestGreen}{-0.4645} & 94.0\% \\
 & TFM            & \textcolor{ForestGreen}{-0.6254 $\pm$ 0.0004} & \textcolor{ForestGreen}{-0.7239} & \textcolor{ForestGreen}{-0.4577} & 94.0\% \\
 & NOISY\text{-}ENC & \textcolor{ForestGreen}{-0.6248 $\pm$ 0.0005} & \textcolor{ForestGreen}{-0.7304} & \textcolor{ForestGreen}{-0.4328} & 92.9\% \\
 & PERCEIVER      & \textcolor{ForestGreen}{-0.5727 $\pm$ 0.0006} & \textcolor{ForestGreen}{-0.7052} & \textcolor{ForestGreen}{-0.3458} & 91.0\% \\
\midrule

\Needspace{12\baselineskip}
\multirow[t]{8}{2.9cm}{\raggedright Train Random Sim $\rightarrow$ Test Pauli Sim (Cross)}
 & MLP            & \textcolor{BrickRed}{0.4115 $\pm$ 0.0024} & \textcolor{ForestGreen}{-0.0015} & \textcolor{BrickRed}{1.0440} & 26.8\% \\
 & MLP\text{-}PRED& \textcolor{BrickRed}{0.3596 $\pm$ 0.0015} & \textcolor{BrickRed}{0.0172} & \textcolor{BrickRed}{0.9375} & 22.2\% \\
 & ENC            & \textcolor{BrickRed}{0.1770 $\pm$ 0.0010} & \textcolor{ForestGreen}{-0.0011} & \textcolor{BrickRed}{0.6146} & 26.2\% \\
 & RNN            & \textcolor{BrickRed}{0.3693 $\pm$ 0.0023} & \textcolor{ForestGreen}{-0.0232} & \textcolor{BrickRed}{1.0173} & 29.2\% \\
 & TF             & \textcolor{ForestGreen}{-0.4589 $\pm$ 0.0006} & \textcolor{ForestGreen}{-0.5819} & \textcolor{ForestGreen}{-0.2208} & 88.7\% \\
 & \textbf{TFM}   & \textcolor{ForestGreen}{-0.5217 $\pm$ 0.0004} & \textcolor{ForestGreen}{-0.6221} & \textcolor{ForestGreen}{-0.3602} & 93.9\% \\
 & NOISY\text{-}ENC & \textcolor{ForestGreen}{-0.5055 $\pm$ 0.0005} & \textcolor{ForestGreen}{-0.6183} & \textcolor{ForestGreen}{-0.3375} & 94.3\% \\
 & PERCEIVER      & \textcolor{ForestGreen}{-0.0610 $\pm$ 0.0008} & \textcolor{ForestGreen}{-0.2509} & \textcolor{BrickRed}{0.0960} & 60.6\% \\
\addlinespace[0.6mm]

\Needspace{12\baselineskip}
\multirow[t]{8}{2.9cm}{\raggedright Algiers Random Simulated (Standard)}
 & MLP            & \textcolor{ForestGreen}{-0.0753 $\pm$ 0.0007} & \textcolor{ForestGreen}{-0.2399} & \textcolor{BrickRed}{0.1038} & 62.9\% \\
 & MLP\text{-}PRED& \textcolor{BrickRed}{0.3961 $\pm$ 0.0011} & \textcolor{BrickRed}{0.1370} & \textcolor{BrickRed}{0.7509} & 11.8\% \\
 & ENC            & \textcolor{BrickRed}{0.2765 $\pm$ 0.0007} & \textcolor{BrickRed}{0.1340} & \textcolor{BrickRed}{0.5161} & 9.5\% \\
 & RNN            & \textcolor{ForestGreen}{-0.0650 $\pm$ 0.0004} & \textcolor{ForestGreen}{-0.1697} & \textcolor{BrickRed}{0.0399} & 67.4\% \\
 & TF             & \textcolor{ForestGreen}{-0.5288 $\pm$ 0.0005} & \textcolor{ForestGreen}{-0.6435} & \textcolor{ForestGreen}{-0.3039} & 89.4\% \\
 & TFM            & \textcolor{ForestGreen}{-0.5230 $\pm$ 0.0005} & \textcolor{ForestGreen}{-0.6381} & \textcolor{ForestGreen}{-0.2956} & 89.2\% \\
 & NOISY\text{-}ENC & \textcolor{ForestGreen}{-0.4740 $\pm$ 0.0006} & \textcolor{ForestGreen}{-0.6022} & \textcolor{ForestGreen}{-0.1900} & 84.7\% \\
 & \textbf{PERCEIVER} & \textcolor{ForestGreen}{-0.5453 $\pm$ 0.0004} & \textcolor{ForestGreen}{-0.6446} & \textcolor{ForestGreen}{-0.3613} & 91.7\% \\
\midrule

\Needspace{12\baselineskip}
\multirow[t]{8}{2.9cm}{\raggedright Train Pauli Sim $\rightarrow$ Test Random Sim (Cross)}
 & MLP            & \textcolor{BrickRed}{0.6421 $\pm$ 0.0014} & \textcolor{BrickRed}{0.3016} & \textcolor{BrickRed}{1.2221} & 6.0\% \\
 & MLP\text{-}PRED& \textcolor{BrickRed}{1.0722 $\pm$ 0.0021} & \textcolor{BrickRed}{0.5587} & \textcolor{BrickRed}{1.8176} & 3.4\% \\
 & ENC            & \textcolor{BrickRed}{0.4898 $\pm$ 0.0017} & \textcolor{BrickRed}{0.1456} & \textcolor{BrickRed}{1.1084} & 8.5\% \\
 & RNN            & \textcolor{ForestGreen}{-0.1626 $\pm$ 0.0012} & \textcolor{ForestGreen}{-0.3899} & \textcolor{BrickRed}{0.1401} & 64.1\% \\
 & TF             & \textcolor{ForestGreen}{-0.1902 $\pm$ 0.0013} & \textcolor{ForestGreen}{-0.4588} & \textcolor{BrickRed}{0.1888} & 63.4\% \\
 & \textbf{TFM}   & \textcolor{ForestGreen}{-0.2004 $\pm$ 0.0013} & \textcolor{ForestGreen}{-0.4601} & \textcolor{BrickRed}{0.1699} & 64.1\% \\
 & NOISY\text{-}ENC & \textcolor{ForestGreen}{-0.1119 $\pm$ 0.0015} & \textcolor{ForestGreen}{-0.4162} & \textcolor{BrickRed}{0.3005} & 58.6\% \\
 & PERCEIVER      & \textcolor{BrickRed}{0.0704 $\pm$ 0.0011} & \textcolor{ForestGreen}{-0.1582} & \textcolor{BrickRed}{0.6600} & 41.6\% \\

\end{longtable}
}

\newpage
{\scriptsize
\setlength{\tabcolsep}{3pt}
\renewcommand{\arraystretch}{0.9}
\setlength{\LTleft}{0pt}
\setlength{\LTright}{0pt}

\begin{longtable}{@{\extracolsep{\fill}}>{\raggedright\arraybackslash}m{2.9cm}lcccc@{}}
\label{tab:cross_testing_l1_results_real}\\

\toprule
Dataset & Model & L1RC Median ($\pm$SE) & P25 & P75 & L1RC \% Improved \\
\midrule
\endfirsthead

\toprule
Dataset & Model & L1RC Median ($\pm$SE) & P25 & P75 & L1RC \% Improved \\
\midrule
\endhead

\bottomrule
\endfoot

\bottomrule
\caption{Standard vs. Cross testing for L1 Relative Change (Real). 'Standard' indicates training and testing on the same class and 'Cross' indicates training on one class and testing on the other. The table reports the median ($\pm$SE) and interquartile range (P25--P75) of the L1 relative change across datasets, with the final column showing the percentage of circuits that are improved, as compared to the noisy baseline. In \textbf{bold} the best performing model per dataset, in green improvements over the noisy baseline, in red worse results than the baseline. }\\
\endlastfoot

\Needspace{12\baselineskip}
\multirow[t]{8}{2.9cm}{\raggedright Algiers Pauli Real (Standard)}
 & MLP            & \textcolor{ForestGreen}{-0.0238 $\pm$ 0.0002} & \textcolor{ForestGreen}{-0.1812} & \textcolor{BrickRed}{0.0128} & 64.5\% \\
 & MLP\text{-}PRED& \textcolor{ForestGreen}{-0.1176 $\pm$ 0.0005} & \textcolor{ForestGreen}{-0.3142} & \textcolor{BrickRed}{0.0140} & 71.9\% \\
 & ENC            & \textcolor{ForestGreen}{-0.0155 $\pm$ 0.0002} & \textcolor{ForestGreen}{-0.1723} & \textcolor{BrickRed}{0.0276} & 58.6\% \\
 & RNN            & \textcolor{ForestGreen}{-0.1891 $\pm$ 0.0006} & \textcolor{ForestGreen}{-0.4060} & \textcolor{ForestGreen}{-0.0298} & 81.2\% \\
 & \textbf{TF}    & \textcolor{ForestGreen}{-0.2039 $\pm$ 0.0007} & \textcolor{ForestGreen}{-0.4471} & \textcolor{ForestGreen}{-0.0324} & 82.5\% \\
 & TFM            & \textcolor{ForestGreen}{-0.2027 $\pm$ 0.0007} & \textcolor{ForestGreen}{-0.4460} & \textcolor{ForestGreen}{-0.0320} & 82.5\% \\
 & NOISY\text{-}ENC & \textcolor{ForestGreen}{-0.1910 $\pm$ 0.0006} & \textcolor{ForestGreen}{-0.4335} & \textcolor{ForestGreen}{-0.0293} & 81.6\% \\
 & PERCEIVER      & \textcolor{ForestGreen}{-0.2033 $\pm$ 0.0006} & \textcolor{ForestGreen}{-0.4533} & \textcolor{ForestGreen}{-0.0326} & 82.9\% \\
\midrule

\Needspace{12\baselineskip}
\multirow[t]{8}{2.9cm}{\raggedright FT Random Real $\rightarrow$ Test Pauli Real (Cross)}
 & MLP            & \textcolor{ForestGreen}{-0.0165 $\pm$ 0.0002} & \textcolor{ForestGreen}{-0.1994} & \textcolor{BrickRed}{0.0203} & 60.4\% \\
 & MLP\text{-}PRED& \textcolor{ForestGreen}{-0.0299 $\pm$ 0.0003} & \textcolor{ForestGreen}{-0.2137} & \textcolor{BrickRed}{0.0527} & 59.0\% \\
 & ENC            & \textcolor{BrickRed}{0.0110 $\pm$ 0.0001} & \textcolor{ForestGreen}{-0.0720} & \textcolor{BrickRed}{0.0456} & 41.4\% \\
 & RNN            & \textcolor{ForestGreen}{-0.0211 $\pm$ 0.0002} & \textcolor{ForestGreen}{-0.1951} & \textcolor{BrickRed}{0.0094} & 65.1\% \\
 & \textbf{TF}    & \textcolor{ForestGreen}{-0.0405 $\pm$ 0.0002} & \textcolor{ForestGreen}{-0.1642} & \textcolor{BrickRed}{0.0026} & 72.6\% \\
 & TFM            & \textcolor{ForestGreen}{-0.0341 $\pm$ 0.0002} & \textcolor{ForestGreen}{-0.1506} & \textcolor{BrickRed}{0.0049} & 70.5\% \\
 & NOISY\text{-}ENC & \textcolor{ForestGreen}{-0.0261 $\pm$ 0.0002} & \textcolor{ForestGreen}{-0.1361} & \textcolor{BrickRed}{0.0131} & 65.0\% \\
 & PERCEIVER      & \textcolor{ForestGreen}{-0.0325 $\pm$ 0.0002} & \textcolor{ForestGreen}{-0.1260} & \textcolor{BrickRed}{0.0046} & 70.8\% \\
\addlinespace[0.6mm]

\Needspace{12\baselineskip}
\multirow[t]{8}{2.9cm}{\raggedright Algiers Random Real (Standard)}
 & MLP            & \textcolor{BrickRed}{0.0066 $\pm$ 0.0001} & \textcolor{ForestGreen}{-0.0178} & \textcolor{BrickRed}{0.0606} & 42.4\% \\
 & MLP\text{-}PRED& \textcolor{BrickRed}{0.0866 $\pm$ 0.0003} & \textcolor{BrickRed}{0.0106} & \textcolor{BrickRed}{0.1944} & 21.8\% \\
 & ENC            & \textcolor{BrickRed}{0.0162 $\pm$ 0.0002} & \textcolor{ForestGreen}{-0.0245} & \textcolor{BrickRed}{0.1078} & 39.9\% \\
 & RNN            & \textcolor{BrickRed}{0.0060 $\pm$ 0.0001} & \textcolor{ForestGreen}{-0.0175} & \textcolor{BrickRed}{0.0463} & 43.1\% \\
 & TF             & \textcolor{ForestGreen}{-0.0166 $\pm$ 0.0001} & \textcolor{ForestGreen}{-0.0812} & \textcolor{BrickRed}{0.0162} & 63.0\% \\
 & TFM            & \textcolor{ForestGreen}{-0.0168 $\pm$ 0.0001} & \textcolor{ForestGreen}{-0.0822} & \textcolor{BrickRed}{0.0169} & 62.8\% \\
 & NOISY\text{-}ENC & \textcolor{ForestGreen}{-0.0182 $\pm$ 0.0001} & \textcolor{ForestGreen}{-0.0929} & \textcolor{BrickRed}{0.0207} & 62.2\% \\
 & \textbf{PERCEIVER} & \textcolor{ForestGreen}{-0.0210 $\pm$ 0.0001} & \textcolor{ForestGreen}{-0.1013} & \textcolor{BrickRed}{0.0156} & 64.6\% \\
\midrule

\Needspace{12\baselineskip}
\multirow[t]{8}{2.9cm}{\raggedright FT Pauli Real $\rightarrow$ Test Random Real (Cross)}
 & MLP            & \textcolor{BrickRed}{0.0127 $\pm$ 0.0001} & \textcolor{ForestGreen}{-0.0170} & \textcolor{BrickRed}{0.1115} & 38.9\% \\
 & MLP\text{-}PRED& \textcolor{BrickRed}{0.3684 $\pm$ 0.0007} & \textcolor{BrickRed}{0.1618} & \textcolor{BrickRed}{0.6454} & 9.7\% \\
 & \textbf{ENC}   & \textcolor{BrickRed}{0.0116 $\pm$ 0.0001} & \textcolor{ForestGreen}{-0.0216} & \textcolor{BrickRed}{0.0732} & 40.6\% \\
 & RNN            & \textcolor{BrickRed}{0.0809 $\pm$ 0.0003} & \textcolor{ForestGreen}{-0.0218} & \textcolor{BrickRed}{0.2532} & 29.8\% \\
 & TF             & \textcolor{BrickRed}{0.0478 $\pm$ 0.0003} & \textcolor{ForestGreen}{-0.0341} & \textcolor{BrickRed}{0.2178} & 34.4\% \\
 & TFM            & \textcolor{BrickRed}{0.0453 $\pm$ 0.0003} & \textcolor{ForestGreen}{-0.0345} & \textcolor{BrickRed}{0.2116} & 34.6\% \\
 & NOISY\text{-}ENC & \textcolor{BrickRed}{0.0634 $\pm$ 0.0003} & \textcolor{ForestGreen}{-0.0264} & \textcolor{BrickRed}{0.2341} & 31.8\% \\
 & PERCEIVER      & \textcolor{BrickRed}{0.0442 $\pm$ 0.0003} & \textcolor{ForestGreen}{-0.0315} & \textcolor{BrickRed}{0.2079} & 34.4\% \\

\end{longtable}
}
\newpage

The data corresponding to \cref{fig:hanoi-multi-boxes} are given in \cref{tab:different_qpus}. 

{\scriptsize
\setlength{\tabcolsep}{3pt}
\renewcommand{\arraystretch}{0.9}
\setlength{\LTleft}{0pt}
\setlength{\LTright}{0pt}

\begin{longtable}{@{\extracolsep{\fill}}>{\raggedright\arraybackslash}m{2.9cm}lcccc@{}}
\label{tab:different_qpus}\\

\toprule
Dataset & Model & L1RC Median ($\pm$SE) & P25 & P75 & L1RC \% Improved \\
\midrule
\endfirsthead

\toprule
Dataset & Model & L1RC Median ($\pm$SE) & P25 & P75 & L1RC \% Improved \\
\midrule
\endhead

\bottomrule
\endfoot

\bottomrule
\caption{Results for models on \Ibmhanoi{} Real data. The table reports the median ($\pm$SE) and interquartile range (P25--P75) of the L1 relative change across datasets, with the final column showing the percentage of circuits that are improved as compared to the noisy baseline. In \textbf{bold} the best performing model per dataset, in green improvements over the noisy baseline, in red worse results than the baseline. The models are all initially pre-trained on \Ibmalgiers{} Simulated data. The `Transfer Learning' case is a double fine-tuning, first using \Ibmalgiers{} Real data followed by \Ibmhanoi{} Real data.}\\
\endlastfoot

\multicolumn{6}{l}{\textbf{Only Predict on Hanoi}}\\
\midrule

\Needspace{12\baselineskip}
\multirow[t]{8}{2.9cm}{\parbox[t]{2.9cm}{\raggedright Pauli}}
 & MLP\text{-}CORRECTION & \textcolor{ForestGreen}{-0.0317 $\pm$ 0.0004} & \textcolor{ForestGreen}{-0.2205} & \textcolor{BrickRed}{0.0401} & 61.1\% \\
 & MLP\text{-}PREDICTION & \textcolor{ForestGreen}{-0.0512 $\pm$ 0.0036} & \textcolor{ForestGreen}{-0.2075} & \textcolor{BrickRed}{0.0482} & 62.6\% \\
 & ENC & \textcolor{ForestGreen}{-0.0115 $\pm$ 0.0014} & \textcolor{ForestGreen}{-0.1951} & \textcolor{BrickRed}{0.0670} & 54.2\% \\
 & RNN & \textcolor{ForestGreen}{-0.2523 $\pm$ 0.0012} & \textcolor{ForestGreen}{-0.4870} & \textcolor{ForestGreen}{-0.0363} & 79.9\% \\
 & TF & \textcolor{ForestGreen}{-0.2679 $\pm$ 0.0033} & \textcolor{ForestGreen}{-0.5077} & \textcolor{ForestGreen}{-0.0488} & 82.0\% \\
 & TFM & \textcolor{ForestGreen}{-0.2665 $\pm$ 0.0015} & \textcolor{ForestGreen}{-0.4959} & \textcolor{ForestGreen}{-0.0469} & 81.6\% \\
 & NOISY\text{-}ENC & \textcolor{ForestGreen}{-0.2636 $\pm$ 0.0065} & \textcolor{ForestGreen}{-0.5099} & \textcolor{ForestGreen}{-0.0420} & 80.8\% \\
 & \textbf{PERCEIVER} & \textcolor{ForestGreen}{-0.2683 $\pm$ 0.0024} & \textcolor{ForestGreen}{-0.5208} & \textcolor{ForestGreen}{-0.0497} & 84.1\% \\
\addlinespace[0.3mm]
\midrule

\Needspace{12\baselineskip}
\multirow[t]{8}{2.9cm}{\parbox[t]{2.9cm}{\raggedright Random}}
 & MLP\text{-}CORRECTION & \textcolor{BrickRed}{0.0157 $\pm$ 0.0004} & \textcolor{ForestGreen}{-0.0139} & \textcolor{BrickRed}{0.1014} & 36.8\% \\
 & MLP\text{-}PREDICTION & \textcolor{BrickRed}{0.1158 $\pm$ 0.0008} & \textcolor{BrickRed}{0.0225} & \textcolor{BrickRed}{0.2519} & 19.5\% \\
 & ENC & \textcolor{BrickRed}{0.0260 $\pm$ 0.0002} & \textcolor{ForestGreen}{-0.0218} & \textcolor{BrickRed}{0.1546} & 37.0\% \\
 & RNN & \textcolor{BrickRed}{0.0150 $\pm$ 0.0002} & \textcolor{ForestGreen}{-0.0122} & \textcolor{BrickRed}{0.0825} & 36.2\% \\
 & TF & \textcolor{ForestGreen}{-0.0130 $\pm$ 0.0008} & \textcolor{ForestGreen}{-0.0779} & \textcolor{BrickRed}{0.0220} & 59.9\% \\
 & TFM & \textcolor{ForestGreen}{-0.0133 $\pm$ 0.0009} & \textcolor{ForestGreen}{-0.0795} & \textcolor{BrickRed}{0.0236} & 59.5\% \\
 & NOISY\text{-}ENC & \textcolor{ForestGreen}{-0.0142 $\pm$ 0.0004} & \textcolor{ForestGreen}{-0.0945} & \textcolor{BrickRed}{0.0279} & 59.4\% \\
 & \textbf{PERCEIVER} & \textcolor{ForestGreen}{-0.0184 $\pm$ 0.0002} & \textcolor{ForestGreen}{-0.1023} & \textcolor{BrickRed}{0.0281} & 60.5\% \\
\addlinespace[0.6mm]
\midrule

\multicolumn{6}{l}{\textbf{Fine-Tuned on Real Hanoi}}\\
\midrule

\Needspace{12\baselineskip}
\multirow[t]{8}{2.9cm}{\parbox[t]{2.9cm}{\raggedright Pauli}}
 & MLP\text{-}CORRECTION & \textcolor{ForestGreen}{-0.0104 $\pm$ 0.0002} & \textcolor{ForestGreen}{-0.1664} & \textcolor{BrickRed}{0.0456} & 56.0\% \\
 & MLP\text{-}PREDICTION & \textcolor{ForestGreen}{-0.1444 $\pm$ 0.0020} & \textcolor{ForestGreen}{-0.3340} & \textcolor{BrickRed}{0.0117} & 73.0\% \\
 & ENC & \textcolor{ForestGreen}{-0.0015 $\pm$ 0.0010} & \textcolor{ForestGreen}{-0.1479} & \textcolor{BrickRed}{0.0557} & 50.9\% \\
 & RNN & \textcolor{ForestGreen}{-0.2422 $\pm$ 0.0005} & \textcolor{ForestGreen}{-0.4618} & \textcolor{ForestGreen}{-0.0457} & 82.0\% \\
 & TF & \textcolor{ForestGreen}{-0.2601 $\pm$ 0.0008} & \textcolor{ForestGreen}{-0.5077} & \textcolor{ForestGreen}{-0.0516} & 85.1\% \\
 & TFM & \textcolor{ForestGreen}{-0.2594 $\pm$ 0.0009} & \textcolor{ForestGreen}{-0.5054} & \textcolor{ForestGreen}{-0.0526} & 85.0\% \\
 & NOISY\text{-}ENC & \textcolor{ForestGreen}{-0.2596 $\pm$ 0.0044} & \textcolor{ForestGreen}{-0.5118} & \textcolor{ForestGreen}{-0.0532} & 84.5\% \\
 & \textbf{PERCEIVER} & \textcolor{ForestGreen}{-0.2623 $\pm$ 0.0009} & \textcolor{ForestGreen}{-0.5155} & \textcolor{ForestGreen}{-0.0554} & 85.6\% \\
\addlinespace[0.3mm]
\midrule

\Needspace{12\baselineskip}
\multirow[t]{8}{2.9cm}{\parbox[t]{2.9cm}{\raggedright Random}}
 & MLP\text{-}CORRECTION & \textcolor{BrickRed}{0.0177 $\pm$ 0.0002} & \textcolor{ForestGreen}{-0.0141} & \textcolor{BrickRed}{0.1095} & 36.2\% \\
 & MLP\text{-}PREDICTION & \textcolor{BrickRed}{0.1064 $\pm$ 0.0008} & \textcolor{BrickRed}{0.0238} & \textcolor{BrickRed}{0.2343} & 18.6\% \\
 & ENC & \textcolor{BrickRed}{0.0257 $\pm$ 0.0002} & \textcolor{ForestGreen}{-0.0214} & \textcolor{BrickRed}{0.1550} & 37.0\% \\
 & RNN & \textcolor{BrickRed}{0.0145 $\pm$ 0.0002} & \textcolor{ForestGreen}{-0.0134} & \textcolor{BrickRed}{0.0898} & 36.8\% \\
 & TF & \textcolor{ForestGreen}{-0.0175 $\pm$ 0.0003} & \textcolor{ForestGreen}{-0.0923} & \textcolor{BrickRed}{0.0239} & 61.2\% \\
 & TFM & \textcolor{ForestGreen}{-0.0172 $\pm$ 0.0008} & \textcolor{ForestGreen}{-0.0913} & \textcolor{BrickRed}{0.0243} & 61.1\% \\
 & NOISY\text{-}ENC & \textcolor{ForestGreen}{-0.0213 $\pm$ 0.0002} & \textcolor{ForestGreen}{-0.1093} & \textcolor{BrickRed}{0.0259} & 61.9\% \\
 & \textbf{PERCEIVER} & \textcolor{ForestGreen}{-0.0221 $\pm$ 0.0001} & \textcolor{ForestGreen}{-0.1070} & \textcolor{BrickRed}{0.0210} & 63.3\% \\
\addlinespace[0.6mm]
\midrule

\multicolumn{6}{l}{\textbf{Transfer Learning Algiers to Hanoi}}\\
\midrule

\Needspace{12\baselineskip}
\multirow[t]{8}{2.9cm}{\parbox[t]{2.9cm}{\raggedright Pauli}}
 & MLP\text{-}CORRECTION & \textcolor{ForestGreen}{-0.0162 $\pm$ 0.0006} & \textcolor{ForestGreen}{-0.1749} & \textcolor{BrickRed}{0.0456} & 57.7\% \\
 & MLP\text{-}PREDICTION & \textcolor{ForestGreen}{-0.1472 $\pm$ 0.0017} & \textcolor{ForestGreen}{-0.3446} & \textcolor{BrickRed}{0.0111} & 73.0\% \\
 & ENC & \textcolor{ForestGreen}{-0.0089 $\pm$ 0.0021} & \textcolor{ForestGreen}{-0.1647} & \textcolor{BrickRed}{0.0475} & 54.9\% \\
 & RNN & \textcolor{ForestGreen}{-0.2430 $\pm$ 0.0009} & \textcolor{ForestGreen}{-0.4646} & \textcolor{ForestGreen}{-0.0458} & 81.9\% \\
 & TF & \textcolor{ForestGreen}{-0.2584 $\pm$ 0.0006} & \textcolor{ForestGreen}{-0.5045} & \textcolor{ForestGreen}{-0.0517} & 84.9\% \\
 & TFM & \textcolor{ForestGreen}{-0.2591 $\pm$ 0.0006} & \textcolor{ForestGreen}{-0.5032} & \textcolor{ForestGreen}{-0.0506} & 85.0\% \\
 & NOISY\text{-}ENC & \textcolor{ForestGreen}{-0.2586 $\pm$ 0.0047} & \textcolor{ForestGreen}{-0.5088} & \textcolor{ForestGreen}{-0.0510} & 83.8\% \\
 & \textbf{PERCEIVER} & \textcolor{ForestGreen}{-0.2660 $\pm$ 0.0053} & \textcolor{ForestGreen}{-0.5189} & \textcolor{ForestGreen}{-0.0535} & 85.5\% \\
\addlinespace[0.3mm]
\midrule

\Needspace{12\baselineskip}
\multirow[t]{8}{2.9cm}{\parbox[t]{2.9cm}{\raggedright Random}}
 & MLP\text{-}CORRECTION & \textcolor{BrickRed}{0.0156 $\pm$ 0.0002} & \textcolor{ForestGreen}{-0.0161} & \textcolor{BrickRed}{0.1065} & 37.7\% \\
 & MLP\text{-}PREDICTION & \textcolor{BrickRed}{0.1015 $\pm$ 0.0011} & \textcolor{BrickRed}{0.0167} & \textcolor{BrickRed}{0.2265} & 20.2\% \\
 & ENC & \textcolor{BrickRed}{0.0326 $\pm$ 0.0010} & \textcolor{ForestGreen}{-0.0231} & \textcolor{BrickRed}{0.1750} & 36.1\% \\
 & RNN & \textcolor{BrickRed}{0.0148 $\pm$ 0.0002} & \textcolor{ForestGreen}{-0.0134} & \textcolor{BrickRed}{0.0911} & 36.9\% \\
 & TF & \textcolor{ForestGreen}{-0.0182 $\pm$ 0.0011} & \textcolor{ForestGreen}{-0.0952} & \textcolor{BrickRed}{0.0211} & 62.2\% \\
 & TFM & \textcolor{ForestGreen}{-0.0182 $\pm$ 0.0004} & \textcolor{ForestGreen}{-0.0958} & \textcolor{BrickRed}{0.0218} & 62.0\% \\
 & NOISY\text{-}ENC & \textcolor{ForestGreen}{-0.0215 $\pm$ 0.0002} & \textcolor{ForestGreen}{-0.1181} & \textcolor{BrickRed}{0.0241} & 62.4\% \\
 & \textbf{PERCEIVER} & \textcolor{ForestGreen}{-0.0208 $\pm$ 0.0007} & \textcolor{ForestGreen}{-0.1064} & \textcolor{BrickRed}{0.0238} & 62.4\% \\

\end{longtable}
}
\newpage

\section{Baseline Error Mitigation Strategies}
In this section we give some further details about the baseline error mitigation strategies discussed in \cref{sec:results-compare-mitigation}.

\subsection{Repolarizer}\label{app:repolariser}
Consider the target state $|\psi\rangle = C_1 C_2...C_t |0\rangle$ and suppose that in its noisy implementation, each unitary $C_i$ in the circuit is followed by a global depolarizing channel $\mathcal{D}$
\begin{equation}
  \mathcal{D}(\rho) = (1-\lambda)\rho + \lambda \frac{\mathbb{I}}{2^{n_{\rm qubits}}}.
\end{equation}
The noisy implementation of our target state will have the form
\begin{equation}
\psi_{\rm noisy} =  \mathcal{D}(C_1 ... \mathcal{D}(C_t|0\rangle\langle 0 | C_t^{\dagger} ) ... C_1^{\dagger}).
\end{equation}
Expanding out the effect of the depolarizing channel results in 
\begin{equation}
\psi_{\rm noisy} = (1-\lambda)^t \psi  + (1- (1-\lambda)^t) \frac{\mathbb{I}}{2^{n_{\rm qubits}}}.
\end{equation} Therefore under the assumption of a global depolarizing channel with a fixed error rate, $\lambda$, that occurs after each unitary, $C_i$ (which may be taken as two-qubit gate or a depth-one layer of gates), the noisy and ideal probability distributions are related by
\begin{equation}
\mathbf{P}_{\rm noisy} = (1-\lambda)^t \mathbf{P}_{\rm ideal} +(1-(1-\lambda)^t )\mathbf{U},
\end{equation}
where $\mathbf{U}$ is the uniform distribution over all $2^{n_{\rm qubits}}$ possible outcomes with $\mathbf{U}^i = \frac{1}{2^{n_{\rm qubits}}}$. This relationship motivates the following error mitigated probability distribution
\begin{equation}
(1-\lambda)^{-t} (\mathbf{P}_{\rm noisy} - (1-(1-\lambda)^t) \mathbf{U}),
\end{equation} when we are given only the noisy distribution, the error rate $\lambda$, the number of layers $t$, and the above promise on the noise model.

In practice, we only get experimental access to the noisy probability distributions through measurement samples, and therefore our empirical distributions come with a statistical finite sampling error. To account for this, and to ensure the mitigated distributions are always positive, we apply a threshold before inverting the above relation. Specifically for each possible outcome $i$ we define the Repolarizer error mitigated distribution via 
\begin{equation}
\mathbf{P}_{\rm repolarizer}^i = 
\begin{cases}
0  &  \text{if} \ \   \mathbf{P}_{\rm noisy}^i < \frac{(1-(1-\lambda)^t}{2^{n_{\rm qubits}}}\\
{\rm const} (1-\lambda)^{-t} \left[\mathbf{P}_{\rm noisy}^i - \frac{1-(1-\lambda)^t)}{2^{n_{\rm qubits}}}\right]   &  \text{otherwise}
\end{cases},
\end{equation} 
for a constant factor that ensures the $\mathbf{P}_{\rm repolarizer}$ is normalized.

\subsection{Thresholding}
\label{app:thresholding}

Thresholding modifies a probability distribution, $\textbf{P}$, by setting all entries below a given threshold $\tau$ to zero, and then renormalizing the distribution:
\begin{equation}
\textbf{P}^{i}_{\rm thresh} =
\begin{cases}
0, & \mathbf{P}_{\rm noisy}^i < \tau,\\[4pt]
\mathbf{P}_{\rm noisy}^i, & \text{otherwise},
\end{cases}
\end{equation}
followed by
\begin{equation}
\textbf{P}_{\rm thresh} \leftarrow \frac{\textbf{P}_{\rm thresh}}{\sum_i \textbf{P}^{i}_{\rm thresh}}.
\end{equation}
For the results in \cref{sec:results-compare-mitigation} we scan over a uniform grid of 30 values in $\tau \in [0,0.5]$, to find the values that give the optimal performance for each dataset. 

\begin{figure}[ht!]
\centering
\includegraphics[width=0.9\linewidth]{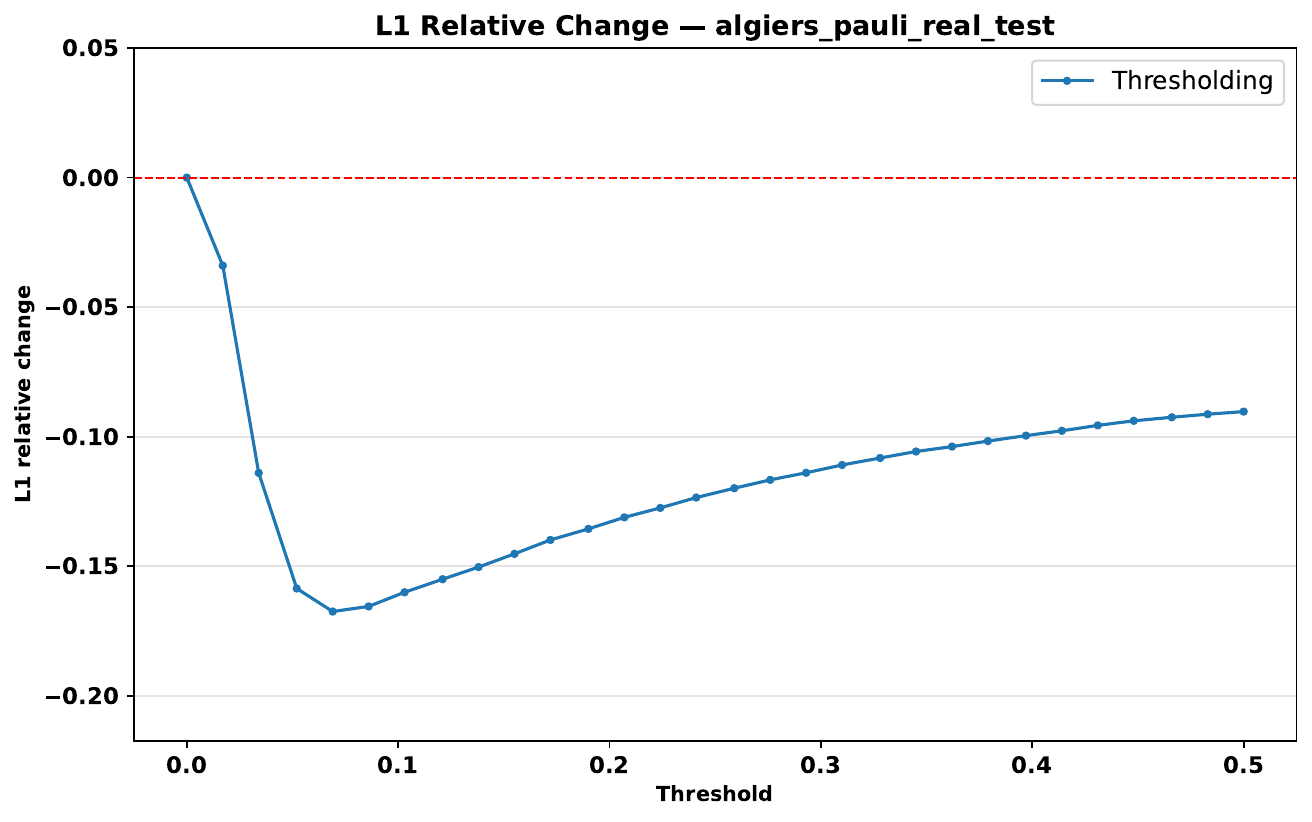}
\caption{
Median L1 Relative Change for the Pauli Real dataset under
thresholding.
Negative values indicate improvement over the noisy baseline.
The dashed horizontal line marks parity with the unmitigated distributions.
}
\label{fig:l1rc_pauli_real}
\end{figure}

\begin{figure}[ht!]
\centering
\includegraphics[width=0.9\linewidth]{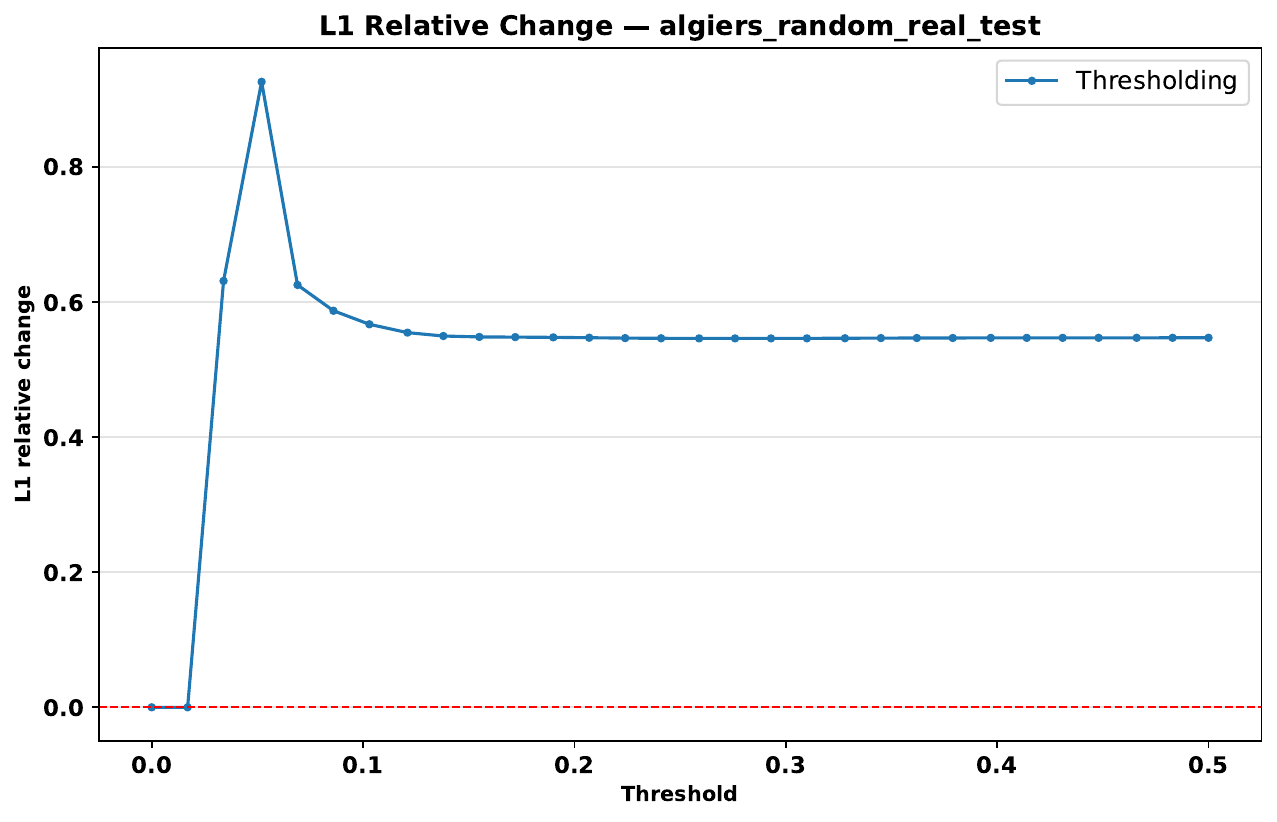}
\caption{
Median L1 Relative Change for the Random Real dataset using
thresholding. Negative values indicate improvement over the noisy baseline.
The dashed horizontal line marks parity with the unmitigated distributions.
}
\label{fig:l1rc_random_real}
\end{figure}

We give an illustrative example of the threshold search procedure, where we sweep over candidate threshold values and evaluate their effect on the L1 Relative Change in \cref{fig:l1rc_pauli_real,fig:l1rc_random_real}. 
Increasing $\tau$ removes low-probability, with the distribution then renormalised. 
For Pauli circuits this has a positive effect on the performance at small $\tau$ values, because it removes non-zero entries arising from gate noise.
However, at larger $\tau$ values, the removal is detrimental, because it starts modifying the real peaks in the distribution.
On Random circuits, where the distributions are less peaked, and the signal is more spread out over multiple bitstring outcomes, any value of $\tau$ will alter features that belong to the signal, rather than the noise.

\section{A Study on the Perceiver Model}
\label[app]{app:perceiver}

To further investigate the behaviour of the best performing architecture identified in our main results (Perceiver), we present a series of ablation studies aimed at understanding how its performance depends on the training formulation, dataset, and circuit depth.  

\subsection{Effect of the Loss Function}
\label{app:perceiver-loss}

In our notation, $\mathbf{P}$, denotes the ideal (target) distribution and, $\mathbf{Q}$, the model-predicted distribution. The loss functions considered were:

\begin{itemize}
    \item \textbf{KL-divergence} as in \cref{eq:kl-divergence}.

    \item \textbf{Reverse KL-divergence}
    \begin{equation}
        \mathrm{KL}(\mathbf{Q},\mathbf{P})
        = \sum_{x \in X} \mathbf{Q}(x)\log\!\left(\frac{\mathbf{Q}(x)}{\mathbf{P}(x)}\right).
        \label{eq:loss-rkl}
    \end{equation}

    \item \textbf{Jensen-Shannon (JS)-divergence}
    \begin{equation}
        \mathrm{JS}(\mathbf{P},\mathbf{Q})
        = \tfrac{1}{2}\mathrm{KL}(\mathbf{P},\mathbf{M})
        + \tfrac{1}{2}\mathrm{KL}(\mathbf{Q},\mathbf{M}),
        \quad
        \mathbf{M}(x)=\tfrac{1}{2}\big(\mathbf{P}(x)+\mathbf{Q}(x)\big).
        \label{eq:loss-js}
    \end{equation}

    \item \textbf{Cross-entropy (soft version)}
    \begin{equation}
        \mathrm{CE}(\mathbf{P},\mathbf{Q})
        = -\sum_{x \in X} \mathbf{P}(x)\log \mathbf{Q}(x).
        \label{eq:loss-ce}
    \end{equation}

    \item \textbf{Mean square error (MSE)}
    \begin{equation}
        \mathrm{MSE}(\mathbf{P},\mathbf{Q})
        = \sum_{x \in X}\big(\mathbf{P}(x)-\mathbf{Q}(x)\big)^2.
        \label{eq:loss-mse}
    \end{equation}

    \item \textbf{Hellinger distance (squared form)}
    \begin{equation}
        \mathrm{HD}(\mathbf{P},\mathbf{Q})
        = \tfrac{1}{2}\sum_{x \in X}\Big(\sqrt{\mathbf{P}(x)}-\sqrt{\mathbf{Q}(x)}\Big)^2.
        \label{eq:loss-hellinger}
    \end{equation}

    \item \textbf{Hybrid KL+L1 loss}
    \begin{equation}
        \mathrm{KLpL1}(\mathbf{P},\mathbf{Q})
        = \mathrm{KL}(\mathbf{P},\mathbf{Q})
        + \lambda \sum_{x \in X}\big|\mathbf{P}(x)-\mathbf{Q}(x)\big|,
        \quad \lambda = 0.1.
        \label{eq:loss-kl-l1}
    \end{equation}
\end{itemize}

Additionally, all variants optionally included an \textbf{entropy bonus term}
\begin{equation}
    \mathcal{L} \leftarrow \mathcal{L} - \beta\,H(\mathbf{Q}),
    \quad
    H(\mathbf{Q}) = -\sum_{x \in X}\mathbf{Q}(x)\log \mathbf{Q}(x),
    \label{eq:loss-entropy-bonus}
\end{equation}
which encourages output diversity and prevents mode collapse during training (although in our experiments $\beta=0$ was generally sufficient).

Each loss embodies a different statistical hypothesis about how the noisy and ideal distributions differ:
KL-based measures assume a consistent support with shifted probabilities, while Hellinger and MSE treat deviations symmetrically and can tolerate partial support mismatch. Cross-entropy provides a compromise between these extremes.  
The hybrid KL+L1 formulation empirically achieves consistent performance across both Pauli and Random datasets.

\begin{figure}
    \centering
    \includegraphics[width=0.9\linewidth]{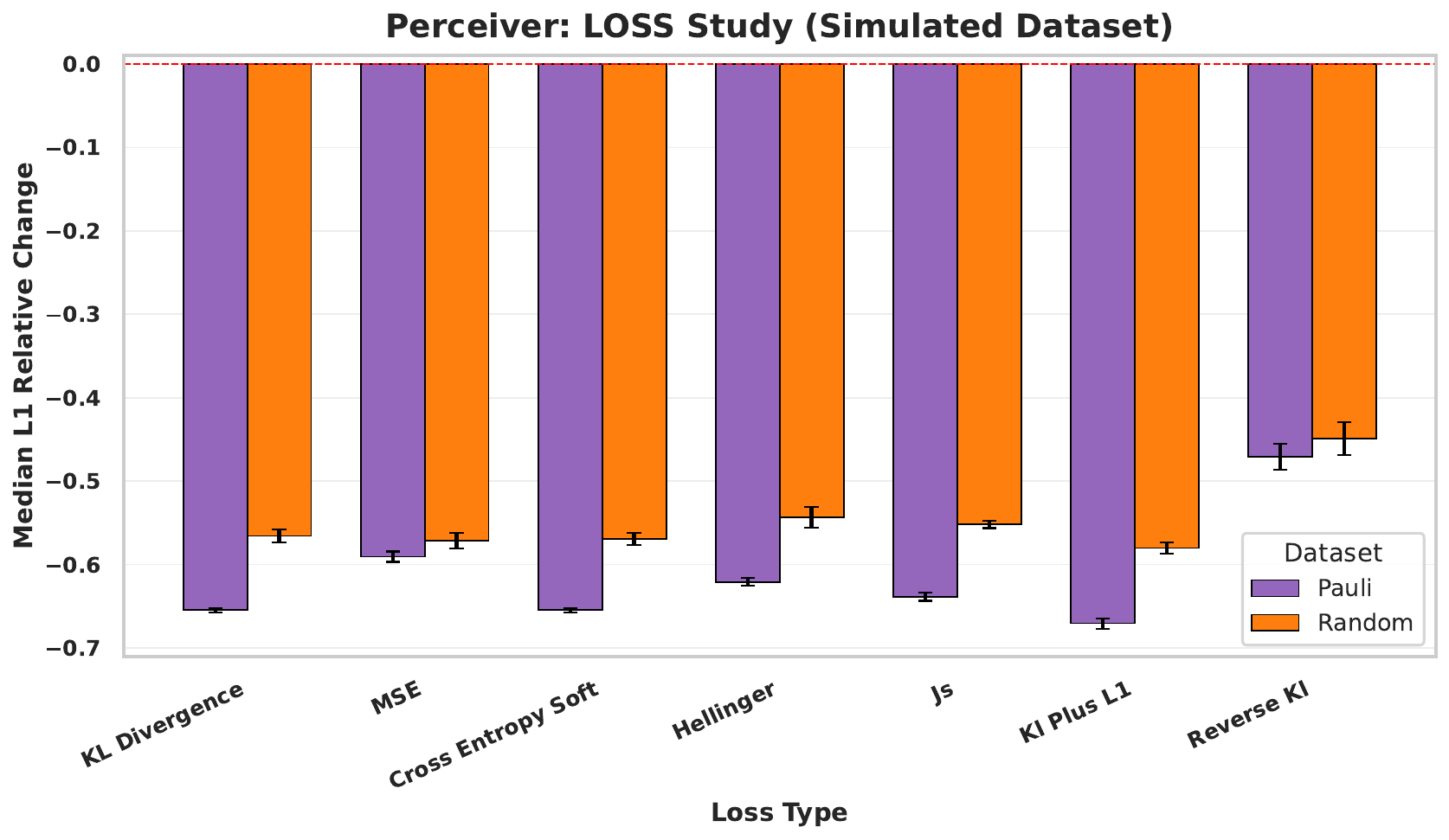}
    \caption{\textbf{Perceiver comparison (Simulated data) for different loss functions across Pauli and Random datasets.} 
    Bars give the median over 5 seeds. Error bars indicating  $\pm 1$ standard deviation over the median performance of the 5 seeds. Lower values indicate better performance.}
    \label{fig:perceiver-loss-sim}
\end{figure}

\begin{figure}
    \centering
    \includegraphics[width=0.9\linewidth]{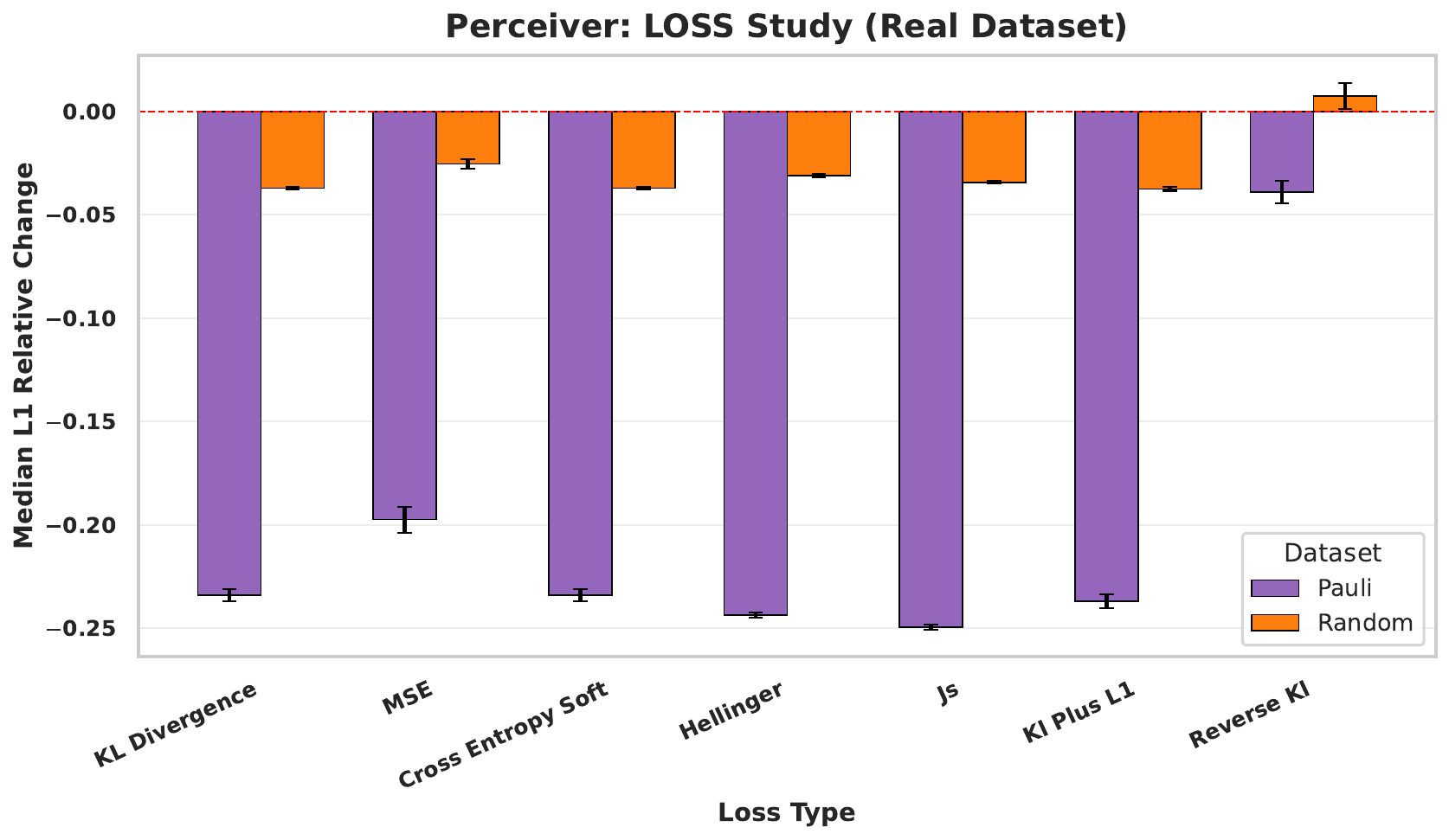}
    \caption{\textbf{Perceiver comparison (Real data) for different loss functions across Pauli and Random datasets.} 
    Bars give the median over 5 seeds. Error bars indicating  $\pm 1$ standard deviation over the median performance of the 5 seeds. Lower values indicate better performance.}
    \label{fig:perceiver-loss-real}
\end{figure}

\Cref{fig:perceiver-loss-sim,fig:perceiver-loss-real} give the resulting median L1 Relative Change for Simulated and Real datasets, respectively. Across all configurations, the Perceiver is insensitive to the type of loss function -- an indication that its latent attention blocks are potentially better at capturing intrinsic correlations.  
Information-theoretic losses (KL-divergence and JS-divergence) offered the best trade-off between convergence speed and final accuracy.

\subsection{Training on Mixed Datasets}
\label{app:perceiver-mix}

\begin{figure}
    \centering
    \includegraphics[width=0.5\linewidth]{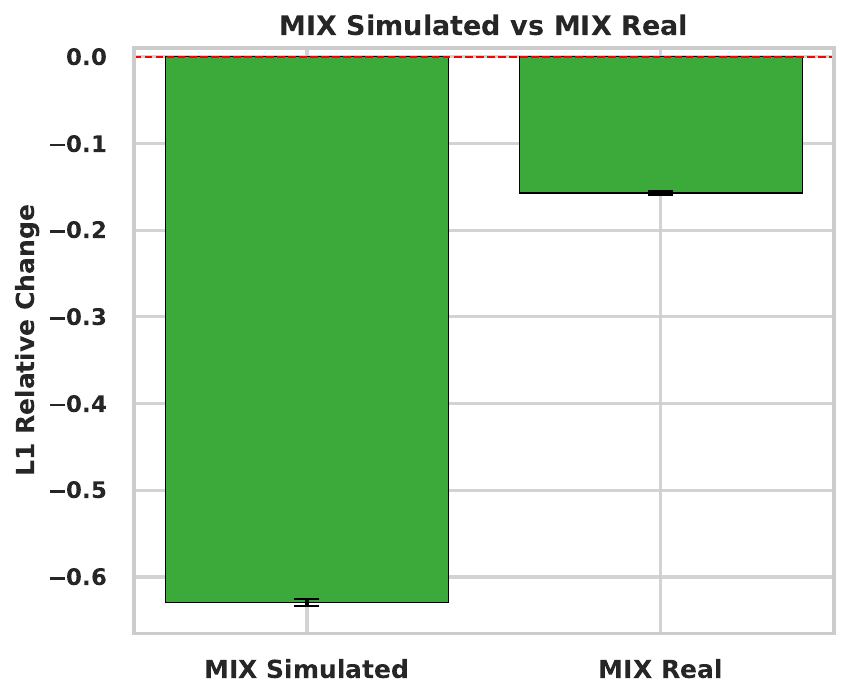}
    \caption{\textbf{Perceiver performance on mixed-dataset training.} 
    Median L1 Relative Change (over 5 seeds each) for the Perceiver trained jointly on both Pauli and Random datasets, then evaluated separately for Simulated and Real data. Error bars indicating  $\pm 1$ standard deviation over the median performance of the 5 seeds. 
    Bars below ($y = 0$) indicate successful mitigation.}
    \label{fig:perceiver-mix}
\end{figure}

We next assessed whether the Perceiver could generalize across datasets of differing circuit structure.
In this experiment, a single model was trained jointly on both Pauli and Random data, mixing the respective training sets.  
The results in \cref{fig:perceiver-mix} show that this mixed-dataset training has comparable performance to the models trained on the separate datasets ($\sim-0.6$ for Simulated circuits as in \cref{tab:simulated-table}, and similar to the $\sim -0.2 / -0.01$ given in \cref{tab:performance_results_algiers} for Real circuits), suggesting that the Perceiver can exploit shared structure between the two circuit classes.  
This indicates potential for future work on unified  models that can learn a device-wide noise representation. Notably, in this example the datasets were not differentiated at input time in a supervised manner, thus future studies might significantly improve on this.

\subsection{Generalization across Circuit Depths}
\label{app:perceiver-depth}

\begin{figure}[!ht]
    \centering
    \includegraphics[width=1\linewidth]{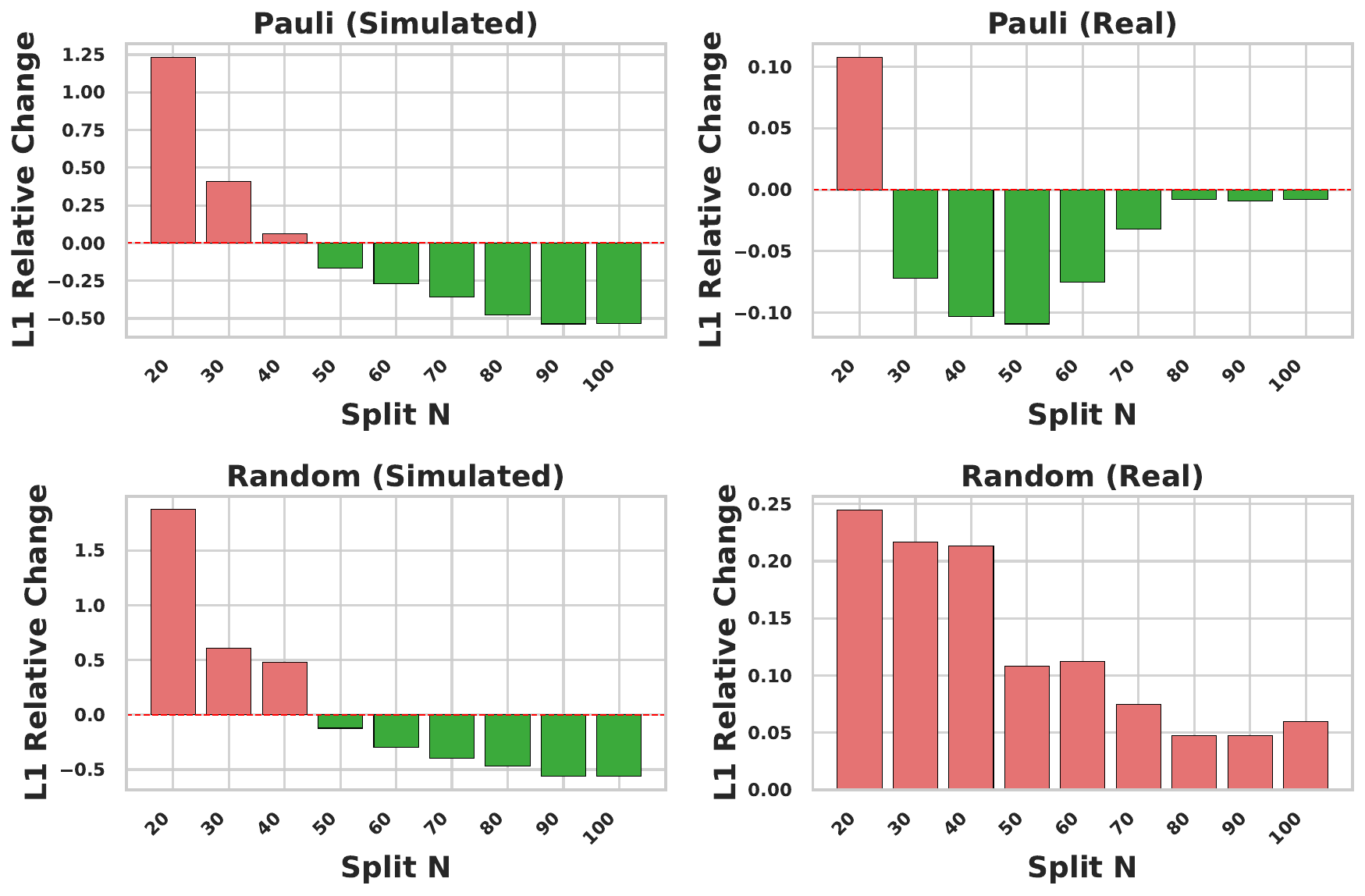}
    \caption{\textbf{Perceiver depth-generalization study.}
    Median L1 Relative Change as a function of maximum training depth (\textit{Split N}) on one fixed seed. 
    Negative values indicate successful mitigation. Splits that are worse than the baseline performance are given in red, whilst those that give better performance are given in green.}
    \label{fig:perceiver-depth}
\end{figure}

Finally, we analysed the effect of restricting training to circuits below a certain depth, $d_\text{train}$, and evaluating on deeper unseen circuits ($d_\text{test} > d_\text{train}$).  
\Cref{fig:perceiver-depth} summarizes the median L1 Relative Change as a function of the training-split threshold, for both Simulated and Real data.

For Pauli data, the model maintains performance even when trained only up to moderate depths ($d_\text{train} \sim 60$), indicating strong depth-wise generalization for Simulated data, while for Real data it peaks at $d_\text{train} \sim 50$, saturating shortly after, likely due to increased hardware noise at higher depths.  
In contrast, Random circuits are poorer at extrapolating, with the performance improving only when the model is exposed to similar-depth data during training. 
This highlights the fact that the Perceiver captures circuit-structure–dependent features that transfer well across depths when those structures are regular (as in Pauli circuits), but less so when distributions contain less inherent structure.

\end{document}